\documentclass[11pt]{article}

\usepackage{hyperref}

\usepackage{booktabs}
\usepackage{amsmath}
\usepackage{amssymb}
\usepackage{hhline}
\usepackage[scale={.75,.75}]{geometry}
\usepackage{url}
\usepackage{caption}
\usepackage[numbers, sort&compress]{natbib}
\usepackage{epsfig}
\usepackage{multirow}
\usepackage{color}
\usepackage{verbatim}
\usepackage{nicefrac}
\usepackage{upgreek}
\usepackage{bbm}

\renewcommand{\Re}{\text{Re}}
\renewcommand{\Im}{\text{Im}}

\newcommand{\p}{\textbf{p}}

\newcommand{\Slash}[1]{\displaystyle{\not}{#1}}
\newcommand{\slashed}[1]{\displaystyle{\not}{#1}}
\newcommand{\rmi}{{\rm I}}
\newcommand{\msol}{m_{\rm sol}}
\newcommand{\matm}{m_{\rm atm}}

\begin{document}
\date{\mbox{ }}
\title{ 
{\normalsize TTK-12-05, TUM-HEP-852/12\hfill\mbox{}\\}
\vspace{0.5cm}
\bf \boldmath 
Dark Matter, Baryogenesis and Neutrino Oscillations from Right Handed
Neutrinos
\\[8mm]}

\author{Laurent Canetti$^{a}$, 
Marco Drewes$^{b,c}$, 
Tibor Frossard$^{d}$,  
Mikhail Shaposhnikov$^{a}$\\[8mm]
{\normalsize \it$^a$ ITP, EPFL,
CH-1015 Lausanne, Switzerland}\\
{\normalsize \it$^b$ Institut f\"ur Theoretische Teilchenphysik und Kosmologie,} \\{\normalsize \it RWTH Aachen, D-52056 Aachen, Germany}\\
{\normalsize \it$^c$ Physik Department T31, Technische Universit\"at M\"unchen,} \\{\normalsize \it James Franck Stra\ss e 1, D-85748 Garching, Germany}\\
{\normalsize \it$^d$ 
Max-Planck-Institut f\"ur Kernphysik, Saupfercheckweg 1, 69117 Heidelberg, Germany
}
}
\maketitle

\thispagestyle{empty}

\begin{abstract}
  \noindent
We show that, leaving aside accelerated cosmic expansion, all
experimental data in high energy physics that  are commonly agreed to
require physics beyond the Standard Model can be explained when
completing it by three right handed neutrinos that can be searched for
using {\em current day experimental techniques}.  The model that
realizes this scenario is known as Neutrino Minimal Standard Model
($\nu$MSM). In this article we give a comprehensive summary of all
known constraints in the $\nu$MSM, along with a pedagogical
introduction to the model. We present the first complete quantitative
study of the parameter space of the model  where no physics beyond the
$\nu$MSM is needed to simultaneously explain neutrino oscillations,
dark matter and the baryon asymmetry of the universe. This requires to
track the time evolution of left and right handed neutrino abundances
from hot big bang initial conditions down to temperatures below the
QCD scale.  We find that the interplay of resonant amplifications,
CP-violating flavor oscillations, scatterings and decays leads to a number of
previously unknown constraints on the sterile neutrino properties. We
furthermore re-analyze bounds from past collider experiments and big
bang nucleosynthesis in the face of recent evidence for a non-zero
neutrino mixing angle $\uptheta_{13}$.  We combine all our results
with existing constraints on dark matter properties from astrophysics
and cosmology. Our results provide a guideline for future experimental
searches for sterile neutrinos.  A summary of the constraints on
sterile neutrino masses and mixings has appeared in
\cite{Canetti:2012vf}. In this article we provide all details of our
calculations and give constraints on other model parameters.
\end{abstract}
\newpage
\begin{small}
\tableofcontents
\end{small}
\newpage

\section{Introduction}
\label{sec:intro}
The Standard Model of particle physics (SM), together with the theory
of general relativity (GR), allows to explain almost all phenomena
observed in nature in terms of a small number of underlying principles
- Poincar\'e invariance, gauge invariance and quantum mechanics - and
a handful of numbers.  In the SM these are $19$ free parameters that
can be chosen as three masses for the charged leptons, six masses,
three mixing angles and one CP violating phase for the quarks, three
gauge couplings, two parameters in the scalar potential and the QCD
vacuum angle. Three leptons, the neutrinos, remain massless in the SM
and appear only with left handed chirality. GR adds another two
parameters to the barcode of nature, the Planck mass and the
cosmological constant.

Despite its enormous success, we know for sure that the above is not a
complete theory of nature for two reasons\footnote{We do not address
theoretical issues of ``aesthetic'' nature such as fine tuning in the
context of the hierarchy problem, the strong CP problem and the
flavor structure. They may be interpreted as hints for new physics,
but could also simply represent nature's choice of parameters.}. On
one hand, it treats gravity as a classical background for the SM,
which is a quantum field theory. Such description necessarily breaks
down at energies near the Planck scale $M_P$ and has to  be replaced
by a theory of quantum gravity. We do not address this problem here,
which is of little relevance for current and near-future experiments. 
On the other hand, the SM fails to explain a number of experimental
facts. These are neutrino oscillations, the observed baryon asymmetry
of the universe (BAU), the observed dark matter (DM) and the
accelerated expansion of the universe today. In addition there is a
number of cosmological problems (e.g. flatness and horizon problem).
These can be explained by cosmic inflation, another phase of
accelerated expansion in the universe's very early history, for which
the SM also cannot provide a mechanism.  To date, these are the only
confirmed empirical proofs of physics beyond the SM\footnote{We leave
aside all experimental and observational anomalies that have not lead
to a claim of detection of new physics, i.e. may be explained within
the SM or by systematic errors.  This includes the long standing
problem of the muon magnetic moment, the inconclusive results of
different direct DM searches as well as various anomalies of limited
statistical significance.}.  In this article we argue that, leaving aside accelerated cosmic expansion, all of them
may be explained by adding three right handed (sterile) neutrinos to
the SM that can be found in experiments.

The model in which this possibility can be realized is known as {\it
Neutrino minimal Standard Model} ($\nu$MSM)
\cite{Asaka:2005an,Asaka:2005pn}. The $\nu$MSM is an extension of the
SM that aims to explain all experimental data with only minimal
modifications. This in particular means that there is no modification
of the gauge group, the number of fermion families remains unchanged
and no new energy scale above the Fermi scale is
introduced\footnote{Because of this the technical hierarchy problem
may be absent in the $\nu$MSM because no new states with energies
between the electroweak and the Planck scale are required
\cite{Bardeen:1995kv,Shaposhnikov:2007nj}.}.  The matter content is,
in comparison to the SM, complemented by three right handed
counterparts to the observed neutrinos. These are singlet under all
gauge interactions.  Over the past years, different aspects of the
$\nu$MSM have been explored using cosmological, astrophysical and
experimental constraints
\cite{Asaka:2005pn,Shaposhnikov:2008pf,Asaka:2005an,Boyarsky:2005us,
Asaka:2006rw,Bezrukov:2005mx,Shaposhnikov:2006nn,Asaka:2006ek,Boyarsky:2006zi,Boyarsky:2006fg,Boyarsky:2006jm,Shaposhnikov:2006xi,
Gorbunov:2007ak,Gorbunov:2007zz,Bezrukov:2007qz,Laine:2008pg,
Anisimov:2008qs,Boyarsky:2009ix,Canetti:2010aw,Asaka:2010kk,
Asaka:2006nq,Gorkavenko:2009vd,Asaka:2011pb,Ruchayskiy:2011aa,
Gorkavenko:2012mj,Ruchayskiy:2012si,Canetti:2012vf,Drewes:2012ma}. 
Moreover, it was suggested that cosmic inflation 
\cite{Bezrukov:2007ep,Bezrukov:2008ut,GarciaBellido:2008ab} and the
current accelerated expansion \cite{Shaposhnikov:2008xb,
Shaposhnikov:2008xi,GarciaBellido:2011de,Bezrukov:2011sz} may also be accommodated in
this framework by modifications in the gravitational sector, which we
will not discuss here\footnote{Inflation can be realized without modification of the gravitational interaction by adding an extra scalar to the $\nu$MSM \cite{Bezrukov:2009yw} (see also \cite{Shaposhnikov:2006xi,Kusenko:2006rh,Petraki:2007gq}). This inflaton can be light enough to be detected in direct searches.}.  
However, though the abundances of dark and baryonic matter have been estimated individually in the framework of the $\nu$MSM,
to date it has not been verified 
that there is a range of right handed neutrino parameters for which they can be explained {\it simultaneously}, in particular for
experimentally accessible sterile neutrinos. In this article
we present detailed results of the first complete quantitative study to identify the range of parameters that allows to simultaneously explain neutrino oscillations, the observed DM density $\Omega_{DM}$ and the observed BAU \cite{Canetti:2012zc}, responsible for today's remnant baryonic density $\Omega_B$.  
We in the following refer to this situation, in which no physics beyond the $\nu$MSM is required to explain these phenomena, as \textbf{scenario I}. In this scenario DM is made of one of the right handed neutrinos, while the other two are responsible for baryogenesis and the generation of active neutrino masses. 
We also study systematically how the constraints relax if one allows the sterile neutrinos that compose DM to be produced by some mechanism beyond the $\nu$MSM (\textbf{scenario II}).
Finally, we briefly comment on a {\textbf{scenario III}, in which the $\nu$MSM is a theory of baryogenesis and neutrino oscillations only, with no relation to DM. A more precise definition of these scenarios is given in section \ref{sec:scenarios}.
Only scenarios I and II are studied in this article, which is devoted to the $\nu$MSM as the common origin of DM, neutrino masses and the BAU.
While scenario II has previously been studied in 
 \cite{Canetti:2010aw}, the constraints coming from the requirement to thermally produce the observed $\Omega_{DM}$ in scenario I are calculated for the first time in this work.
We combine our results with bounds coming from big bang nucleosynthesis (BBN) and direct searches for sterile neutrinos, which we re-derived in the face of recent data from neutrino experiments (in particular $\uptheta_{13}\neq 0$).

Centerpiece of our analysis is the study of all lepton numbers
throughout the evolution of the early universe. As will be explained
below, in the $\nu$MSM lepton asymmetries are crucial for both,
baryogenesis and DM production.  We determine the time evolution of
left and right handed neutrino abundances for a wide range of sterile
neutrino parameters from hot big bang initial conditions at
temperatures $T\gg T_{EW}\sim 200$ GeV down to temperatures below the
QCD scale by means of effective kinetic equations. They incorporate
various effects, including thermal production of sterile neutrinos
from the primordial plasma, coherent oscillations, back reaction,
washouts, resonant amplifications, decoherence, finite temperature
corrections to the neutrino properties and the change in effective number of
degrees of freedom in the SM background.  Many of these were only
roughly estimated or completely neglected in previous studies. The
various different time scales appearing in the problem make an
analytic treatment or the use of a single CP-violating parameter
impossible in most of the parameter space. Most of our results are obtained numerically. However, the
parametric dependence on the experimentally relevant parameters
(sterile neutrino masses and mixings) can be understood in a simple
way. Furthermore, we discover a number of tuning conditions that can
be understood analytically and allow to reduce the dimensionality of
the parameter space. 

We find that there exists a considerable fraction of the $\nu$MSM
parameter space in which the model can simultaneously explain neutrino
oscillations, dark matter and the baryon asymmetry of the universe.
This includes a range of masses and couplings for which the right
handed neutrinos can be found in laboratory experiments
\cite{Gorbunov:2007ak}.
The main results of our study, constraints on sterile neutrino masses and mixings, have previously been presented in \cite{Canetti:2012vf}. In this article we give details of our
calculation and constraints on other model parameters, which are not discussed in \cite{Canetti:2012vf}. 

The remainder of this article is organized as follows. In Section
\ref{section:numsm} we overview the $\nu$MSM, its parametrization, and
describe the universe history in its framework, including baryogenesis
and dark matter production. In Section \ref{ExistingBounds} we discuss
different experimental and cosmological bounds on the properties of
right-handed neutrinos in the $\nu$MSM. In Section  \ref{SectionKinEq}
we formulate the kinetic equations which are used to follow the time
evolution of sterile neutrinos and active neutrino flavors in the
early universe. 
In Section \ref{baryogenesissection} we present our results on baryogenesis in scenario II.
In Section \ref{DMproductionSection} we study the generation of lepton asymmetries at late times, essential for thermal dark matter production in the $\nu$MSM. 
In Section \ref{Combined} we combine the constraints of the two previous Sections and define the region of parameters where scenario I can be realized, i.e. the $\nu$MSM
explains simultaneously neutrino masses and oscillations, dark matter,
and baryon asymmetry of the universe. In Section \ref{section:concl}
we present our conclusions. In a number of  appendices we give 
technical details on kinetic equations (\ref{HeffSec}), on the parametrization of the $\nu$MSM Lagrangian (\ref{DiracConnection}), on different notations to describe lepton asymmetries (\ref{asymmconv})  and on the decay rates of
sterile neutrinos (\ref{DecayRatesAppendix}).


\section{The $\nu$MSM}
\label{section:numsm}
The $\nu$MSM is described by the Lagrangian
\begin{eqnarray}
	\label{nuMSM_lagrangian}
	\mathcal{L}_{\nu MSM} =\mathcal{L}_{SM}+ 
	i \overline{\nu_{R}}\slashed{\partial}\nu_{R}-
	\overline{L_{L}}F\nu_{R}\tilde{\Phi} -
	\overline{\nu_{R}}F^{\dagger}L_L\tilde{\Phi}^{\dagger} \nonumber\\ 
	-{\rm \frac{1}{2}}(\overline{\nu_R^c}M_{M}\nu_{R} 
	+\overline{\nu_{R}}M_{M}^{\dagger}\nu^c_{R}). 
	\end{eqnarray}
Here we have suppressed flavor and isospin indices.
$\mathcal{L}_{SM}$ is the Lagrangian of the SM. $F$ is a matrix of
Yukawa couplings and $M_{M}$ a Majorana mass term for the right handed
neutrinos $\nu_{R}$. $L_{L}=(\nu_{L},e_{L})^{T}$ are the left handed
lepton doublets in the SM and $\Phi$ is the Higgs doublet. We chose a
basis where the charged lepton Yukawa couplings and $M_{M}$ are
diagonal. The Lagrangian (\ref{nuMSM_lagrangian}) is well-known in the
context of the seesaw mechanism for neutrino masses
\cite{Minkowski:1977sc} and leptogenesis \cite{Fukugita:1986hr}. While
the eigenvalues of $M_M$ in most models are related to an energy scale
far above the electroweak scale, it is a defining assumption of the
$\nu$MSM that the observational data can be explained without
involvement of any new scale above the Fermi one.


\subsection{Mass- and Flavor Eigenstates}
For temperatures $T<M_W$ below the mass of the W-boson we can in good
approximation replace the Higgs field $\Phi$ by its vacuum expectation
value $v=174$ GeV. Then (\ref{nuMSM_lagrangian}) can be written as
\begin{eqnarray}
	\label{Lagrangian_flavour_base}
	\mathcal{L} = \mathcal{L}_{SM}+i\overline{\nu_{R}}_{,I}
	\slashed{\partial}\nu_{R,I} - (m_D)_{\alpha I} \overline{\nu_{L}}_{,\alpha} \nu_{R, I} - (m_D^{*})_{\alpha I}
	\overline{\nu_{R}}_{,I} \nu_{L, \alpha} \nonumber\\- {\small
	\frac{1}{2}}((M_{M})_{IJ} \overline{\nu_{R}^c}_{,I} \nu_{R,J}
	+  (M_{M})^{*}_{IJ} \overline{\nu_{R}}_{,I} \nu_{R,J}^c)
	\end{eqnarray}
with the Dirac mass matrix $m_{D}=Fv$.  When the eigenvalues of $M_M$
are much larger than those of $m_{D}$, the seesaw mechanism naturally
leads to light active and heavy sterile neutrinos. This hierarchy is
realized in the $\nu$MSM.

In vacuum there are two sets of mass eigenstates; on one hand  {\it
active neutrinos} $\upnu_i$ with masses $m_i$, which are mainly
mixings of the SU(2) charged fields $\nu_L$, 
\begin{equation}
P_L\upnu_{i}=\left(U_\nu^{\dagger}\left(\left(1-
\frac{1}{2}\theta\theta^{\dagger}\right)\nu_L-
\theta\nu_{R}^c\right)\right)_i,
\end{equation}
with $\theta_{\alpha I}=(m_D M_M^{-1})_{\alpha I}$, and on the other
hand {\it sterile neutrinos}\footnote{In \cite{Shaposhnikov:2008pf}
the notation is slightly different and the letter ``$N_I$'' does not
denote mass eigenstates.}  $N_{I}$ with masses $M_I$, which are mainly
mixings of the singlet fields $\nu_R$,
\begin{equation}
\label{NIdef}
P_{R}N_{I}=\left(U_N^{\dagger}\left(\left(\mathbbm{1}-
\frac{1}{2}\theta^{T}\theta^{*}\right)\nu_{R}+
\theta^{T}\nu_{L}^{c}\right)\right)_{I}.
\end{equation}
Here $P_{R,L}$ are chiral projectors and $N_I$ ($\upnu_i$) are
Majorana spinors, the left chiral (right chiral) part of which is
fixed by the Majorana relations $N_I^c=N_I$ and $\upnu_i=\upnu_i^c$. The
matrix $U_N$ diagonalises the sterile neutrino mass matrix $M_N$
defined below. The entries of the matrix $\theta$ determine the
active-sterile mixing angles.

The neutrino mass matrix can be block diagonalized. At leading order
in the Yukawa couplings $F$ one obtains the mass matrices
\begin{eqnarray}
		 m_{\nu} &=& -\theta M_M \theta^T~,
\label{activeneutrinomasses}\\  M_N &=& M_M + \frac{1}{2}
\big(\theta^{\dagger} \theta M_M + M_M^T \theta^T \theta^{*}
\big)~.
\label{sterileneutrinomassesAll}
\end{eqnarray}
The mass matrices $m_{\nu}$ and $M_N$ are not diagonal and lead to
neutrino oscillations. While there is very little mixing between
active and sterile flavors at all temperatures of interest, the
oscillations between sterile neutrinos can be essential for the
generation of a lepton asymmetry.  $m_{\nu}$ can be parameterized in
the usual way by active neutrino masses, mixing angles and phases,
$m_{\nu}=U_\nu{\rm diag}(m_1,m_2,m_3)U_\nu^{T}$. In the basis where
the charged lepton Yukawas are diagonal, $U_\nu$ is identical to the
Pontecorvo-Maki-Nakagawa-Sakata (PMNS) lepton mixing matrix.

The physical sterile neutrino masses $M_I$ are given by the eigenvalues of $M_N^\dagger M_N$.
In the seesaw limit $M_N$ is almost diagonal and they are very close to the entries of $M_M$. We nevertheless
need to keep terms $\mathcal{O}(\theta^2)$ because the masses $M_2$
and $M_3$ are degenerate in the $\nu$MSM, see section
\ref{finetunings}, and the mixing of the sterile neutrinos $N_{2,3}$
amongst each other may be large despite the
seesaw-hierarchy\footnote{It turns out that the region where $U_N$ is
close to identity phenomenologically is the most interesting, see
section \ref{finetunings}.}. This mixing is given by the matrix $U_N$, which can be seen as
analogue to $U_\nu$. It is worth noting that due to
(\ref{sterileneutrinomassesAll}) the matrix $U_N$ is real at this
order in $F$.  The experimentally relevant coupling between active and
sterile species is given by the matrix $\Theta$ with\footnote{The fact
that matrix appearing in (\ref{NIdef}) is
$U_N^\dagger\theta^T=(\theta^* U_N)^\dagger$ rather than
$\Theta^\dagger=U_N^\dagger\theta^\dagger$ is due to the fact that the
$N_I$ couple to $\nu_{L,\alpha}$, but overlap with
$\nu_{L,\alpha}^c$.} 
\begin{equation}
\label{ThetaDef}
\Theta_{\alpha I}\equiv(\theta U_N)_{\alpha I} = (m_D M_M^{-1}
U_N)_{\alpha I}.
\end{equation}
In practice, experiments to date cannot distinguish the sterile
flavors and are only sensitive to the quantities
\begin{equation}
U_\alpha^2\equiv\sum_I \Theta_{\alpha I}\Theta_{\alpha I}^*=\sum_I
\theta_{\alpha I}\theta_{\alpha I}^*.\label{UalphaDef}
\end{equation} 
Therefore $U_N$, and hence the sterile-sterile mixing and the coupling of individual sterile flavors to the SM, cannot be probed in direct searches.

\subsection{Benchmark Scenarios}\label{sec:scenarios}
The notation introduced above allows to define the scenarios I-III introduced in the introduction more precisely.
\begin{itemize}
\item In \textbf{scenario I} no physics beyond the $\nu$MSM is needed to explain the observed $\Omega_{DM}$, neutrino masses and $\Omega_B$.
DM is composed of thermally produced sterile neutrinos $N_1$. $N_2$ and $N_3$ generate active neutrino masses via the seesaw mechanism, and their CP-violating oscillations produce lepton asymmetries in the early universe. 
The effect of $N_1$ on neutrino masses and lepton asymmetry generation is negligible because its Yukawa couplings $F_{\alpha 1}$ are constrained to be tiny by the requirement to be a viable DM candidate, c.f. section \ref{subsubDmbounds}.
The lepton asymmetries produced by $N_{2,3}$ are crucial on two occasions in the history of the universe: On one hand the asymmetries generated at early times ($T\gtrsim 140$ GeV) are responsible for the generation of a BAU via flavored leptogenesis, on the other hand the late time asymmetries ($T\sim 100$ MeV) strongly affect the rate of thermal $N_1$ production.
Due to the latter the requirement to produce the observed $\Omega_{DM}$ imposes indirect constraints on the particles $N_{2,3}$.
There are determined in sections \ref{DMproductionSection} and \ref{Combined} and form the main result of our study.
\item In \textbf{scenario II} the roles of $N_{2,3}$ and $N_1$ are the same as in scenario I, but we assume that DM was produced by some unknown mechanism beyond the $\nu$MSM.
The astrophysical constraints on the $N_1$ mass and coupling equal those in scenario I. $N_{2,3}$ are again required to generate the active neutrino masses via the seesaw mechanism and to produce sufficient flavored lepton asymmetries at $T\sim 140$ MeV to explain the BAU. However, there is no need for a large late time asymmetry. This considerably relaxes the bounds on $N_{2,3}$.
Scenario II is studied in detail in section \ref{baryogenesissection}.
\item In \textbf{scenario III} the $\nu$MSM is not required to explain DM, i.e. it is considered to be a theory of neutrino masses and low energy leptogenesis only. Then all three $N_I$ can participate in the generation of lepton asymmetries.
This makes the parameter space for baryogenesis considerably bigger than in scenarios I and II, including new sources of CP violation. We do not study scenario III in this work, some aspects are discussed in \cite{Drewes:2012ma}.
\end{itemize}

\subsection{Effective Theory of Lepton Number Generation}
In scenarios I and II the lightest sterile neutrino $N_1$ is a DM candidate.
In this article we focus on those two scenarios.
If $N_1$ is required to compose all
observed DM, its mass $M_1$ and mixing are constrained by
observational data, see section \ref{ExistingBounds}. Its mixing is so
small that its effect on the active neutrino masses is negligible.  Note
that this implies that one active neutrino is much lighter than the
others (with mass smaller than ${\cal O}(10^{-5})$ eV
\cite{Asaka:2005an}). Finding three massive active neutrinos with
degenerate spectrum would exclude the $\nu$MSM with three sterile
neutrinos as common and only origin of active neutrino oscillations,
dark matter and baryogenesis.  $N_1$ also does not contribute
significantly to the production of a lepton asymmetry at any time. 
This process can therefore be described in an effective theory with
only two sterile flavors $N_{2,3}$.  In the following we will almost
exclusively work in this framework.  To simplify the notation, we will
use the symbols $M_N$ and $U_N$ for both, the full ($3\times3$) mass
matrix and mixing matrices defined above and the ($2\times2$ and
$3\times2$) sub-matrices that only involve the sterile flavors
$I=2,3$, which appear in the effective theory. The mixing between
$N_1$ and $N_{2,3}$ is negligible due to the smallness of $F_{\alpha 1}$, which is enforced by the seesaw relation (\ref{activeneutrinomasses}) and the observational bounds on $M_1$ summarized in Section \ref{subsubDmbounds}.  The effective $N_{2,3}$ mass matrix can be written as
\begin{eqnarray}
M_N&=&M\mathbbm{1}_{2 \times 2} + \Delta M\sigma_3+M^{-1}{\rm
Re}(m_{D}^{\dagger}m_{D})\label{sterileneutrinomasses},
\end{eqnarray}
where $\sigma_3$ is the third Pauli matrix and we chose the
parameterization $M_M={\rm diag}(M-\Delta M,M+\Delta M)$. This
equality holds because we chose $M_M$ real and diagonal.  The physical
masses $M_2$ and $M_3$ are given by the eigenvalues of $M_N$. They
read
\begin{eqnarray}
M_{2,3}&=&\bar{M}\pm\delta M\label{M23}\\
\bar{M}&=&M+\frac{1}{2M}{\rm Re}\left({\rm tr}\left(m_{D}^{\dagger}m_{D}\right)\right)\label{Mbar}\\
(\delta M)^2 &=&  
\left( \frac{1}{2M} \left({\rm Re}\left(m^{\dagger}_{D} m_{D}\right)_{33} - {\rm Re}\left(m^{\dagger}_D m_D\right)_{22} \right) + \Delta M \right)^2 
+ \frac{1}{M^2} {\rm Re}\left(m^{\dagger}_D m_D\right)^{2}_{23}
.\nonumber\\\label{deltaM}
\end{eqnarray}
For all parameter choices we are interested in $\bar{M}\simeq M$ holds
in very good approximation.  The masses $M_{2,3}$ are too big to be
sensitive to loop corrections. In contrast, the splitting $\delta M$
can be considerably smaller than the size of radiative corrections to
$M_{2,3}$ \cite{Roy:2010xq}. The above expressions have a different
shape than those given in \cite{Shaposhnikov:2008pf} because we use a
different base in flavor space, see appendix \ref{DiracConnection}.

These above formulae hold for the (zero temperature) masses in the
microscopic theory. At finite temperature the system is described by a
thermodynamical ensemble, the properties of which can usually be
described in terms of quasiparticles with temperature dependent
dispersion relations. We approximate these by temperature dependent
``thermal masses''. 

\subsection{Thermal History of the Universe in the $\nu$MSM}
\label{ThermalHistory}
\begin{figure}
  \centering
    \includegraphics[width=12cm]{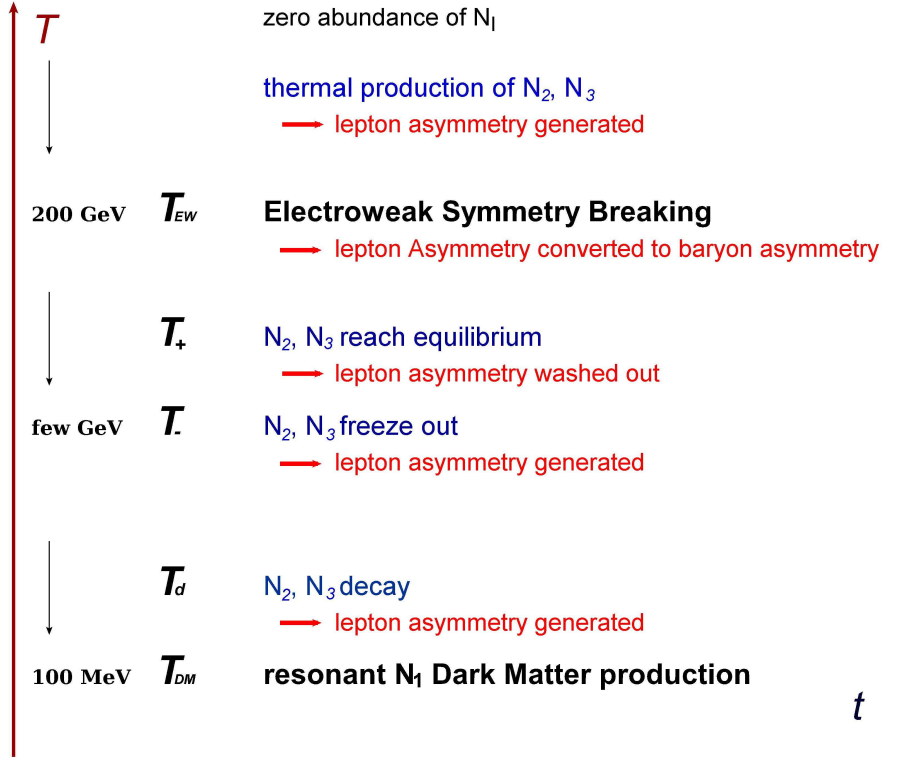}
    \caption{The thermal history of the universe in the
$\nu$MSM.\label{ThermalHistoryPic}}
\end{figure} 
Apart from the very weakly coupled sterile neutrinos, the matter
content of the $\nu$MSM is the same as that of the SM.  Therefore the
thermal history of the universe during the radiation dominated era is
similar in both models. Here we only point out the differences that
arise due to the presence of the fields $\nu_R$, see figure \ref{ThermalHistoryPic}. They couple to the SM
only via the Yukawa matrices $F$, which are constrained by the seesaw
relation. For sterile neutrino masses below the electroweak scale, the
abundances are too small to affect the entropy during the
radiation dominated era significantly.    However, the additional
sources of CP-violation contained in them have a huge effect on the
lepton chemical potentials in the plasma.

\paragraph{Baryogenesis}
The $\nu$MSM adds no new degrees  of freedom to the SM above the
electroweak scale.  As a consequence of the smallness of the Yukawa
couplings $F$, the $N_I$ are produced only in negligible amounts
during reheating \cite{Bezrukov:2008ut}.  Therefore the thermal
history for $T\gg T_{EW}$ closely resembles that in the SM\footnote{If
a non-minimal coupling of the Higgs field $\Phi$ to gravity is
introduced in the $\nu$MSM, $\Phi$ can play the role of the inflaton.
Though this way no fields are added, the thermal
history at very early times (during reheating) changes due to a
non-minimal coupling to curvature, see \cite{Bezrukov:2008ut}. Here we
assume an initial state without $N_I$ at $T\gg T_{EW}$.}. The sterile
neutrinos have to be produced thermally from the primordial plasma in
the radiation dominated epoch.  During this non-equilibirum process,
all Sakharov conditions \cite{Sakharov:1967dj} are fulfilled:  Baryon number is
violated by SM sphalerons \cite{Kuzmin:1985mm}, and the oscillations
amongst the sterile neutrinos violate CP \cite{Akhmedov:1998qx}.
Source of this CP-violation are the complex phases in the
Yukawa couplings $F_{\alpha I}$. Due to the Majorana mass $M_M$ neither
the individual (active) leptonic currents, defined in (\ref{JI}) and
(\ref{Jalpha}), nor the total lepton number are strictly conserved.
However, for $T\gg M$ the effect of the Majorana masses is negligible.
Though the neutrinos are Majorana particles, one can define
\textit{neutrinos} and \textit{antineutrinos} as the two helicity
states, transitions between which are suppressed at $T\gg M$. We will
in the following always use the terms ``neutrinos'' and
``antineutrinos'' in this sense.

In scenarios I and II the abundance of $N_{1}$ remains negligible until $T\sim 100$ MeV because of the smallness of its coupling that is required to be in accord with astrophysical bounds on DM, see Section \ref{subsubDmbounds}.
$N_{2,3}$, on the other hand, are produced efficiently in the early universe. During
this process flavored ``lepton
asymmetries'' can be generated \cite{Akhmedov:1998qx}. $N_{2,3}$ reach
equilibrium at a temperature $T_{+}$  \cite{Shaposhnikov:2008pf}. 
Though the total lepton number (\ref{JI}) at $T_+\gg M$ is very small,
there are asymmetries in the above helicity sense in the
individual active and sterile flavors. Sphalerons,
which only couple to the left chiral fields, can convert them into a
baryon asymmetry.  
The
washout of lepton asymmetries becomes efficient at $T\lesssim T_+$. It
is a necessary condition for baryogenesis that this washout has not
erased all asymmetries at $T_{EW}$, which is fulfilled for $T_+\gtrsim
T_{EW}$.
The BAU at $T\sim T_{EW}$ can be estimated by
today's baryon to photon ratio, see \cite{Canetti:2012zc} for a recent
review.  A precise value can be obtained by combining data from the
cosmic microwave background and large scale structure
\cite{Komatsu:2010fb}, 
\begin{equation}
\eta_B=(6.160\pm0.148) \cdot 10^{-10}.\label{BAUvalue}
\end{equation}
The parameter $\eta_B$ is related to the remnant density of baryons
$\Omega_B$, in units of the critical density, by
$\Omega_B\simeq\eta_B/(2.739\cdot10^{-8}h^2)$, where $h$ parameterizes
today's Hubble rate $H_0=100 h$ (km/s)/Mpc. In order to generate this
asymmetry, the effective (thermal) masses $M_2(T)$ and $M_3(T)$ of the
sterile neutrinos in the plasma need to be quasi-degenerate at
$T\gtrsim T_{EW}$, see section \ref{finetunings}.

After $N_2$ and $N_3$ reach equilibrium, the lepton asymmetries are
washed out. This washout takes longer than the kinetic equilibration,
but it was been estimated in \cite{Shaposhnikov:2008pf} that no
asymmetries survive until $N_{2,3}$-freezeout at $T=T_-$.  In
\cite{Boyarsky:2011uy} it has been suggested that some asymmetry may
be protected from this washout by the chiral anomaly, which transfers
them into magnetic fields.   Here we take the most conservative
approach and assume that no asymmetry survives between $T_+$ and
$T_-$.  Around $T=T_{-}$, the interactions that keep $N_{2,3}$ in
equilibrium become inefficient. During the resulting freezeout the
Sakharov conditions are again fulfilled and a new asymmetries are
generated. Even later, a final contribution to the lepton asymmetries
are added when the unstable particles $N_{2,3}$ decay at a temperature
$T_d$. 

\paragraph{DM production}
The abundance of the third sterile neutrino $N_1$ in scenario I remains below
equilibrium at all times due to its small Yukawa coupling.  In absence
of chemical potentials, the thermal production of these particles
(\textit{Dodelson-Widrow mechanism} \cite{Dodelson:1993je}) is not
sufficient to explain all dark matter as relic $N_1$ abundance if the
observational bounds summarized in section \ref{ExistingBounds} are
taken into consideration.  However, in the presence of a lepton
asymmetry in the primordial plasma, the dispersion relations of active
and sterile neutrinos are modified by the Mikheyev-Smirnov-Wolfenstein
effect (MSW effect) \cite{Wolfenstein:1977ue}.  The thermal mass of
the active neutrinos can be large enough to cause a level crossing
between the dispersion relation for active and sterile flavors at
$T_{DM}$, resulting in a resonantly enhanced production of $N_1$
\cite{Laine:2008pg} (resonant or \textit{Shi-Fuller mechanism}
\cite{Shi:1998km}). This mechanism requires a lepton asymmetry
$|\mu_\alpha|\gtrsim 8\cdot 10^{-6}$ to be efficient enough to explain
the entire observed dark matter density $\Omega_{DM}$ in terms of
$N_1$ relic neutrinos \cite{Laine:2008pg}. Here we have characterized
the asymmetry by\footnote{Note that $\mu_\alpha$ is not a chemical
potential, but an abundance (or yield). We chose the symbol $\mu$ for
notational consistency with \cite{Shaposhnikov:2008pf}. The relation of $\mu_\alpha$ to the lepton chemical potential $\upmu_\upalpha$ is given in appendix \ref{asymmconv}.}
\begin{equation}
\label{muDef}
\mu_\alpha=\frac{n_\alpha}{s},
\end{equation}
where $s$ is the entropy density of the universe and $n_\alpha$ the
total number density (particles minus antiparticles) of active (SM)
leptons of flavor $\alpha$.  The relations between $\mu_\alpha$
defined in (\ref{muDef}) and other ways to characterize the asymmetry
(e.g. the chemical potential) are given in appendix \ref{asymmconv}.

\paragraph{Cosmological constraints}
Thus, in scenario I there are two cosmological requirements related to the lepton
asymmetry that have to be fulfilled to produce the correct $\Omega_B$
and $\Omega_{DM}$ within the $\nu$MSM: 
\begin{enumerate}
\item[i)] $\mu_\alpha\sim 10^{-10}$ at $T_{EW}\sim 200$ GeV for successful baryogenesis  and 
\item[ii)] $|\mu_{\alpha}|> 8\cdot 10^{-6}$ at $T_{DM}$ for dark matter production. 
\end{enumerate}
In scenarios I and II the asymmetry generation in both cases relies on a resonant
amplification and quasi-degeneracy of $M_2$ and $M_3$, which we
discuss in section \ref{finetunings}. This may be considered as {\it
fine tuning}. On the other hand, the fact that the BAU (and thus the
baryonic matter density $\Omega_B$) and DM production in the SM both
rely on essentially the same mechanism may be considered as a hint for
an explanation for the apparent coincidence $\Omega_B\sim\Omega_{DM}$, though the connection is not obvious as $\Omega_B$ and $\Omega_{DM}$ also depend on other parameters.

In scenario II only the condition i) applies. The resulting constraints on the $N_{2,3}$ properties have
been studied in detail in \cite{Canetti:2010aw}.  In section
\ref{baryogenesissection} we update this analysis in the face of
recent data from neutrino experiments, in particular evidence for an
active neutrino mixing angle $\uptheta_{13}\neq 0$
\cite{Abe:2011sj,Adamson:2011qu,An:2012eh}.  In section
\ref{DMproductionSection} we include the second condition and study
which additional constraints come from the requirement
$|\mu_\alpha|>8\cdot 10^{-6} $ at $T_{DM}$.   Previous estimates
suggest $T_{DM}\sim 100$ MeV $\lesssim T_{QCD}$ \cite{Laine:2008pg}
and $T_{-}<M_{W}$ \cite{Shaposhnikov:2008pf,Boyarsky:2009ix}, where
$M_W$ is the mass of the $W$-boson and $T_{QCD}$ the temperature at
which quarks form hadrons.  

Though we are concerned with the conditions under which $N_1$ can
explain all observed dark matter, the $N_1$  will not directly enter
our analysis because the lepton asymmetry that is necessary for
resonant $N_1$ production in scenario I is created by $N_{2,3}$. Instead, we derive constraints on the properties of $N_{2,3}$, which can be searched for
in particle colliders. $N_1$, in contrast, cannot be detected directly
in the laboratory due to its small coupling. However, the $N_1$
parameter space is constrained from all sides by indirect observations
including structure formation, Ly$_\alpha$ forest,  X-rays and phase
space analysis, see section \ref{ExistingBounds}.

\subsection{Parameterization}
\label{Parametrisationsection}
Adding $k$ flavors of right handed neutrinos to the SM with three
active neutrinos extends the parameter space of the model by $7k-3$
parameters. In the $\nu$MSM $k=3$, thus there are $18$ parameters in
addition to those of the SM. These can be chosen as the masses $m_i$
and $M_I$ of the three active and three sterile neutrinos,
respectively, and three mixing angles as well as three phases in each
of the mixing matrices $U_\nu$ and $U_N$ that diagonalize $m_\nu$ and
$M_N$, respectively.

In the following we consider an effective theory with only two right
handed neutrinos, which is appropriate to describe the generation of lepton asymmetries in scenarios I and II. After dropping $N_1$ from the Lagrangian
(\ref{Lagrangian_flavour_base}), the effective Lagrangian contains
$11$ new parameters in addition to the SM.  $7$ of them are related to
the active neutrinos. In the standard parametrization they are two
masses $m_i$ (one active neutrino has a negligible mass), three mixing
angles $\uptheta_{ij}$, a Dirac phase $\delta$ and a Majorana phase
$\phi$. They can at least in principle be measured in active neutrino
experiments. The remaining four are related to sterile neutrino
properties. In the common Casas-Ibarra parametrization
\cite{Casas:2001sr} two of them are chosen as $M_2$, $M_3$. The last
two are the real and imaginary part of a complex angle $\upomega$
\footnote{Note that $F$ as a polynomial in $z=e^{i\upomega}$ only
contains terms of the powers $z$ and $1/z$.}. The Yukawa coupling is
written as
\begin{equation}
\label{CasasIbarraDef}
F=U_\nu \sqrt{m_\nu^{\rm diag}} \mathcal{R} \sqrt{M_M},
\end{equation}
where $m_{\nu}^{\rm diag}={\rm diag}(m_1,m_2,m_3)$. For normal
hierarchy of active neutrino masses ($m_1\simeq 0$) $\mathcal{R}$ is
given by 
\begin{eqnarray}
\mathcal{R}=\left(
\begin{tabular}{c c}
$0$ & $0$ \\
$\cos(\upomega)$ & $\sin(\upomega)$\\
$-\xi \sin(\upomega)$ & $\xi \cos(\upomega)$ 
\end{tabular}
\right) \ {\rm normal}\ {\rm hierarchy}
\end{eqnarray}
while for inverted hierarchy ($m_3\simeq 0$) it reads
\begin{eqnarray}
\mathcal{R}=\left(
\begin{tabular}{c c}
$\cos(\upomega)$ & $\sin(\upomega)$\\
$-\xi \sin(\upomega)$ & $\xi \cos(\upomega)$ \\
$0$ & $0$
\end{tabular}
\right) \  {\rm inverted}\ {\rm hierarchy},
\end{eqnarray}
where $\xi=\pm 1$. The matrix $U_{\nu}$ can be parameterized as
\begin{equation}
U_\nu=V_{23}U_{\delta}V_{13}U_{-\delta}V_{12}{\rm diag}(e^{i\alpha_1
/2},e^{i\alpha_2 /2},1)
\end{equation}
with $U_{\pm\delta}={\rm diag}(e^{\mp i\delta/2},1,e^{\pm i\delta/2})$ and 
\begin{eqnarray}
V_{23}=\left(\begin{tabular}{ccc}
$1$ & $0$ & $0$ \\ $0$ & $c_{23}$ & $s_{23}$ \\ $0$ & $-s_{23}$ & $c_{23}$
\end{tabular},\
\right), \
V_{13}=\left(
\begin{tabular}{ccc}
$c_{13}$ & 0 & $s_{13}$ \\ 0 & 1 & 0 \\ $-s_{13}$ & 0 & $c_{13}$
\end{tabular}
\right),\
V_{12}=\left(
\begin{tabular}{ccc}
$c_{12}$ & $s_{12}$ &  0\\ $-s_{12}$ & $c_{12}$ & 0 \\ 0 & 0 & 1
\end{tabular}
\right), 
\end{eqnarray}
where $c_{ij}$ and $s_{ij}$ stand for $\cos(\uptheta_{ij})$ and
$\sin(\uptheta_{ij})$, respectively, and $\alpha_1$, $\alpha_2$ and
$\delta$ are the CP-violating phases.  For normal hierarchy the Yukawa
matrix $F$ only depends on the phases $\alpha_2$ and $\delta$, for the
inverted hierarchy, it depends on $\delta$ and the difference
$\alpha_1-\alpha_2$. This is because $N_1$ has no measurable effect on
neutrino masses due to $M_1\ll M_{2,3}$.
\begin{table}
\begin{center}
\begin{tabular}{| c | c | c | c | c |}
\hline
$\msol^2 [{\rm eV}^2]$ & $\matm^2 [{\rm eV}^2]$ & $\sin^2\uptheta_{12}$ & $\sin^2\uptheta_{13}$ & $\sin^2\uptheta_{23}$\\
\hline
 $7.58\cdot 10^{-5}$ & $ 2.35\cdot10^{-3}$ & $0.306$ & $0.021$ & $0.42$\\ 
\hline
\end{tabular}
\caption{Neutrino  masses and mixings as found in 
\cite{Fogli:2011qn}. We parameterize the masses $m_i$ according to
$m_1=0$, $m_2^2=\msol^2$, $m_3^2=\matm^2+\msol^2/2$ for normal
hierarchy and $m_1^2=\matm^2-\msol^2/2$, $m_2^2=\matm^2+\msol^2/2$,
$m_3=0$ for inverted hierarchy. Using the values for $\uptheta_{13}$
found more recently in \cite{An:2012eh,Ahn:2012nd} 
has no visible
effect on our results.\label{NeutrinoParameters}}
\end{center}
\end{table}
In practice we will use the following parameters: two active neutrino
masses $m_i$, five parameters in the active mixing matrix (three
angles, one Dirac phase, one Majorana phase), the average physical
sterile neutrino mass $\bar{M}=(M_1+M_2)/2\simeq M$, the mass
splitting $\Delta M$.

The masses and mixing angles of active neutrinos have been measured
(the absolute mass scale is fixed as the lightest active neutrino is
almost massless in scenarios I and II). We use the experimental values
obtained from the global fit published in reference
\cite{Fogli:2011qn} in all calculations, which are summarized in table
\ref{NeutrinoParameters}. Shortly after we finished our numerical
studies, the mixing angle $\uptheta_{13}$ was measured by the Daya Bay
\cite{An:2012eh} and RENO \cite{Ahn:2012nd}  collaborations. The
values found there slightly differ from the one given in
\cite{Fogli:2011qn}, see also \cite{Fogli:2012ua}. We checked that the
effect of using one or the other value on the generated asymmetries in
negligible, which justifies to use the self-consistent set of
parameters given in table \ref{NeutrinoParameters}.  The remaining
parameters can be constrained in decays of sterile neutrinos in the
laboratory.

It is one of the main goals of this article to impose bounds on them to
provide a guideline for experimental searches.  In order to identify
the interesting regions in parameter space we proceed as follows. We
neglect $\Delta M$ in (\ref{CasasIbarraDef}), but of course keep it in
the effective Hamiltonian introduced in section \ref{SectionKinEq}.
This is allowed in the region $\Delta M \ll M$, which we consider in
this work. Unless stated differently, we always allow the CP-violating
Majorana and Dirac phases to vary. We then numerically determine the
values that maximize the asymmetry and fix them to those. 
In section \ref{baryogenesissection}, where we study the condition i) for
baryogenesis, we apply the same procedure to $\upomega$.  On the other
hand,  the requirement ii), necessary to explain $\Omega_{DM}$ in scenario I, almost fixes the parameter ${\rm Re}\upomega$ to a multiple of $\pi/2$ \footnote{This is explained in  section \ref{finetunings}.}.
In section \ref{DMproductionSection} we
therefore fix ${\rm Re}\upomega=\pi/2$.

The remaining parameters $\xi$, $M$, $\Delta M$ and ${\rm Im}\upomega$
contain a redundancy.  For $\Delta M\ll M$  changing simultaneously
the signs of $\xi$, $\Delta M$  and ${\rm Im}\upomega$ along with the
transformation ${\rm Re}\upomega \leftrightarrow \pi - {\rm
Re}\upomega$ corresponds to swapping the names of $N_2$ and $N_3$. To
be definite, we always chose $\xi=1$ and consider both signs of ${\rm
Im}\upomega$. Our main results consist of bounds on the parameters
$M$, ${\rm Im}\upomega$ and $\Delta M$.

For experimental searches the most relevant properties of the sterile
neutrinos are the mass $\bar{M}\simeq M$ and their mixing with active
neutrinos.   We therefore also present our results in terms of $M$,
the physical mass splitting $\delta M$ and 
\begin{equation}
	\label{Mixing_U}
U^2\equiv{\rm tr}(\Theta^\dagger\Theta) = {\rm
tr}(\theta^\dagger\theta)=\sum_\alpha U_\alpha^2,
\end{equation}
where $\Theta$ and $U_\alpha^2$ are given by (\ref{ThetaDef}) and
(\ref{UalphaDef}), respectively. $U^2$ measures the mixing between
active and sterile species. $\delta M$ and $U^2$ can, however, not be
mapped on parameters in the Lagrangian in a unique way; there exists
more than one choice of $\upomega$ leading to the same $U^2$.

\subsection{``Fine Tunings'' and the {\it Constrained $\nu$MSM}}
\label{finetunings}
In most models that incorporate the seesaw mechanism the eigenvalues
of $M_M$ are much larger than the scale of electroweak symmetry
breaking. It is a defining feature of the $\nu$MSM that all
experimental data can be explained without introduction of such a new
scale. In order to keep the sterile neutrino masses below the
electroweak scale and the active neutrino masses in agreement with
experimental constraints, the Yukawa couplings $F$ have to be very
small.  As a consequence of this, the thermal production rates for
lepton asymmetries are also very small unless they are resonantly
amplified.
In scenarios I and II this requires a small mass splitting between $M_2$ and $M_3$. This can either be viewed as ``fine tuning'' or be related to a new symmetry
\cite{Shaposhnikov:2006nn,Shaposhnikov:2008pf}. 
In the following we focus on these two scenarios, I and II.
We do not discuss the origin of the small mass splitting here, but only list the implications \footnote{As far as this work is concerned the sterile neutrino mass spectrum in the $\nu$MSM follows from the requirement to simultaneously explain $\Omega_B$ and $\Omega_{DM}$. It is in accord with the principle of minimality and the idea to explain new physics without introduction of a new scale (above the electroweak scale). In this work we do not discuss a possible origin of the mass spectrum and flavor structure in the SM and $\nu$MSM, which to date is purely speculative. Some ideas on the origin of a low seesaw scale can be found in \cite{Harigaya:2012bw,Nemevsek:2012cd,Araki:2011zg,Adulpravitchai:2011rq,Merle:2011yv,Barry:2011wb,Lindner:2010wr,McDonald:2010jm,Kusenko:2010ik,Bezrukov:2009th,Kadota:2007mv,Chen:2006hn}, see also \cite{Abazajian:2012ys}. Some speculations on the small mass splitting have been made in \cite{Shaposhnikov:2006nn,Shaposhnikov:2008pf,Roy:2010xq,He:2009pt}. A similar spectrum has been considered in a supersymmetric theory in \cite{Mazumdar:2012qk}.}.

Fermionic dispersion relations in a medium can have a complicated
momentum dependence. In the following we make the simplifying
assumption that all neutrinos have hard spacial momenta $\bar{p} \sim
T$ and parameterize the effect of the medium by a temperature
dependent quasiparticle mass matrix $M_N(T)$,\footnote{See e.g.
\cite{Drewes:2010pf,Beneke:2010wd,Garny:2011hg,Drewes:2012qw} for a discussion of
the quasiparticle description.} which we define as
$M_N(T)^2=H^2-\bar{p}^2$ at $|\textbf{p}|=\bar{p}\sim T$. Here $H$ is
the dispersive part of the temperature dependent effective Hamiltonian
given in the appendix, cf. (\ref{Heff3}). The general structure of
$M_N(T)$ is rather complicated, but we are only interested in the
regimes $T\lesssim M$ (DM production) and $T>T_{EW}$ (baryogenesis).
Analogue to the vacuum notation in (\ref{M23})-(\ref{deltaM}), we
refer to the temperature dependent eigenvalues of $M_N(T)$ as $M_1(T)$
and $M_2(T)$, their average as $\bar{M}(T)$ and their splitting as
$\delta M(T)$. Though $N_I$ are the fields whose excitations
correspond to mass eigenstates in the microscopic theory, the mass
matrix $M_N(T)$ in the effective quasiparticle description is not
necessarily diagonal in the $N_I$-basis for $T\neq0$. The effective
physical mass splitting $\delta M(T)$ depends on $T$ in a non-trivial
way. This dependence is essential in the regime $\bar{M}(T)\gg \delta
M(T)$, which we are mainly interested in. In principle also
$\bar{M}(T)$ depends on temperature, but this dependence is
practically irrelevant and replacing $\bar{M}(T)$ by $M$ at all
temperatures of consideration does not cause a significant error. 

There are three contributions to the temperature dependent physical
mass splitting: The splitting $\Delta M$ that appears in the
Lagrangian, the Dirac mass $m_D(T)=F v(T)$ that is generated by the
coupling to the Higgs condensate and thermal masses due to forward
scattering in the plasma, including Higgs particle
exchange\footnote{We ignore the running of the mass parameters, which
has been studied in \cite{Roy:2010xq}.}.  The interplay between the
different contributions leads to non-trivial effects as the
temperature changes. 

\subsubsection{Baryogenesis}
For successful baryogenesis it is necessary to produce a lepton
asymmetry of $\mu_\alpha\sim 10^{-10}$ at  $T\gtrsim T_+$ that
survives until $T_{EW}$ and is partly converted into a baryon
asymmetry by sphalerons, see condition i).  In this work we focus on
scenarios I and II, in which only two sterile neutrinos $N_{2,3}$ are involved in baryogenesis. 
In these scenarios baryogenesis is only possible if the physical mass splitting 
is sufficiently small ($\delta M(T)\ll M$) and leads to a resonant
amplification 
.  
On the other hand it
should be large enough for the sterile neutrinos to perform at least
one oscillation. Thus, baryogenesis is most efficient if it is of the
same order of magnitude as the relaxation rate (or thermal damping rate) at $T\gtrsim T_+$, 
\begin{equation}
\delta M(T)\sim (\Gamma_N)(T)_{IJ}\label{resoncond}.
\end{equation} 
Here $\Gamma_N$ is the temperature dependent dissipative part of the
effective Hamiltonian that appears in the kinetic equations given in
section \ref{SectionKinEq}; it is defined in appendix
\ref{DissipativePartSec} and calculated in section \ref{ratecompsec}. It is essentially given by the sterile neutrino thermal width. 
However, (\ref{resoncond}) only provides a rule of thumb to identify
the region where baryogenesis is most efficient. Numerical studies in
section \ref{baryogenesissection} show that the observed BAU can be
explained even far away from this point, for $M\gg \delta M(T)\gg
\Gamma_N(T)$. Thus, the mass degeneracy $\delta M(T)\ll M$ is the only
serious tuning required in scenario II.  In
\cite{Drewes:2012ma} it has been found that no such mass degeneracy is
required in scenario III.

\subsubsection{Dark Matter Production}
\label{DMfinefunings}
In scenario I $N_1$ dark matter has to be produced thermally from the
primordial plasma \cite{Dodelson:1993je}.  In absence of chemical potentials, the resulting
spectrum of $N_1$ momenta has been determined in \cite{Asaka:2006nq}.  State of the art X-ray observations,
structure formations and Ly$_\alpha$ forest observations suggest that
this production mechanism is not sufficient to explain $\Omega_{DM}$
because the required $N_1$ mass and mixing are astrophysically
excluded \cite{Boyarsky:2008mt,Boyarsky:2009ix}.  However, in the
presence of a lepton chemical potential, the dispersion relation for
active neutrinos is modified due to the MSW effect.   If the chemical
potential is large enough, this can lead to a level crossing between
active and sterile neutrinos, resulting in a resonant amplification of
the $N_1$ production rate \cite{Shi:1998km}.  The full dark matter
spectrum is a superposition of a smooth distribution from the non-resonant
production and a non-thermal spectrum with distinct peaks at low momenta from
resonant mechanism.  In order to explain all observed dark matter by
$N_1$ neutrinos, lepton asymmetries $|\mu_\alpha|\sim 8\cdot 10^{-6}$
are required at $T_{DM}\sim 100$ MeV \cite{Laine:2008pg}. This is the
origin of the condition ii) already formulated in section
\ref{ThermalHistory}. Again the resonance condition (\ref{resoncond})
indicates the region where the asymmetry production is most
efficient. For $T_d,T_-\ll T_{EW}$ it imposes a much stronger
constraint on the mass splitting than during baryogenesis because the
thermal rates $\Gamma_N$ are much smaller.

The asymmetries $\mu_{\alpha}$ can be created in two different ways,
either during the freezeout of $N_{2,3}$ around $T\sim T_-$ or in
their decay at $T\sim T_d$. During these processes we can use the
vacuum value for $v$.  As discussed in appendix \ref{DispersivePart},
the temperature dependence of $\delta M(T)$ is weak for $T< T_-$.  The
rates, on the other hand, still depend rather strongly on temperature,
thus it is usually not possible to fulfill the requirement
(\ref{resoncond}) at $T=T_-$ and $T=T_d$ simultaneously. Therefore one
can distinguish two scenarios: the asymmetry generation is efficient
either during freezeout (\textit{freezeout scenario}) or during decay
(\textit{decay scenario}). On the other hand, (\ref{resoncond}) can be
fulfilled simultaneously at $T=T_+$ and $T=T_d$ or at $T=T_+$ and
$T=T_-$ because at $T=T_+$ also the mass splitting depends on
temperature. The strongest ``fine tuning'' requirement in the $\nu$MSM
is therefore\footnote{It is in fact sufficient for baryogensis if $\delta M(T)\sim (\Gamma_N)_{IJ}(T)$ at some temperature $T>T_+$ as long as some flavor asymmetries survive until $T_{EW}$. The washout of the $\mu_\alpha$ typically becomes efficient around $T_+$, but chemical equilibration can take long if one active flavour couples to the sterile neutrinos much weaker than the others.} 
\begin{eqnarray}\label{ResonanceCondition}
\begin{tabular}{c c c c}
 & $\delta M(T_+)\sim (\Gamma_N)_{IJ}(T_+)$ & and & $\delta M(T_-)\sim
(\Gamma_N)_{IJ}(T_-)$\\ or & $\delta M(T_+)\sim (\Gamma_N)_{IJ}(T_+)$
& and & $\delta M(T_d)\sim (\Gamma_N)_{IJ}(T_d)$. 
\end{tabular}
\end{eqnarray} 
From (\ref{deltaM}) it is clear that during the decay $\delta
M(T_d)\approx \delta M(T=0)$ and
\begin{eqnarray}
\delta M &\geq& \frac{1}{2M} \left({\rm Re}\left(m^{\dagger}_{D}
m_{D}\right)_{33} - {\rm Re}\left(m^{\dagger}_D m_D\right)_{22}
\right) + \Delta M \label{weakreq}\\ |\delta M| &\geq&\frac{1}{M} {\rm
Re}\left(m^{\dagger}_D m_D\right)_{23}.\label{strongreq}
\end{eqnarray} 
Fulfilling the resonance condition (\ref{resoncond}) at low
temperature requires a precise cancellation of the parameters in
(\ref{weakreq}) and (\ref{strongreq}), both of which have to be
fulfilled individually. The condition (\ref{strongreq}) imposes a
strong constraint on the active neutrino mass matrix (\ref{activeneutrinomasses}). It can be
fulfilled when real part of the off-diagonal elements is small.  Note
that this due to (\ref{sterileneutrinomasses}) implies that $U_N$ is
close to unity. This is certainly the case when the real part of
complex angle $\upomega$ in $\mathcal{R}$ is a multiple of $\pi/2$. In
sections \ref{DMproductionSection} and \ref{Combined} we will focus on
this region and always choose ${\rm Re}\upomega=\pi/2$. It should be
clear that this is a conservative approach, since the production of
lepton asymmetries can also be efficient away from the maximally
resonant regions defined by (\ref{ResonanceCondition}). The lower
bound (\ref{weakreq}) can always be made consistent with
(\ref{resoncond}) by adjusting the otherwise unconstrained parameter
$\Delta M$.  At tree level this parameter is effectively fixed by
\begin{equation}
	\Delta M=-\frac{1}{4M}\left({\rm Re}\left(m^{\dagger}_{D}
m_{D}\right)_{33} - {\rm Re}\left(m^{\dagger}_D m_D\right)_{22}
\right)\pm\delta M,\label{finetunningDeltaM}
\end{equation} 
where the dependence of the RHS on $\Delta M$ is weak. The range of
values for $\Delta M$ dictated by this condition is extremely narrow;
it requires a tuning of order $\sim 10^{-11}$ (in units of $M$).
Quantum corrections are of order $\sim m_i$ \cite{Roy:2010xq}, i.e.
much bigger than $\delta M(T_-)$.  The high degree of tuning,
necessary to explain the observed $\Omega_{DM}$, is not understood
theoretically. Some speculations can be found in 
\cite{Shaposhnikov:2006nn,Shaposhnikov:2008pf,Roy:2010xq,He:2009pt}. However,
the origin of this fine-tuning plays no role for the present work.

In the following we will refer to the $\nu$MSM with the condition
${\rm Re}\upomega=\frac{1}{2}$ and the fixing of $\Delta M$ as {\it
constrained $\nu$MSM}. Since the first term in the square root in
(\ref{deltaM}) also depends on ${\rm Re}\upomega$, fixing this
parameter exactly to a multiple of $\pi/2$ usually does not exactly
give the minimal $\delta M$. However, it considerable simplifies the
analysis, and deviations from such a value can in any case only be
small due to the above considerations.

\section{Experimental Searches and Astrophysical Bounds}
\label{ExistingBounds}
The experimental, astrophysical and BBN bounds presented in this
section and in the figures in sections
\ref{baryogenesissection}-\ref{Combined} are derived under the premise
that the mass and mixing of $N_1$ qualify it as a DM candidate, while
$N_{2,3}$ are responsible for baryogenesis (scenarios I and II). Some of them loosen if one drops the DM requirement and considers the $\nu$MSM as a theory of baryogenesis and neutrino oscillations only in scenario III.

\subsection{Existing Bounds}
A detailed discussion of the existing experimental and observational
bounds on the $\nu$MSM can be found in \cite{Boyarsky:2009ix}. Some
updates that incorporate the effect of recent measurements of the
active neutrino mixing matrix $U_\nu$, in particular
$\uptheta_{13}\neq0$, have been published in
\cite{Asaka:2011pb,Ruchayskiy:2011aa,Ruchayskiy:2012si}.  In the
following we re-analyze all relevant constraints on the seesaw
partners $N_{2,3}$ from direct search experiments and BBN in the face
of these experimental results.  We also briefly review existing
constraints on the dark matter candidate $N_1$.

As far as the known (active) neutrinos are concerned, the main
prediction of the $\nu$MSM is that one of them is (almost) massless.
This fixes the absolute mass scale of the remaining two neutrinos 
\cite{Asaka:2005an}. Currently there is neither a clear prediction for
the phases in $U_\nu$ in the $\nu$MSM nor an experimental
determination, though the experimental value for $\uptheta_{13}$
\cite{An:2012eh,Ahn:2012nd} suggests that a measurement in principle
might be possible.  Regarding the sterile neutrinos, one has to
distinguish between $N_{2,3}$ and $N_1$.

\subsubsection{Seesaw Partners $N_2$ and $N_3$}\label{N23searches}}
\paragraph{LHC} The small values of $M_I\ll v$ in principle make it
possible to produce them in the laboratory. However, the smallness of
the Yukawa couplings $F$ implies that the branching ratios are very
small. Therefore the number of collisions (rather than the required
collision energy) is the main obstacle in direct searches for the
sterile neutrinos.     In particular, they cannot be seen in high
energy experiments such as ATLAS or CMS.  It is therefore a prediction
of the $\nu$MSM that they see nothing but the Higgs boson. Vice versa,
the lack of findings of new physics beyond the SM at the LHC  can be
viewed as indirect support for the model (though this prediction is of
course relaxed if nature happens to be described by the $\nu$MSM plus
something else).

\paragraph{Direct Searches} The sterile neutrinos participate in all
processes that involve active neutrinos, but with a probability that is
suppressed by the small mixings $U_\alpha^2$. The mixing of $N_{2,3}$
to the SM is large enough that they can be found experimentally
\cite{Gorbunov:2007ak}. A number of experiments that allow to
constrain the sterile neutrino properties has been carried out in the
past \cite{Yamazaki:1984sj,Daum:2000ac}, in particular CERN PS191 \cite{Bernardi:1985ny,Bernardi:1987ek},
NuTeV \cite{Vaitaitis:1999wq}, CHARM \cite{Bergsma:1985is}, NOMAD
\cite{Astier:2001ck} and BEBC \cite{CooperSarkar:1985nh} (see
\cite{Atre:2009rg} for a review).  These can be grouped into {\it beam
dump experiments} and {\it peak searches}.

Peak search experiments look for the decay of charged mesons into
charged leptons ($e^\pm$ or $\mu^\pm$) and neutrinos. Due to the
mixing of the active neutrino flavor eigenstates with the sterile
neutrinos, the final state in a fraction of decays suppressed by
$U_e^2$ (or $U_\mu^2$) is $e^\pm+N_I$ (or $\mu^\pm+N_I$).  The
kinematics of the two body decay can be reconstructed from the
measured charged lepton, but the sterile flavor cannot be determined
because of the $N_I$ mass degeneracy. Therefore these experiments are
only sensitive to the inclusive mixing $U_\alpha^2$ defined in
(\ref{UalphaDef}), where $_\alpha$ is the flavor of the charged
lepton. 

In beam dump experiments, sterile neutrinos are also created in the
decay of mesons, which have been produced by sending a proton beam
onto a fixed target. A second detector is placed near the beamline to
detect the decay of the sterile neutrinos into charged particles. Also
in beam dump experiments, the sterile flavors cannot be
distinguished. In this case, the expected signal is of the order
$U_\alpha^2 U_\beta^2$ because creation and decay of the $N_I$ each
involve one active-sterile mixing.  For instance, the CERN PS191
experiment constrains the combinations $(U_e^2)^2$, $(U_\mu^2)^2$,
$U_e^2 U_\mu^2$ and\footnote{There are also constraints on $U_\tau^2$
which are, however, too weak to be of practical relevance.}
\begin{equation}
\sum_\beta U_\alpha^2\left(c_\beta U_\beta^2\right),
\end{equation}
where 
\begin{equation}
c_e=\frac{1+4\sin^2\theta_W+8\sin^2\theta_W}{4},\ 
c_\mu=c_\tau=\frac{1-4\sin^2\theta_W+8\sin^2\theta_W}{4}.
\end{equation}
This set differs from the quantities considered by the experimental
group \cite{Bernardi:1985ny,Bernardi:1987ek}. It has been pointed out
in \cite{Ruchayskiy:2011aa} that the original interpretation of the
PS191 (and also CHARM) data cannot be directly applied to the seesaw
Lagrangian (\ref{nuMSM_lagrangian}).  The authors of
\cite{Ruchayskiy:2011aa} translate the bounds on active-sterile
neutrino mixing published by the PS191 and CHARM collaborations into
bounds that apply to the $\nu$MSM and kindly provided us with their
data.  We use these bounds, along with the NuTeV constraints, as an
input to constrain the region in the $\nu$MSM parameter space that is
compatible with experiments. 

Our results are displayed as green lines of different shade in the
summary plots in figures \ref{BA_Max_mixingangle}, \ref{BA_DM_U} and
\ref{BA_SP_U}  in sections \ref{baryogenesissection} and
\ref{Combined}. The different lines have to be interpreted as follows.
Each shade of green corresponds to one experiment. For each
experiment, there is a solid and a dashed line. The solid line is an
{\it exclusion bound}. That means that there exists no choice of
$\nu$MSM parameters that leads to a combination of $U^2$ and $M$ above
this line and is consistent with table \ref{NeutrinoParameters} and
the experiment in question.  In order to obtain the exclusion bound
from an experiment for a particular choice of $M$ we varied the
CP-violating phases and ${\rm Im}\upomega$.\footnote{The mixings
$U_\alpha^2$ do not depend on ${\rm Re}\upomega$ and the dependence on
$\Delta M$ is negligible.}  We checked for each choice whether the
resulting $U_\alpha^2$ are compatible with the experiment in question.
The exclusion bound in the $M-U^2$ plane is obtained from the set of
parameters that leads to the maximal $U^2$ for given $M$ amongst all
choices that are in accord with experiment. The exclusion plots are
independent of the other lines in the summary plots. The dashed lines
(in the same shade as the exclusion plots) represent the bounds
imposed by each experiment if the CP-violating phases are
self-consistently fixed to the values that we used to produce the red
and blue lines in the summary plots, which encircle the regions in
which enough asymmetry is created to explain the BAU and DM. The NuTeV
experiment puts bounds only on the mixing angle $U_\mu^2$. This
induces a much weaker constraint in the $M-U^2$ plane for inverted
mass hierarchy than the other experiments. Our results differ from
those of \cite{Canetti:2010aw}. In the latter, the experimental
constraints on the individual $U_\alpha^2$ were directly reported in
the $M-U^2$ plane plotted in figure 3 of \cite{Canetti:2010aw}.
Moreover, only the PS191 exclusion bound was computed by
distinguishing between mass hierarchies.

\paragraph{Active Neutrino Oscillation Experiments} The region below
the "seesaw" line in figures \ref{BA_Max_mixingangle}, \ref{BA_DM_U},
\ref{DM_Max_mixingangle} and \ref{BA_SP_U} is excluded because for the
experimental values listed in table \ref{NeutrinoParameters}, there
exists no choice of $\nu$MSM parameters that would lead to this
combination of $M$ and $U^2$.

\paragraph{Big Bang Nucleosynthesis} It is a necessary requirement
that $N_{2,3}$ have decayed sufficiently long before BBN that their
decay products do not affect the predicted abundances of light
elements, which are in good agreement with observation.   The total
increase of entropy due to the $N_{2,3}$ decay is small, but the decay
products have energies in the GeV range and even a small number of
them can dissociate enough nuclei to modify the light element
abundances. Since the sterile neutrinos are created as flavor
eigenstates, they oscillate rapidly around the time of BBN. On
average, they spend roughly half the time in each flavor state,  and
not the individual lifetimes of each flavor determine the relaxation
time, but their average.   This allows to estimate the inverse
$N_{2,3}$ lifetime $\tau$ by as $\tau^{-1}\simeq\frac{1}{2}{\rm
tr}\Gamma_N$ at $T=1$ MeV. For $\tau< 0.1$s the decay products and all
secondary particles have lost their excess energy to the plasma in
collisions and reached equilibrium by the time of BBN \footnote{A more
precise analysis of this condition for $M<140$ MeV has been performed
in \cite{Ruchayskiy:2012si}.}.  We impose the condition $\tau<0.1$s
and vary all free parameters to identify the region in the $M$-$U^2$
plane consistent with this condition.  The BBN exclusion bounds in
figures  \ref{BA_Max_mixingangle}, \ref{BA_DM_U} and \ref{BA_SP_U}
represent the region in which no choice of $\nu$MSM parameters exists
that is consistent with table \ref{NeutrinoParameters} and the above
condition.  Note that $\tau^{-1}\simeq \frac{1}{2}{\rm tr}\Gamma_N$
and the condition $\tau<0.1$s are both rough estimates; the BBN bound
we plot may change by a factor of order one when a detailed
computation is performed.

\subsubsection{Dark Matter Candidate $N_1$}\label{subsubDmbounds} 
The coupling of the DM candidate $N_1$ is too weak to be constrained by any past laboratory experiment.
However, different indirect methods have been used to identify the allowed region in the $\theta_{\alpha 1}^2-M_1$ plane. 
The possibility of sterile neutrino DM has been studied by many authors in the past, see \cite{Kusenko:2009up,Boyarsky:2009ix} for reviews. In the following we summarize the most important constraints.

As a decaying dark matter candidate $N_1$ particles produce a
distinct X-ray line in the sky that can be searched for. 
These pose an \emph{upper
bound} $M_{1}\lesssim 3-4$~keV~(see
e.g.~\cite{Abazajian:2006yn,Boyarsky:2007ay,Boyarsky:2008xj}).  If
this were the only mechanism, the tension between these bounds would
rule out the $\nu$MSM as the common source of BAU, DM and neutrino
oscillations. 

There are two different mechanisms for DM production in the $\nu$MSM.
The first one, common thermal (non-resonant) production
\cite{Dodelson:1993je},  leads to a smooth distribution of momenta. 
The second one, which relies on a resonance produced by a level crossing in active and sterile neutrino dispersion relations (see below), creates a highly non-thermal spectrum \cite{Shi:1998km,Laine:2008pg}.
Observations of the matter distribution in the universe constrain the DM free
streaming length.
Without resonant production, the distribution reconstructed from Ly$_\alpha$ forest observations suggests a \emph{lower bound} on the mass
$M_{1} \gtrsim 8$~keV~\cite{Boyarsky:2008xj}, see
also~\cite{Viel:2005ha,Viel:2006kd,Seljak:2006qw}\footnote{There it
has been assumed that baryonic feedback on matter distribution, see
e.g. \cite{Semboloni:2011fe} for a discussion, has negligible
effect.}.
In combination with the X-ray bound, this would make resonant production necessary. 
In a realistic
scenario involving both production mechanisms
($|\mu_\alpha|\gtrsim10^{-5}$) this bound relaxes and has been
estimated as $M_1>2$ keV \cite{Boyarsky:2008mt}. 
In our analysis we take these results for granted, though some uncertainties remain to be clarified, see section \ref{futureN1}.

The DM production rate can be resonantly amplified by the
presence of a lepton chemical potential in the plasma
\cite{Shi:1998km}. The resonance occurs due to a level crossing
between active and sterile neutrino dispersion relations, caused by
the MSW effect. This mechanism enhances the production rate for
particular momenta as they pass through the resonance, resulting in a
non-thermal DM momentum distribution that is dominated by low momenta
and thus ``colder'' \footnote{The results quoted
in~\cite{Lovell:2011rd} demonstrate that the resonantly produced
sterile is ``warm enough'' to change the number of substructure of a
Galaxy-size halo, but ``cold enough'' to be in agreement with ${\rm
Ly}_\alpha$ bounds~\cite{Boyarsky:2008mt}.}.  Effectively, this
mechanism ``converts'' lepton asymmetries into DM abundance, as the
asymmetries are erased while DM is produced. The full DM spectrum in
the $\nu$MSM is a superposition of the two components. The
dependence on $\mu_\alpha$ is, however, rather complicated. In
particular, the naive expectation that the largest $|\mu_\alpha|$,
which maximized the efficiency of the resonant production mechanism,
leads to the lowest average momentum (``coldest DM'') is not true
because $\mu_\alpha$ does not only affect the efficiency of the
resonance, but also the momentum distribution of the produced
particles.

The $N_1$-abundance must correctly reproduce the observed DM density
$\Omega_{DM}$.  This requirement defines a line in the
$M_1-\sum_\alpha|\theta_{\alpha 1}|^2$-plane, the production curve.
All combinations of $M_1$ and $\sum_\alpha|\theta_{\alpha 1}|^2$ along
the production curve lead to the observed DM abundance. Due to the
resonant contribution, the production curve depends on $\mu_\alpha$.
This dependence has been studied in \cite{Laine:2008pg}, where it was
assumed that $\mu_e=\mu_\mu=\mu_\tau$.
\begin{figure}
  \centering
    \includegraphics[width=10cm]{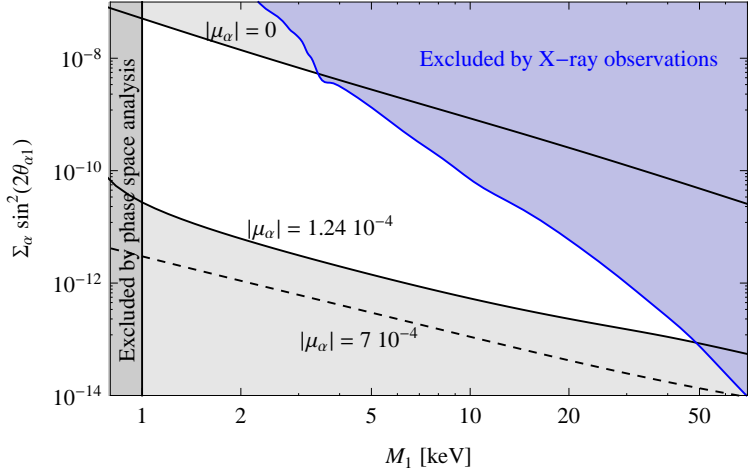}
    \caption{Different constraints on $N_1$ mass and mixing. The
blue region is excluded by X-ray observations, the dark gray region
$M_1< 1$ keV by the Tremaine-Gunn bound
\cite{Tremaine:1979we,Boyarsky:2008ju,Gorbunov:2008ka}.  The points on
the upper solid black line correspond to observed $\Omega_{DM}$ 
produced in scenario I in the absence of lepton asymmetries (for $\mu_{\alpha}=0$) 
\cite{Laine:2008pg}; points on the lower solid black line give the
correct $\Omega_{DM}$ for $|\mu_{\alpha}|=1.24\cdot10^{-4}$, the
maximal asymmetry we found.  The region between these lines is
accessible for $0\leq |\mu_\alpha|\leq 1.24\cdot10^{-4}$. 
We do not display bounds derived from Ly$_\alpha$ forest observations  because it depends on $\mu_\alpha$ in a complicated
way and the calculation currently includes considerable uncertainties
\cite{Boyarsky:2008mt}.
\label{DMexclusion}}
\end{figure} 

Finally, DM sterile neutrinos may have interesting effects for supernova explosions \cite{Fuller:2009zz,Fuller:2003gy,Kusenko:1998bk,Hidaka:2007se,Hidaka:2006sg}. 

Figure \ref{DMexclusion} summarizes a number of bounds on the
properties of $N_1$. The two thick black lines are the production
curves for $\mu_\alpha=0$ and $|\mu_\alpha|=1.24\cdot 10^{-4}$, the
maximal asymmetry we found at $T=100$ MeV in our analysis, see figure
\ref{DM_Max_asymmetry}.  The allowed region lies between these lines;
above the $\mu_\alpha=0$ line, the non-resonant production alone would
already overproduce DM, below the production curve for maximal
asymmetry  $N_{2,3}$   fail to produce the required asymmetry for all
choices of parameters. The maximal asymmetry has been estimated as
$\sim 7\cdot 10^{-4}$ in \cite{Shaposhnikov:2008pf,Laine:2008pg},
which agrees with our estimate shown in figure \ref{DM_Max_asymmetry}
up to a factor $\sim 5$.  The corresponding production curve is shown
as a dotted line in figure \ref{DMexclusion}. Our result is smaller
and imposes a stronger lower bound on the $N_1$-mixing, which makes it
easier to find this particle (or exclude it as the only constituent of
the observed $\Omega_{DM}$) in X-ray observations. However, though our
calculation is considerably more precise than the previous estimate,
the exclusion bound displayed in figure \ref{DM_Max_asymmetry}  still
suffers from uncertainties of order one due to the issues discussed in
appendix \ref{Uncertainties} and the strong
assumption\footnote{Indeed, we find that this assumption does not hold
in most of the parameter space. The asymmetries in individual flavors
can be very different and even have opposite signs. The reason is that
asymmetries generated at $T>M$ are mainly ``flavored'' asymmetries,
i.e. the total lepton number violation is small, but the individual
flavors can carry asymmetries.} $\mu_e=\mu_\mu=\mu_\tau$, made in
\cite{Shaposhnikov:2008pf,Laine:2008pg} to find the dependence of the
production curve on the asymmetry. In order to determine the precise
exclusion bound, the dependence of the production curve on individual
flavor asymmetries has to be determined.

The shadowed (blue) region is excluded by the non-observation of an
X-ray line from $N_1$ decay in DM dense regions
\cite{Abazajian:2001vt,Dolgov:2000ew,Boyarsky:2005us,%
Boyarsky:2006fg,Boyarsky:2006ag,Watson:2006qb,Abazajian:2006yn,%
Boyarsky:2007ay,Abazajian:2006jc,RiemerSorensen:2006fh,Boyarsky:2007ge,%
Loewenstein:2008yi}.   A lower limit on the mass of DM sterile
neutrino below $M_{1} \sim 1$~keV  can be obtained when applying the
Tremaine-Gunn bound on phase space densities \cite{Tremaine:1979we} to
the Milky way's dwarf spheroidal
galaxies~\cite{Boyarsky:2008ju,Gorbunov:2008ka}.   The ${\rm
Ly}_\alpha$ method yields similar bounds (not displayed in figure
\ref{DMexclusion}).

\subsection{Future Searches}
\subsubsection{Dark Matter Candidate $N_{1}$}\label{futureN1}
\paragraph{Indirect Detection}
The DM candidate $N_1$ can be searched for
astrophysically, using high resolution X-ray spectrometers to look for the emission line from its decay in DM dense regions. For
details and references see the proposal submitted to European Strategy
Preparatory Group by Boyarsky et al., \cite{ESDark}.

\paragraph{Structure Formation}
Model-independent constraints on $N_1$ can be
derived from consideration of dynamics of dwarf galaxies \cite{Tremaine:1979we,Boyarsky:2008ju,Gorbunov:2008ka}.
The existing small scale structures in the universe, such as galaxy subhalos, provides another probe that is sensitive to $N_1$ properties because such structures would be erased if the mean free path of DM particles is too long.
It can be exploited by comparing numerical simulations of structure formation to the distribution of matter in the universe that is reconstructed from Ly$_\alpha$ forest observations.  
However, the momentum distribution of resonantly produced $N_1$ particles can be complicated \cite{Shi:1998km,Laine:2008pg}, leading to a complicated dependence of the allowed mixing angle on the $N_1$ mass and lepton asymmetries $\mu_\alpha$ in the plasma \cite{Boyarsky:2008xj,Boyarsky:2008mt}. 
A reliable quantitative analysis would involve numerical simulations that use the non-thermal $N_1$ momentum distribution predicted in scenario I as input.
While for Cold Darm Matter extensive studies have been performed, see e.g. \cite{Springel:2008cc}, simulations for other spectra have only been done for certain benchmark scenarios; a model of Warm Dark Matter has been studied in \cite{Lovell:2011rd}.

\paragraph{Direct Detection}
As the solar system passes through the interstellar medium, the DM particles $N_1$ can interact with atomic nuclei in the laboratory via the $\theta_{\alpha 1}\theta_{\alpha 1}^*$ suppressed weak interaction. This in principle opens the possibility of direct detection \cite{Ando:2010ye}. Such detection is extremely challenging due to the small mixing angle and the background from solar and stellar active neutrinos. It has, however, been argued that it may be possible \cite{Li:2010vy,Liao:2010yx}.

\paragraph{BBN}
The primordial abundances of light elements are sensitive to the number of relativistic particle species in the primordial plasma during BBN because these affect the energy budget, which determines the expansion rate and temperature evolution. Any deviation from the SM prediction is usually parameterized in terms of the effective number of neutrino species $N_{eff}$. 
At temperatures around $2$ MeV, most $N_1$ particles are relativistic. However, the occupation numbers are far below their equilibrium value, and the effect of the $N_1$ on $N_{eff}$ is very small. Given the error bars in current measurements \cite{Hamann:2010bk,Steigman:2010zz,Steigman:2012ve}, the $\nu$MSM predicts a value for $N_{eff}$ that is practically not distinguishable from $N_{eff}=3$.

In principle the late time asymmetry in active neutrinos predicted by the $\nu$MSM also affects BBN because the chemical potential modifies the momentum distribution of neutrinos in the plasma. However, the predicted asymmetry is several orders of magnitude smaller than existing bounds \cite{Canetti:2012zc} and it is extremely unlikely that this effect can be observed in the foreseeable future.

\subsubsection{Seesaw Partners $N_{2,3}$} 
The singlet fermions participate in all
the reactions the ordinary neutrinos do with a probability suppressed
roughly by a factor $U^2$. However, due to their masses, the kinematic changes when an ordinary neutrino is replaced by $N_I$.
The $N_{2,3}$ particles can be found in the laboratory \cite{Gorbunov:2007ak} using the strategies outlined in section \ref{N23searches}, which have been applied in past searches. 

One strategy, used in peak searches, is the study of kinematics of rare $K,~D$, and $B$
meson decays can constrain the strength of the $N_I$ masses and mixings. This includes two body decays (e.g. $K^\pm \rightarrow \mu^\pm N$, $K^\pm \rightarrow
e^\pm N$) or three-body decays (e.g. $K_{L,S}\rightarrow \pi^\pm + e^\mp +
N_{2,3}$). 
The precise study of the kinematics 
is possible in $\Upphi$ (like KLOE), charm, and B factories, or in experiments with kaons where the initial 4-momentum is well known. For $3 {\rm MeV}< M_I < 414$ MeV this possibility has recently been discussed in \cite{Lello:2012gi}. 

The second strategy aims at observing the decay of the $N_I$ themselves (``nothing" $\rightarrow$ leptons and hadrons) in proton beam dump experiments.
The $N_I$ are created in the decay  $K,~ D$ or $B$ mesons emitted by a fixed target, into which the proton beam is dumped. The detector must be placed in some distance along the beamline.
Several existing or planned neutrino facilities (related, e.g., to CERN SPS, MiniBooNE, MINOS or J-PARC), could be complemented by a {\em dedicated} near detector for these searches. 
Finally, these two strategies can be unified, so that the production and the
decay occurs inside the same detector \cite{Achard:2001qw}.

For the mass interval $M < m_K$, both strategies can be used.
An upgrade of the NA62 experiment at CERN
would allow to search in the mass region below the Kaon mass $m_K$. 
For $m_K < M < m_D$ it is unlikely that a peak search for missing energy at beauty, charm, and $\tau$ factories will gain the necessary statistics. 
Thus, in this region the search for $N_{2,3}$ decays is the most promising strategy. 
Dedicated experiments using the SPS proton
beam at CERN can completely explore the very interesting parameter
range for $M < 2$ GeV. 
This has been outlined in detail in the European Strategy Preparatory Group by Gorbunov et al., \cite{ESbau}. 

An upgrade of the LHCb experiment could allow to combine
both strategies. 
This would allow to constrain the cosmologically interesting region in the $M-U^2$ plane..
With existing or planned proton beams and B-factories the mass region between the $D$-mass and $B$-meson thresholds is in principle accessible, but such experiments would be extremely challenging.
A search in the cosmologically interesting parameter
space would require an increase in the present intensity of the
CERN SPS beam by two orders of magnitude or to produce and study the
kinematics of more than $10^{10}$ B-mesons \cite{ESbau}. 

\section{Kinetic Equations}
\label{SectionKinEq}
Production, freezeout and decay of the sterile neutrinos are
nonequilibrium processes in the hot primordial plasma.  We describe
these by effective kinetic equations of the type used in
\cite{Sigl:1992fn} and further elaborated in
\cite{Akhmedov:1998qx,Asaka:2005pn,Asaka:2006rw,Shaposhnikov:2008pf}. 
These equations are similar to those commonly used to describe the
propagation of active neutrinos in a medium. They rely on a number of
assumptions and may require corrections when memory effects or
off-shell contributions are relevant. These assumptions are discussed
in appendix \ref{HeffSec}.  
We postpone a more refined study to the
time when such precision is required from the experimental side. In
the following we briefly sketch the derivation of the kinetic
equations we use. More details are given in appendix \ref{HeffSec}.

\subsection{Short derivation of the Kinetic Equations}
We describe the early universe as a thermodynamical ensemble. In
quantum field theory, any such ensemble - may it be in equilibrium or
not - can be described by a density matrix $\hat{\rho}$.  The
expectation value of any operator $\mathcal{A}$ at any time can be
computed as
\begin{equation}\label{expvalue}
\langle\mathcal{A}\rangle={\rm tr}(\hat{\rho}\mathcal{A}).
\end{equation}
As there are infinitely many states in which the world can be,
infinitely many numbers are necessary to exactly characterize
$\hat{\rho}$. These can either be given by all matrix elements of
$\hat{\rho}$ or, equivalently, by all $n$-point correlation functions
for all quantum fields. Either way, any practically computable
description requires truncation. 

The leptonic charges can be expressed in terms of field bilinears,
thus it is sufficient to concentrate on the two-point functions.
Instead of bilinears in the field operators themselves we consider
bilinears in the ladder operators $a_{I}$, $a_I^{\dagger}$ for sterile
and $a_{\alpha}$, $a_\alpha^{\dagger}$ for active neutrinos.  In
principle there is a large number of bilinears, but only few of them
are relevant for our purpose. 

For each momentum mode of sterile neutrinos we consider two $2\times
2$ matrices formed by products of ladder operators $a_I^\dagger a_J$,
one for positive and one for negative helicity\footnote{This
description is similar to the one commonly used in neutrino physics
and could also be formulated in terms of ``polarization vectors''.}. 
Since $M_N$ is diagonal in the $N_I$-basis, $a_{I}^{\dagger}a_{I}$ can
be interpreted as a number operator for physical sterile neutrinos
while $a_{I}^{\dagger}a_{J}$ with $I\neq J$ correspond to coherences.
$N_I$ are Majorana fields, but we can define a notion of ``particle''
and ``antiparticle'' by their helicity states. In the limit $T\gg M$,
i.e. for a negligibly small Majorana mass term, the total lepton number (sum over $_\alpha$ and $_I$)
defined this way are conserved. All other bilinears in the ladder
operators for sterile neutrinos are either of higher order in $F$ or
quickly oscillating and can be neglected.  Practically we are not
interested in the time evolution of individual modes, but only in the
total asymmetries. We therefore describe the sterile neutrinos by
momentum integrated abundance matrices $\rho_N$ for ``particles'' and
$\rho_{\bar{N}}$ for ``antiparticles''. The precise definitions are
given in appendix \ref{howtocharaceriseasymmetries}. 

The active leptons are close to thermal equilibrium at all times of
consideration. This is because kinetic equilibration is driven by fast
gauge interactions, while the relaxation rates for the asymmetries are
of second order in the small Yukawa couplings $F$. We thus describe
the active sector by four numbers\footnote{It has been found in \cite{Garbrecht:2012pq} that mixing amongst the different SM lepton doublets can occur due to their coupling to the right handed neutrinos and affect leptogenesis. However, in the $\nu$MSM the Yukawa couplings $F$ are too tiny to lead to a sizable effect.}, the temperature and three
asymmetries (one for each flavor, integrated over momentum). More
precisely, the asymmetry in the SM leptons of flavor $\alpha$ is
given by the difference between lepton and antilepton abundance, which
we denote by $\mu_{\alpha}$, see (\ref{muDef}) \footnote{Note that the
$\mu_{\alpha}$ here are abundances, not chemical potentials, which
could alternatively be used to characterize the asymmetries. The
relations between different characterizations of the asymmetries are
given in appendix \ref{asymmconv}.}.  We study the time
evolution of each flavor separately and find that they can differ
significantly from each other. Following the steps sketched in
appendix \ref{HeffSec}, one can find the effective kinetic ``rate equations''  
\begin{eqnarray}
\label{kinequ1}
i \frac{d\rho_{N}}{d X}&=&[H, \rho_{N}]-\frac{i}{2}\{\Gamma_N, \rho_{N} - \rho^{eq}\} +\frac{i}{2} \mu_\alpha{\tilde\Gamma^\alpha}_N~,\\
i \frac{d\rho_{\bar{N}}}{d X}&=& [H^*, \rho_{\bar{N}}]-\frac{i}{2}\{\Gamma^*_N, \rho_{\bar{N}} - \rho^{eq}\} -\frac{i}{2} \mu_\alpha{\tilde\Gamma^{\alpha *}}_N~,\label{kinequ2}\\
i \frac{d\mu_\alpha}{d X}&=&-i\Gamma^\alpha_L \mu_\alpha +
i {\rm tr}\left[{\tilde \Gamma^\alpha}_L(\rho_{N} -\rho^{eq})\right] -
i {\rm tr}\left[{\tilde \Gamma^{\alpha*}}_L(\rho_{\bar{N}}  -\rho^{eq})\right]
~.
\label{kinequ3}
\end{eqnarray}
Here $X=M/T$, $\rho^{eq}$ is the common equilibrium value of the
matrices $\rho_N$ and $\rho_{\bar{N}}$, $H$ is the dispersive part of
the effective Hamiltonian for sterile neutrinos that is responsible
for oscillations and rates $\Gamma_N$, $\tilde{\Gamma}_N^{\alpha}$ and
$\Gamma_L^\alpha$ form the dissipative part of the effective
Hamiltonian.

It is convenient to describe the sterile sector by $\rho_+$ and
$\rho_-$, the CP-even and CP-odd deviations from equilibrium, rather
than $\rho_N$ and $\rho_{\bar{N}}$,
\begin{eqnarray}
\begin{tabular}{c c}
$\rho_N-\rho^{eq}=\rho_+ +\frac{\rho_-}{2}$  ,&  $\rho_{\bar{N}}-\rho^{eq}=\rho_+ -\frac{\rho_-}{2}$.
\end{tabular}
\end{eqnarray}
In terms of $\rho_+$ and $\rho_-$, (\ref{kinequ1})-(\ref{kinequ3}) read
\begin{eqnarray}
i \frac{d\rho_{+}}{d X}&=& [{\rm Re}H, \rho_{+}] -\frac{i}{2}\{{\rm Re}\Gamma_N, \rho_{+} \} + S_+~\label{kinequ1new},\\
i \frac{d\rho_{-}}{d X}&=& [{\rm Re}H, \rho_{-}] -\frac{i}{2}\{{\rm Re}\Gamma_N, \rho_{-} \} +S_- ,\label{kinequ2new}~\\
i\frac{d\mu_\alpha}{dX}&=& -i\Gamma^\alpha_L \mu_\alpha + S_\mu ,\label{kinequ3new}
\end{eqnarray}
with
\begin{eqnarray}
S_+&=&-i\frac{d\rho^{eq}}{dX}+ \frac{i}{2}[{\rm Im}H, \rho_{-}] +\frac{1}{4}\{{\rm Im}\Gamma_N, \rho_{-} \}- \frac{1}{2} \mu_\alpha{\rm Im}{\tilde\Gamma^\alpha}_N ,\\
S_-&=& 2i[{\rm Im}H, \rho_{+}] +\{{\rm Im}\Gamma_N, \rho_{+} \}+i \mu_\alpha{\rm Re}{\tilde\Gamma^\alpha}_N ,\\
S_\mu&=& i{\rm tr}\left[{\rm Re}({\tilde \Gamma^\alpha}_L)\rho_{-}\right] -
2 {\rm tr}\left[{\rm Im}({\tilde \Gamma^{\alpha}}_L)\rho_{+}\right].\label{musource}
\end{eqnarray}
Equations (\ref{kinequ1new})-(\ref{musource}) are the basis of our
numerical studies.


\subsection{Computation of the Rates}
\label{ratecompsec}
\begin{figure}
  \centering
    \includegraphics[width=10cm]{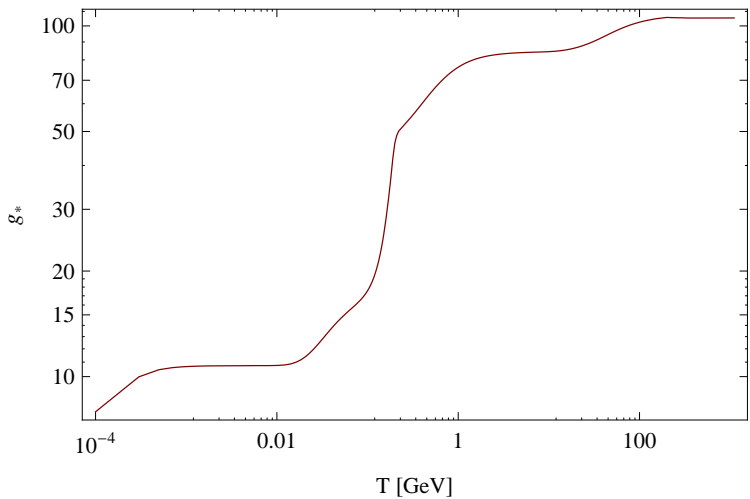}
    \caption{Number of effective relativistic degrees of freedom $g_*$
as a function of temperature \cite{Laine:2006cp}.
\label{gstar}}
\end{figure}

The rates appearing in (\ref{kinequ1new})-(\ref{musource}) can be
expressed as
\begin{eqnarray}
\Gamma_N&=&\uptau\phantom{i} \sum_{\alpha}\big(\tilde{F}_{\alpha I}\tilde{F}^*_{\alpha J}R(T,M)_{\alpha\alpha}\nonumber\\&&+\tilde{F}^*_{\alpha I}\tilde{F}_{\alpha J}R_M(T,M)_{\alpha\alpha}\big),\label{rates1}\\
(\tilde{\Gamma}_L^\alpha)_{IJ} \simeq (\tilde{\Gamma}_N^\alpha)_{IJ}&=& \uptau\phantom{i} \big( \tilde{F}_{\alpha I}\tilde{F}^*_{\alpha J}R(T,M)_{\alpha\alpha}\nonumber\\&&-\tilde{F}^*_{\alpha I}\tilde{F}_{\alpha J}R_M(T,M)_{\alpha\alpha}\big),\label{rates2} \\
\Gamma_L^\alpha
&=&\uptau\phantom{i} \big((FF^\dagger)_{\alpha\alpha}\left(R(T,M)_{\alpha\alpha}+R_M(T,M)_{\alpha\alpha}\right)\big)\label{rates3} 
,\end{eqnarray}
with no sum over $\alpha$ in (\ref{rates2}) and (\ref{rates3}).  The
flavor matrices $R$ and $R_M$ are defined in appendix
\ref{HamiltonianAppendix}, see  (\ref{Rdef}) and (\ref{RMdef}). Here
$\tilde{F}=F U_N$ and 
\begin{equation}
\uptau\phantom{i}=\frac{\partial t}{\partial X}=-\frac{T^2}{M}\frac{\partial}{\partial T}\frac{M_0}{2T^2},
\end{equation} 
where $T^2/M_0$ is the Hubble rate and
$M_0=M_P(45)^{1/2}/(4\pi^3g_*)^{1/2}$ with the effective number of
relativistic degrees of freedom $g_*$ computed in \cite{Laine:2006cp}
and shown in figure \ref{gstar}.

The flavor matrices $R(T,M)$ and $R_M(T,M)$ are almost diagonal since
off-diagonal elements of $\rho_{\alpha\beta}(p)$ include active
neutrino oscillations, which are at least suppressed by $m_i/T$. We
will always neglect the off-diagonal elements. $R(T,M)$ and $R_M(T,M)$
contain contributions from decays and scatterings. In finite
temperature field theory these can be associated with different cuts 
\cite{Bedaque:1996af} through the $N_I$ self-energy shown in figure
\ref{Nselfenergies}. The scatterings keep the $N_I$ in thermal
equilibrium for $T>T_-$. At $T\simeq T_-$ they become inefficient and
the sterile neutrinos freeze out. Due to their small coupling they are long-lived, but unstable and decay at a
temperature $T_d$. For $T_d\ll T_-$, decay and freezeout are two
separate processes and can be treated independently. This is the case
in the interesting part of the $\nu$MSM parameter space.

\subsubsection{Baryogenesis}
\label{baryogenrates}
For $T>v$ the SM fields are light and $M_M$ is negligible. We can
therefore neglect $R_M(T,M)$ as well as the flavor dependence of
$R(T,M)$.  Dispersion relations as well as relaxation rates are
dominated by contributions from scatterings between $\nu_R$ and Higgs
and lepton particles. The corresponding rates can be extracted from
cuts through the diagrams shown in figure \ref{Nselfenergies}$b)$.
They have been computed in \cite{Asaka:2005pn,Canetti:2010aw} 
\begin{eqnarray}
\label{R_BAU}
R(T,M)=0.02\frac{T}{4}\mathbbm{1}_{2\times 2}, \ \phantom{x} R_M(T,M)=0 \ 
{\rm for} \ T\gtrsim T_{EW}.
\end{eqnarray}

\subsubsection{Dark Matter Production}
\label{DMproductionrates}
For $T_d\ll T_-$, which is the case in the interesting part of the
$\nu$MSM parameter space, freezeout and decay happen in different
temperature regimes. At temperatures $T\gtrsim T_-$ the processes that
keep the plasma in equilibrium are scatterings mediated by the weak
interaction. Furthermore, the lepton masses are in first approximation
negligible\footnote{For some parameter choices this assumption can be
violated for the $\tau$ mass, introducing a small error.}. Thus,
$R(T,M)$ and $R_M(T,M)$ are proportional to the unit matrix and can be
described by two scalar functions $R^{(S)}(T,M)$ and
$R_M^{(S)}(T,M)$.  Around $T_-\sim M$, $R_M^{(S)}(T,M)$ gives a small
correction, which we account for in our analysis.
\begin{figure}
  \centering
    \includegraphics[width=10cm]{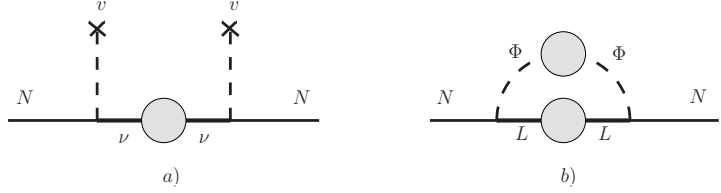}
    \caption{Contributions to the $N_I$ self energies. Diagram $a)$
dominates for $T<v$, diagram $b)$ for $T>v$. $\Gamma_N$ is obtained
from the discontinuity of the diagrams \cite{Weldon:1983jn}, which can
be computed by cutting it in various ways \cite{Bedaque:1996af}. The
gray self energy blobs indicate that dressed lepton and Higgs
propagators have to be used. Cuts through them reveal a large number
of processes, which are summarized in \cite{Asaka:2006nq,Asaka:2006rw}
and appendix A of \cite{Shaposhnikov:2008pf}. Recently it has been
pointed out that current estimates suffer from an error
$\mathcal{O}(1)$ due to infrared and collinear enhancements at high temperature.
Systematic approaches to include these effects can be found in  \cite{Anisimov:2010gy,Besak:2012qm} for $T> M$ (relevant for baryogenesis) and \cite{Laine:2011xm,Laine:2011pq,Salvio:2011sf} for $M>T$ (relevant for late time asymmetries).   
We ignore this effect in our
current study as it is comparable to other uncertainties in the
kinetic equations and would only slightly change the results for the
relevant regions in the $\nu$MSM parameter
space.\label{Nselfenergies}}
\end{figure} 
In practice they have to be computed numerically. They are displayed
in figure \ref{RRMplots}.
\begin{figure}
\centering
  \begin{tabular}{c c}
    \includegraphics[width=6.5cm]{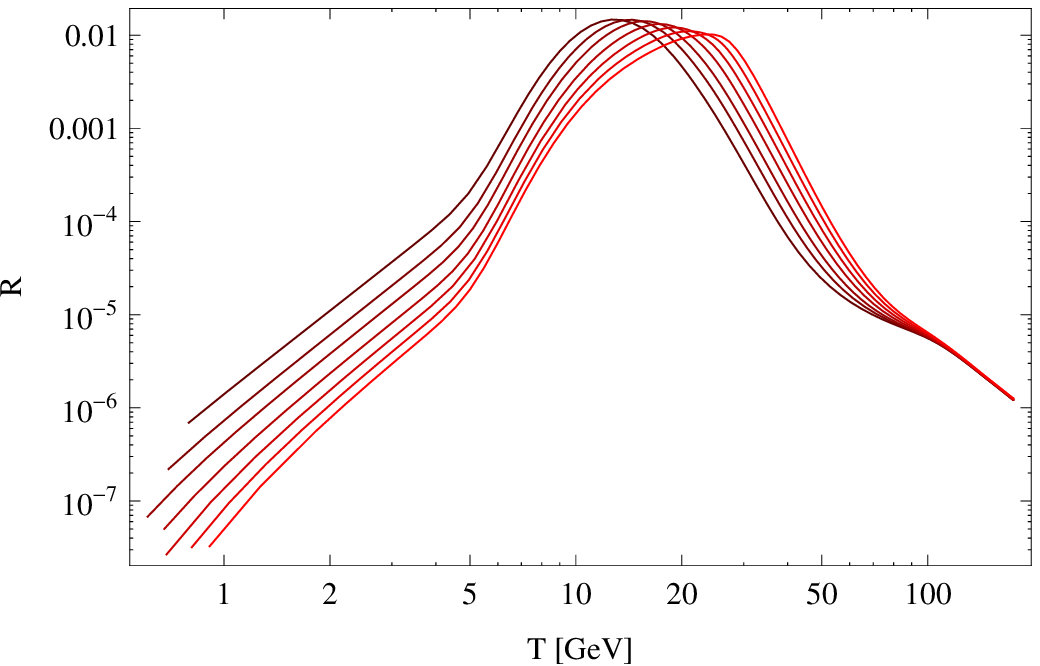}
&
\includegraphics[width=6.5cm]{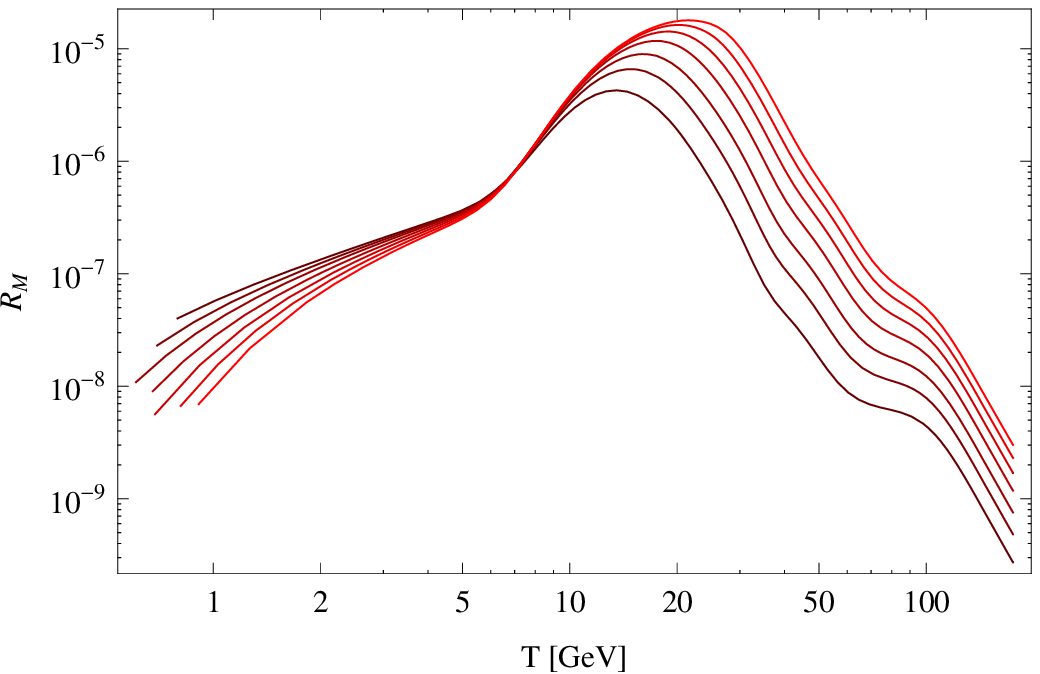}
\end{tabular}
    \caption{The functions $R^{(S)}(T,M)$ and $R_M^{(S)}(T,M)$ for
$M=1.2$ GeV (darkest curve), $M=1.6$ GeV, $M=2$ GeV, $M=2.5$ GeV,
$M=3$ GeV, $M=3.5$ GeV, $M=4$ GeV (lightest curve). We would like to
thank Mikko Laine for providing us numerical data.\label{RRMplots}}
\end{figure}

To be specific, in the high temperature limit, when all lepton masses
are negligible, (\ref{rates1}) simplifies to   
\begin{equation}
\Gamma_N\simeq\uptau U_N^T\left((F^\dagger F)^*R^{(S)}(T,M)+F^\dagger
F R_M^{(S)}(T,M)\right)U_N^* \ \phantom{X} {\rm at} \ T\sim
T_-\label{GammaNhighT}.
\end{equation}
In the low temperature regime, where $R\simeq R_M$, one finds
\begin{equation}
\Gamma_N\simeq\uptau\sum_{\alpha}{\rm Re}\left(\tilde{F}_{\alpha
I}\tilde{F}^*_{\alpha J}\right)R^{(D)}_{\alpha}(M)  \ \phantom{X} {\rm at} \ T\sim
T_d .
\end{equation}
The indices $^{(S)}$ and $^{(D)}$ indicate that the dominant
contribution to the rate comes from scatterings or decays,
respectively. The functions $R^{(S)}(T,M)$ and $R_M^{(S)}(T,M)$ can be
obtained from the discontinuity of the $N_I$ self energies shown in
figure \ref{Nselfenergies} at finite temperature.  At $T< M$ the
different SM lepton masses become relevant.

When the sterile neutrinos decay around $T_d\ll M$,  the flavors are
distinguishable. The decaying particle is non-relativistic and the
density of the surrounding plasma low.  For $T=0$, $R(0,M)=R_M(0,M)$
take the same values and their elements are given by 
$R(0,M)_{\alpha\alpha}=R_M(0,M)_{\alpha\alpha}=R^{(D)}_\alpha(M)$,
where $R^{(D)}_\alpha(M)$ are functions that can be computed from the
vacuum decay rates for sterile neutrinos. The $N_I$ can decay into
various different final states, see appendix \ref{DecayRatesAppendix}.
There are leptonic and semi-leptonic channels, depending on the
temperature either with quarks (before hadronization) or mesons (after
hadronization) in the final state. Let $\Upgamma_{N_I\rightarrow
\psi_\alpha}$ be the rate at which $N_I$ decays into a final state
$\psi_\alpha$ of flavor $\alpha$ (e.g. $\nu_{\alpha} q \bar{q}$ or
$\nu_\alpha \bar{L}_\beta L_\beta$). Then 
\begin{equation}
R^{(D)}_\alpha(M)=2\sum_{\psi_\alpha}\frac{\Upgamma_{N_I\rightarrow
\psi_\alpha}}{|\tilde{F}_{\alpha I}|^2}\label{zerotemperatureR},
\end{equation}
where the sum runs over all possible final states that have flavor
$\alpha$. The factor $2$ is due to the equal probabilities for decay
into particles and antiparticles at tree level. The simple form of
(\ref{zerotemperatureR}) is a result of the fact that the Yukawa
couplings can be factored out of the corresponding amplitudes and the
kinematics of $N_2$ and $N_3$ is the same due to their degenerate
mass. Most of the rates required for our study have been computed in
\cite{Gorbunov:2007ak}, the remaining ones are given in appendix
\ref{DecayRatesAppendix}. For $T\sim T_d\neq 0$ with $T_d\ll M$ the sterile neutrinos are non-relativistic and one can estimate
\begin{eqnarray}
R(T\ll M,M)_{\alpha\alpha}&\simeq& \frac{M}{E}R^{(D)}_\alpha(M),\\
R_M(T\ll M,M)_{\alpha\alpha}&\simeq&
\frac{E-\bar{p}}{E+\bar{p}}\frac{M}{E}R^{(D)}_\alpha(M),\label{RMT0}
\end{eqnarray}
with $E=(M^2+\bar{p}^2)^{1/2}$. Here $\bar{p}\sim T$ is the average
sterile neutrino momentum.\\
\hspace{0.5cm}
We therefore do not need all elements of the matrices $R(T,M)$ and
$R_M(T,M)$ at all temperatures, but only two functions $R^{(S)}(T,M)$
and $R_M^{(S)}(T,M)$ for $T\gtrsim T_-$ and three other functions
$R^{(D)}_\alpha(M)$ for $T\lesssim T_d\ll M$. In practice we can
simply add these contributions at all temperatures, though in
principle we do not known the scattering contribution outside the
range plotted in figure \ref{RRMplots} and the decay contribution is
obtained from vacuum rates. This is justified because for $T\gtrsim
T_-$ our expressions for the decay rates are incorrect, but 
$R^{(D)}_\alpha(M)\ll R^{(S)}(T,M)$. On the other hand
$R^{(D)}_\alpha(M)\gg R_M^{(S)}(T,M)$ in the regime $T\ll M,T_-$,
which is not covered by the data shown in figure \ref{RRMplots}. For
$T_d<T<T_-$ our expressions for both, decay and scattering rates, are
incorrect, but they are both smaller than the rate of Hubble expansion and have negligible effect. 


\section{Baryogenesis from Sterile Neutrino Oscillations}
\label{baryogenesissection}
The BAU in the $\nu$MSM is produced during the thermal production of
sterile neutrinos $N_I$. This is in contrast to most other (thermal)
leptogenesis scenarios, where decays and inverse decays play the
central role. The violation of total fermion number by the Majorana
mass term $M_M$ is negligible at $T_{EW}\gg M$, but asymmetries in the
helicity states of the individual flavors can be created. The sum
over these vanishes up to terms suppressed by $M/T_{EW}$, but because
sphaleron processes only act on left chiral fields, the generated BAU
can be much bigger. In this sense baryogenesis in the $\nu$MSM can be
regarded as a version of ``flavored leptogenesis''.

In this section we explore the part of the $\nu$MSM parameter space
where a BAU consistent with (\ref{BAUvalue}), i.e.  $\eta\sim 10^{-10}$,
can be generated. 
We assume two sterile
neutrinos $N_{2,3}$ participate in baryogenesis, as required in scenarios I and II. This assumption is
motivated by the premise that $N_1$ should be a valuable DM candidate,
with masses and mixing consistent with astrophysical bounds. These
require its Yukawa interaction to be too small to be relevant for
baryogenesis, see section \ref{ExistingBounds}. In this sense, we
consider the $\nu$MSM as a model of both, baryogenesis and DM
production, but are not concerned with the DM
production mechanism, which is discussed in the following section
\ref{DMproductionSection}. This corresponds to scenario II.
The requirement to explain $\Omega_{DM}$
only enters implicitly, as we demand the $N_1$ mass and mixing to be
consistent with astrophysical observations. If one completely drops
the requirement to explain the observed DM and study the $\nu$MSM as a
theory of baryogenesis and neutrino oscillations only (scenario III), 
the resulting bounds on the parameters weaken considerably. In particular, it was found in \cite{Drewes:2012ma} that no mass degeneracy between the
sterile neutrino masses is needed.

We extend the analysis performed in \cite{Canetti:2010aw}, but take
into account two additional aspects. First, we use the non-zero value
for the active neutrino mixing angle $\uptheta_{13}$ given in table
\ref{NeutrinoParameters}, which brings in a new source of CP-violation
through the phase $\delta$.  Second, we include the contribution from
the temperature dependent Higgs expectation value $v(T)$ to the
effective Hamiltonian, coming from the real part of the diagram in
figure \ref{Nselfenergies}a). It is relevant for temperatures close to
the electroweak scale.

We solve numerically the system of equations
(\ref{kinequ1new})-(\ref{musource}) to find the lepton asymmetries at
$T\sim T_{EW}$, assuming that there is no initial asymmetry. The
effective Hamiltonian is given by (\ref{Hamiltonian_highT}) and
(\ref{rates1})-(\ref{rates3}) with (\ref{R_BAU}). We fix the active
neutrino masses and mixing angles according to table
\ref{NeutrinoParameters}  and choose the phases $\delta$, $\alpha_1$
and $\alpha_2$ as well as $\Re\upomega$ to maximize the asymmetry.
Interestingly, for normal hierarchy of neutrino masses, the value of
${\rm Re}\upomega$ that maximizes the asymmetry is close to
$\frac{\pi}{2}$, as required in the constrained $\nu$MSM.  This allows
to identify the region in the remaining three-dimensional parameter
space consisting of $M$, $\Delta M$ and $\Im\upomega$ where an
asymmetry $\gtrsim 10^{-10}$ can be created. Deep inside this region,
the asymmetry generated for this choice of phases can be much too
large, but it can always be reduced by choosing different phases. Thus,
any choice of $M$, $\Delta M$ and $\Im\upomega$ inside this region can
reproduce the observed BAU. 

In practice it is difficult to find the phases that maximize the
asymmetry in each single point, as we are dealing with a
seven-dimensional parameter space. However, the analysis can be
simplified.  First, the choice of phases that maximize the asymmetry
practically does not depend on $\Delta M$ because the dependence of
the Yukawa coupling (\ref{CasasIbarraDef}) on $\Delta M$ is very
weak.  Second, our numerical studies reveal in most of the parameter
space ${\rm Im}\upomega$ is the main source of CP-violation. The other
phases have comparably little effect on the final asymmetry, except
for the region around ${\rm Im}\upomega=0$. Surprisingly, the values
for $\delta$, $\alpha_1$ and $\alpha_2$ that maximize it vary only
very little and are always close to zero. One possible interpretation
is that ${\rm Im}\upomega$ provides the main source for the asymmetry
generation, while $\delta$, $\alpha_1$ and $\alpha_2$ contribute
stronger to the washout. However, due to the various different time
scales involved we cannot extract a single CP-violating parameter at
this point, which is commonly used in thermal leptogenesis scenarios
to study such connections analytically. The above seems to be valid
everywhere except in the region ${\rm Im}\upomega\sim 0$, where
$\delta$, $\alpha_1$ and $\alpha_2$ are the only sources of
CP-violation.

We therefore split the parameter space into two regions. In the region
$0.5<e^{{\rm Im}\upomega}<1.5$ we chose  $\alpha_2 = \pi$, $\delta =
0$, ${\rm Re}\upomega= \frac{7}{10}\pi$ for normal hierarchy and 
$\alpha_2-\alpha_1=\pi$,  $\delta = \pi$, ${\rm Re}\upomega=
\frac{3}{4}\pi$ for inverted hierarchy. Everywhere else we chose 
$\alpha_2 = \frac{7\pi}{5}$ and $\delta = \frac{3}{20}\pi$, ${\rm
Re}\upomega= \frac{1}{2}\pi$ for normal hierarchy and
$\alpha_2-\alpha_1=\frac{11}{10}\pi$, $\delta = \frac{11}{20}\pi$,
${\rm Re}\upomega= \frac{4}{5}\pi$ for inverted hierarchy. Note that
for normal hierarchy $F$ only depends on $\alpha_2$ and $\delta$,
while for inverted hierarchy it depends on $\alpha_2-\alpha_1$ and
$\delta$ because one neutrino is massless.   In order to determine
$v(T)$ one needs to fix the Higgs mass $m_H$. We used the value $m_H =
126$ GeV suggested by recent LHC data \cite{:2012gk,:2012gu},
corresponding to electroweak scale of $T_{EW} \sim 140$ GeV. This is
consistent with the $\nu$MSM being a valid description of nature up to
the Planck scale \cite{Bezrukov:2012sa}.

\begin{figure}[!h]
\centering
\begin{tabular}{cccc}
\includegraphics[width=10cm]{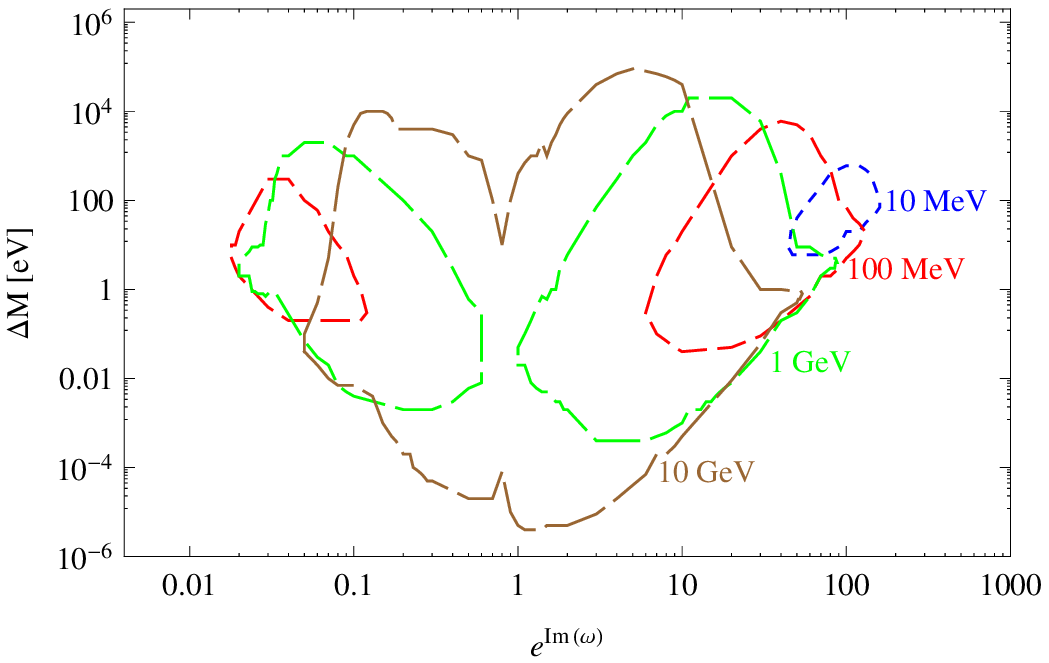} \\
\includegraphics[width=10cm]{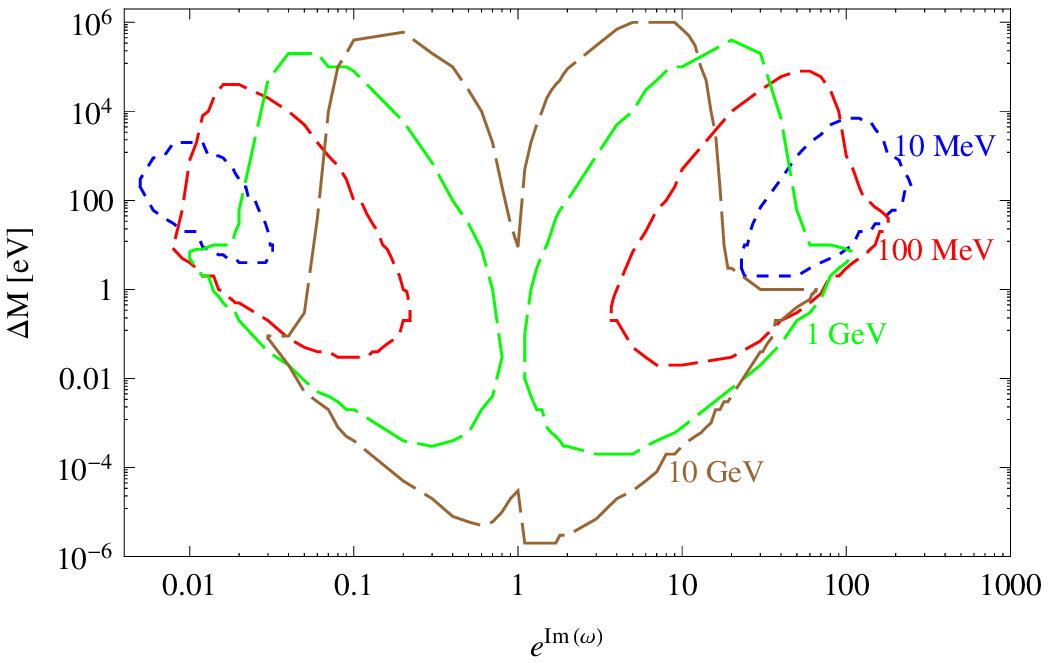}
\end{tabular} \caption{Values of $\Delta M$ and ${\rm Im}\upomega$
that lead to the observed baryon asymmetry in scenarios I and II for different sterile
neutrino masses $M = 10$, $100$ MeV, $1$ and $10$ GeV.  The blue
(shortest dashed) line corresponds to $M = 10$ MeV, red (short dashed)
to $M = 100$ MeV, brown (long dashed) to $M=1$ GeV and green (longest
dashed) to $M = 10$ GeV.   The phases that maximize the asymmetry
differ significantly for ${\rm Im}\upomega \approx 0$ and away from
that region. In the region $0.5<e^{{\rm Im}\upomega}<1.5$ we chose 
$\alpha_2 = \pi$, $\delta = 0$, ${\rm Re}\upomega= \frac{7}{10}\pi$
for normal hierarchy and  $\alpha_2-\alpha_1=\pi$,  $\delta = \pi$,
${\rm Re}\upomega= \frac{3}{4}\pi$ for inverted hierarchy. Everywhere
else we chose 
$\delta = \frac{3}{20}\pi$, ${\rm Re}\upomega= \frac{1}{2}\pi$ for
normal hierarchy and $\alpha_2-\alpha_1=\frac{11}{10}\pi$, $\delta =
\frac{11}{20}\pi$, ${\rm Re}\upomega= \frac{4}{5}\pi$ for inverted
hierarchy.  The upper panel shows the results for normal hierarchy,
the lower panel for inverted hierarchy. \label{BA_Max_contour}}
\end{figure}

We present our results in figure \ref{BA_Max_contour}, which shows the
allowed region in the $\Delta M - \Im\upomega$ plane for several
masses $M$. The lines correspond to the exact observed asymmetry,
inside more asymmetry is generated. As pointed out above, any point
inside the lines is consistent with observation because the asymmetry
can be reduced by choosing different phases. 
Figure \ref{BA_Max_contour} shows that even for small masses around
$10$ MeV enough asymmetry can be created. However, for small masses
the CP-violation contained in $\delta$, $\alpha_1$ and $\alpha_2$ is
not sufficient, and the allowed region consists of two disjoint parts
that are separated by the ${\rm Im}\upomega\simeq 0$ region. The area
of these increases with $M$. For masses of a few GeV, the CP-violation
from $\delta$, $\alpha_1$ and $\alpha_2$ alone is sufficient and the
regions join. Interestingly, there appear to be mass-independent
diagonal lines in the $\Delta M - e^{{\rm Im}\upomega}$ plane that
confine the region where enough asymmetry can be generated. We
currently have not understood the origin of these lines
parametrically. The inverted hierarchy generally allows to produce
more asymmetry than the normal hierarchy.  There is an approximate
symmetry between regions with positive and negative ${\rm
Im}\upomega$.  It would be exact when simultaneously changing $\xi$
and is related to the symmetry of the Lagrangian under exchange of
$N_2$ and $N_3$. As expected, these results are close to those
obtained in \cite{Canetti:2010aw}, which provides a good consistency
check. The slightly bigger asymmetry is due to the additional source
of CP-violation for $\uptheta_{13}\neq 0$.

\begin{figure}[!h]
\centering
\includegraphics[width=10cm]{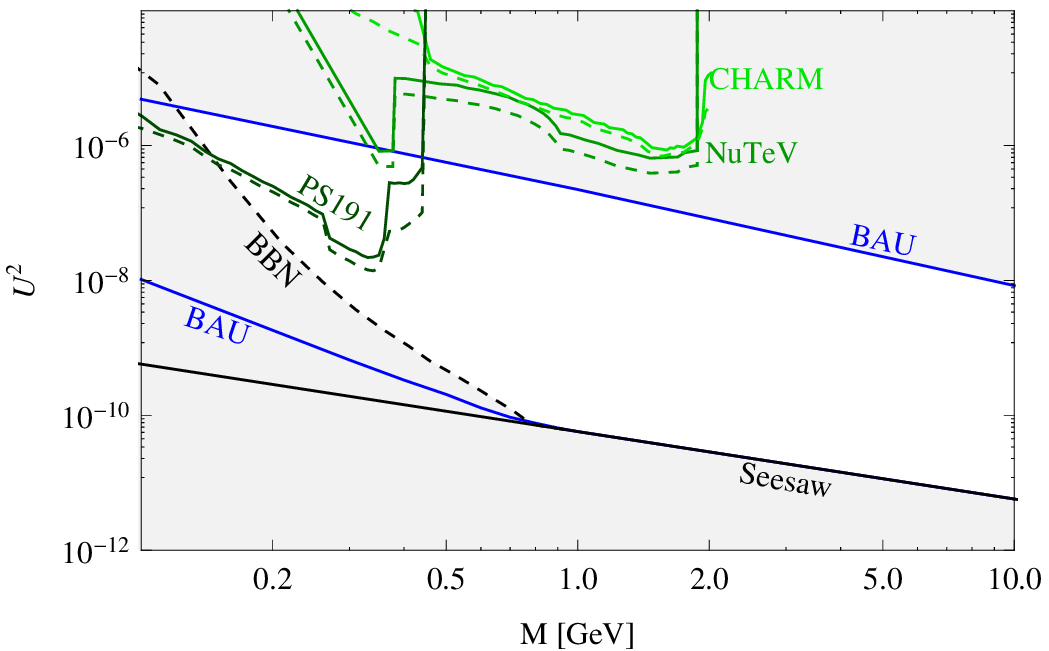} 
\includegraphics[width=10cm]{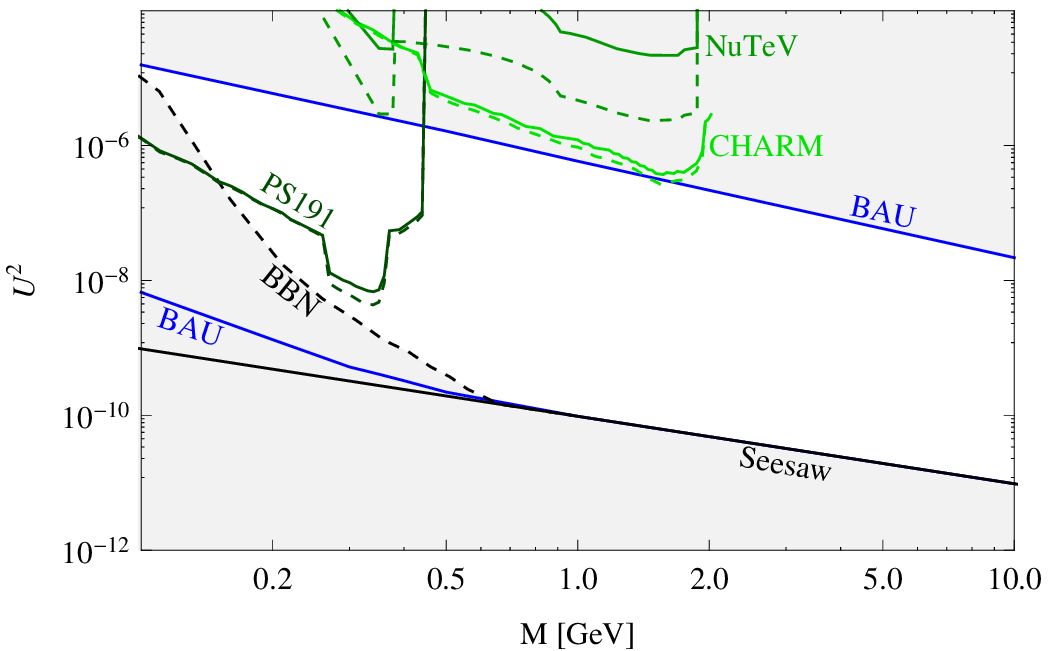}
\caption{Constraints on the $N_{2,3}$ masses $M_{2,3}\simeq M$ and
mixing $U^2={\rm tr}(\theta^\dagger\theta)$ from baryogenesis in scenarios I and II; upper panel - normal
hierarchy, lower panel - inverted hierarchy. In the region between the
solid blue ``BAU'' lines, the observed BAU can be generated. The
regions below the solid black ``seesaw'' line and dashed black ``BBN''
line are excluded by neutrino oscillation experiments and BBN,
respectively. The areas above the green lines of different shade are
excluded by direct search experiments, as indicated in the plot. The
solid lines are {\it exclusion plots} for all choices of $\nu$MSM
parameters, for the dashed lines the phases were chosen to maximize
the BAU, consistent with the blue lines.
 \label{BA_Max_mixingangle}}
\end{figure}

For experimental searches, the most relevant parameters are the mass
$M$ and the mixing between active and sterile neutrinos. In figure
\ref{BA_Max_mixingangle} we translate our results into bounds on the
flavor independent mixing parameter $U^2$ defined in equation
(\ref{Mixing_U}). Using the results displayed in figure
\ref{BA_Max_contour}, we chose $\delta M$ to maximize the asymmetry
and find the region in the $U^2-M$ plane within which baryogenesis is
possible. The plot has to be read as follows: For each point in the
region between the blue lines there exists at least one choice of
$\nu$MSM parameters that allows for successful baryogenesis. 
The plots in figure \ref{BA_Max_mixingangle} are similar to the ones
of figure $3$ in \cite{Canetti:2010aw}, but the allowed region is
slightly bigger due to the effect the new source of CP-violation for
$\uptheta_{13}\neq 0$. 

\begin{figure}[!h]
\centering
\includegraphics[width=10cm]{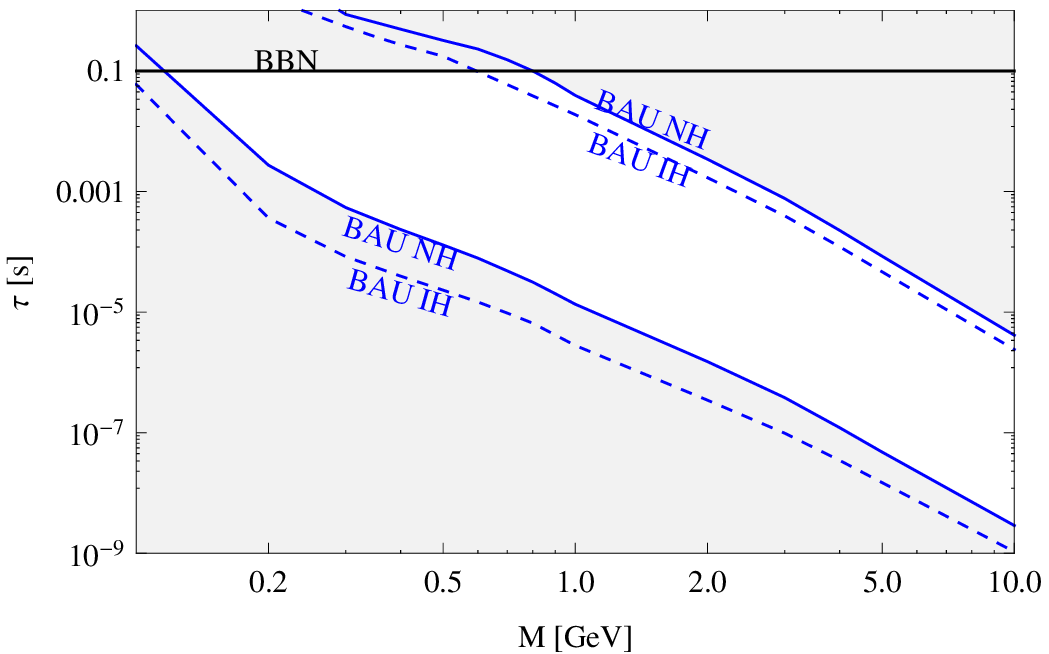}
\caption{Constraints on the $N_{2,3}$ masses $M_{2,3}\simeq M$ and
lifetime $\tau^{-1}\simeq\frac{1}{2}{\rm tr}\Gamma_N|_{T=1 {\rm MeV}}$ from baryogenesis in scenarios I and II.
In the region between the blue ``BAU'' lines, the observed BAU can be
generated (solid line - normal hierarchy, dotted line - inverted
hierarchy). The region above the solid black ``BBN'' line is excluded
by BBN.
\label{BA_Max_lifetime}}
\end{figure}

The constraints on the mixing angle $U^2$ shown in figure
\ref{BA_Max_mixingangle} can be translated into constraints on the
neutrino lifetime $\tau^{-1}\simeq\frac{1}{2}{\rm tr}\Gamma_N$ (at
$T=1$ MeV) shown in figure \ref{BA_Max_lifetime}.
The plot in figure \ref{BA_Max_lifetime} is similar to the ones of
figure $4$ in \cite{Canetti:2010aw}

\section{Late Time Lepton Asymmetry and Dark Matter Production}
\label{DMproductionSection}
The lepton asymmetry at temperatures of a few hundred MeV is of
crucial importance for the dark matter production in scenario I. 
Resonant dark matter production requires a lepton asymmetry
$|\mu_\alpha|\sim 8\cdot 10^{-6}$ in the plasma, much larger than the
baryon asymmetry. The details of this process have been outlined in
\cite{Asaka:2006rw}. Here we are not concerned with the dark matter
production itself, but with the mechanisms that generate the required
lepton asymmetry. This asymmetry must come from a source that is
different from that of the baryon asymmetry because $N_{2,3}$ reach
chemical equilibrium at some temperature $T_{+}<T_{EW}$ and the
asymmetry in the leptonic sector is washed out (while the baryon
asymmetry remains as sphalerons are inefficient at
$T<T_{EW}$)\footnote{It has been suggested that some asymmetry may be
preserved in magnetic fields down to temperatures $T<T_-$ due to the
chiral anomaly \cite{Boyarsky:2011uy}. Here we take the most
conservative approach and do not take into account this possibility.}.
There are two distinct mechanisms that contribute to the late time
asymmetry, the freezeout of $N_{2,3}$ at $T\sim T_{-}$ and their decay
at $T\sim T_d$.

The requirement that these two mechanisms produce enough asymmetry put
severe constraints on the parameters of the model, described in
section \ref{finetunings}. The value of $\Re\upomega$ is fixed to
values near $\pi/2$.  The mass splitting $\Delta M$ is limited to a
very narrow range by equation (\ref{finetunningDeltaM}).  Therefore we
will use the mass splitting in vacuum $\delta M$ instead of $\Delta M$
as a free parameter in the following. All experimentally known
parameters are fixed to the values given in table
\ref{NeutrinoParameters}.  The phases $\delta$, $\alpha_1$ and
$\alpha_2$ are chosen to maximize the asymmetry. As in section
\ref{baryogenesissection} we observe that in most of the parameter
space ${\rm Im}\upomega$ is the main source of CP-violation.  We again
find that it is convenient to split the parameter space into the
region $0.5<e^{{\rm Im}\upomega}<1.5$ and the complement. For normal
hierarchy we chose the phases  $\alpha_2=\frac{\pi}{2}$,
$\delta=\frac{3}{2}\pi$ in the region $0.5<e^{{\rm Im}\upomega}<1.5$
and  $\alpha_2=\frac{\pi}{5}$ and $\delta=0$ everywhere else.  For
inverted hierarchy we chose  $\alpha_2-\alpha_1=\frac{7}{5}$ and
$\delta=\frac{3}{5}\pi$ in the region $0.5<e^{{\rm Im}\upomega}<1.5$
and  $\alpha_2-\alpha_1=0$,  $\delta=\frac{9}{10}\pi$ everywhere else.
Note that for normal hierarchy $F$ only depends on $\alpha_2$ and
$\delta$, while for inverted hierarchy it depends on
$\alpha_2-\alpha_1$ and $\delta$ because one neutrino is massless. We
then study the parameter space spanned by $M$, $\delta M$ and
$\Im\upomega$.

As in section \ref{baryogenesissection}, we use the kinetic equations
(\ref{kinequ1new})-(\ref{musource}) in order to calculate the lepton
asymmetries as a function of $T$. The effective Hamiltonian is
calculated from (\ref{hamiltonian_lowT}) and
(\ref{rates1})-(\ref{rates3}) with (\ref{GammaNhighT})-(\ref{RMT0}).
We impose thermal equilibrium with vanishing chemical potentials as
initial condition at a temperature $T>T_-$ and look for the parameter
region where $\sum_\alpha|\mu_\alpha|>8\cdot 10^{-6}$ at $T=100$
MeV.\footnote{We solve the kinetic equations down to $T=50$ MeV in
order to avoid numerical artifacts at the boundary.}

\begin{figure}[!h]
\centering
\begin{tabular}{cccc}
\includegraphics[width=6.5cm]{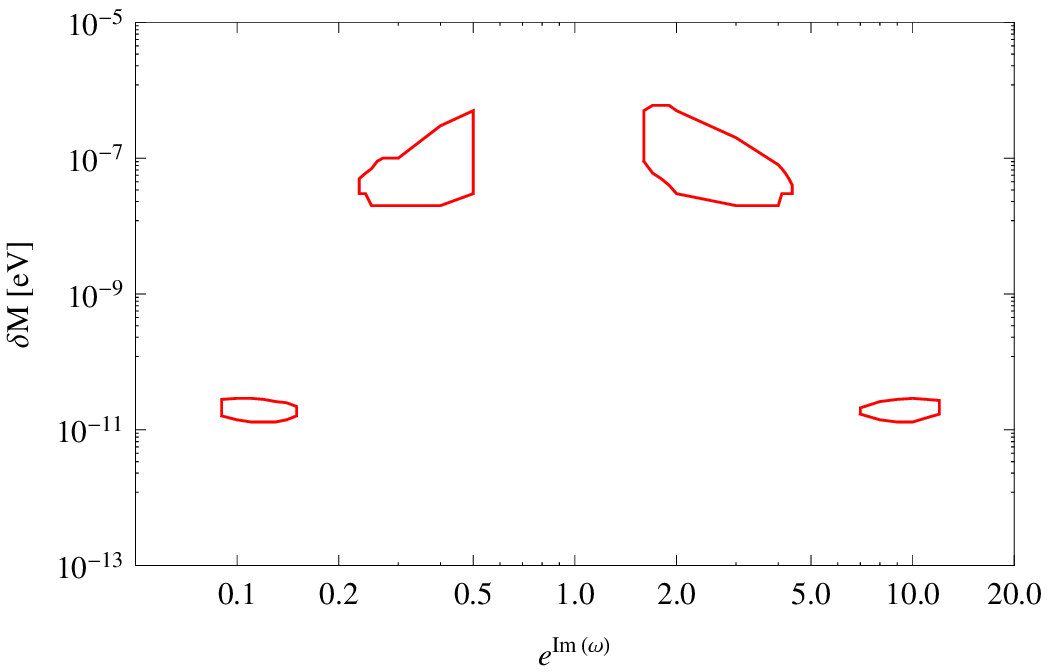} &
\includegraphics[width=6.5cm]{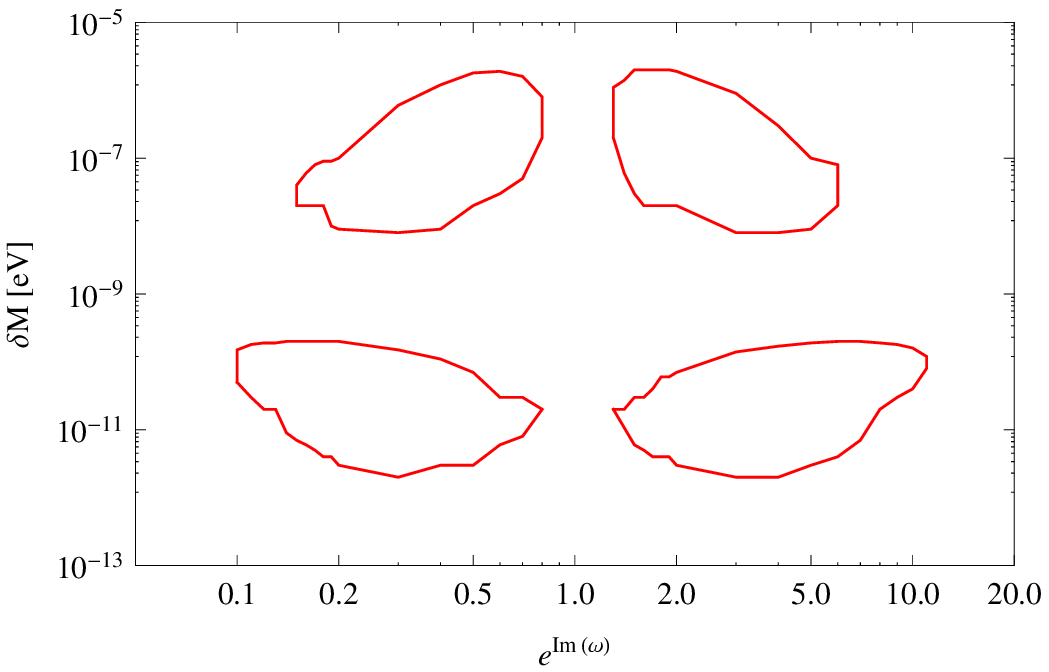} \\
\includegraphics[width=6.5cm]{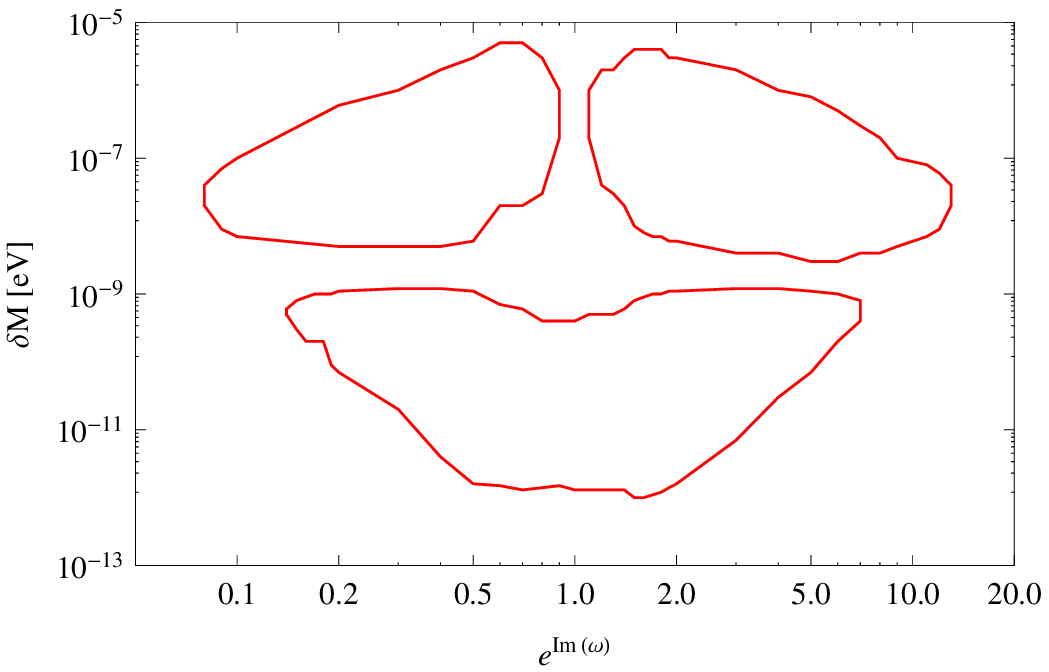} &
\includegraphics[width=6.5cm]{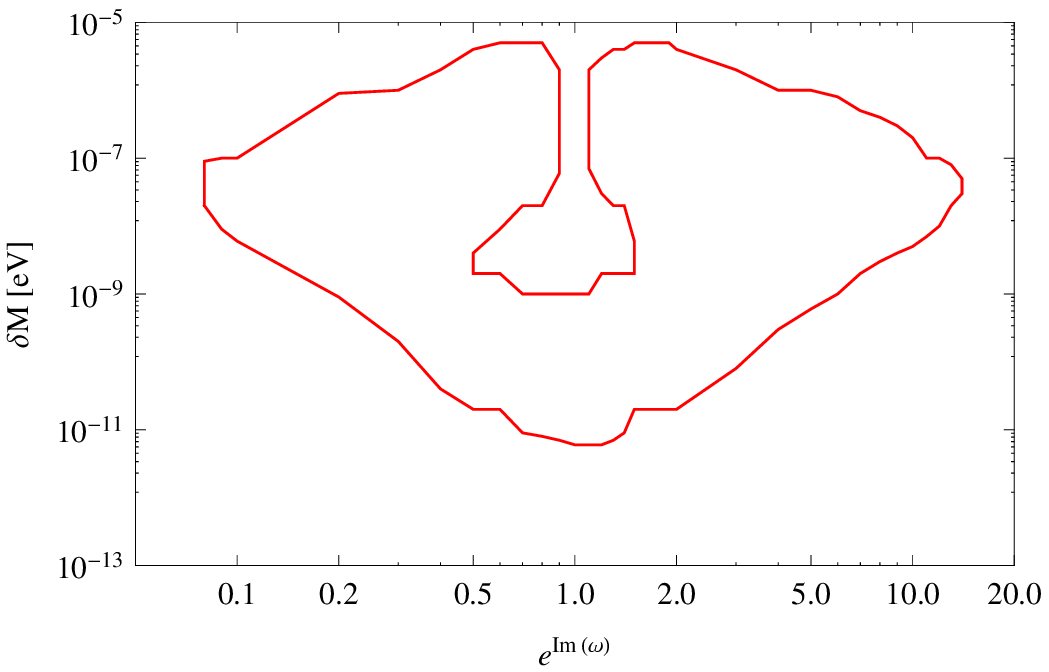}
\end{tabular}
\caption{Values of $\delta M$ and ${\rm Im}\upomega$ that lead to the
lepton asymmetry required for dark matter production in scenario I for
different singlet fermion masses, $M = 2.5$, $4$, $7$ and $10$ GeV and
for normal hierarchy. The upper left panel corresponds to $M = 2.5$
GeV, the upper right panel to $M = 4$ GeV, the lower left panel to
$M=7$ GeV and the lower right panel to $M = 10$ GeV. The phases that
maximize the asymmetry differ significantly for $Im\upomega \approx 0$
and away from that region.  We chose the phases 
$\alpha_2=\frac{\pi}{2}$, $\delta=\frac{3}{2}\pi$ in the region
$0.5<e^{{\rm Im}\upomega}<1.5$ and  $\alpha_2=\frac{\pi}{5}$ and
$\delta=0$ everywhere else. 
\label{DM_Max_contour_nor}}
\end{figure}

\begin{figure}[!h]
\centering
\begin{tabular}{cccc}
\includegraphics[width=6.5cm]{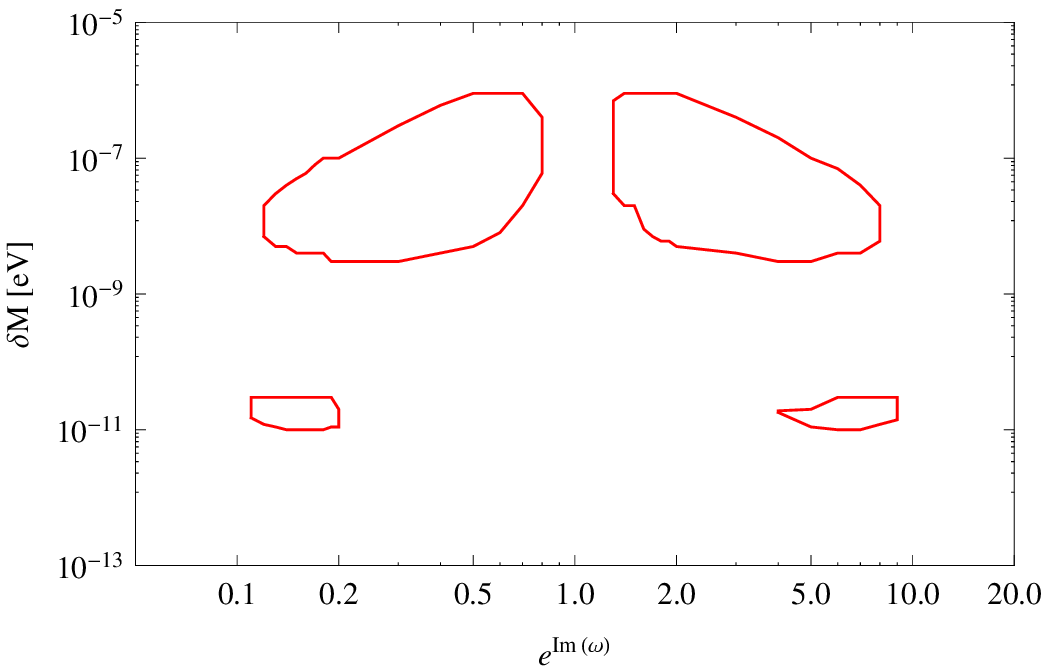} &
\includegraphics[width=6.5cm]{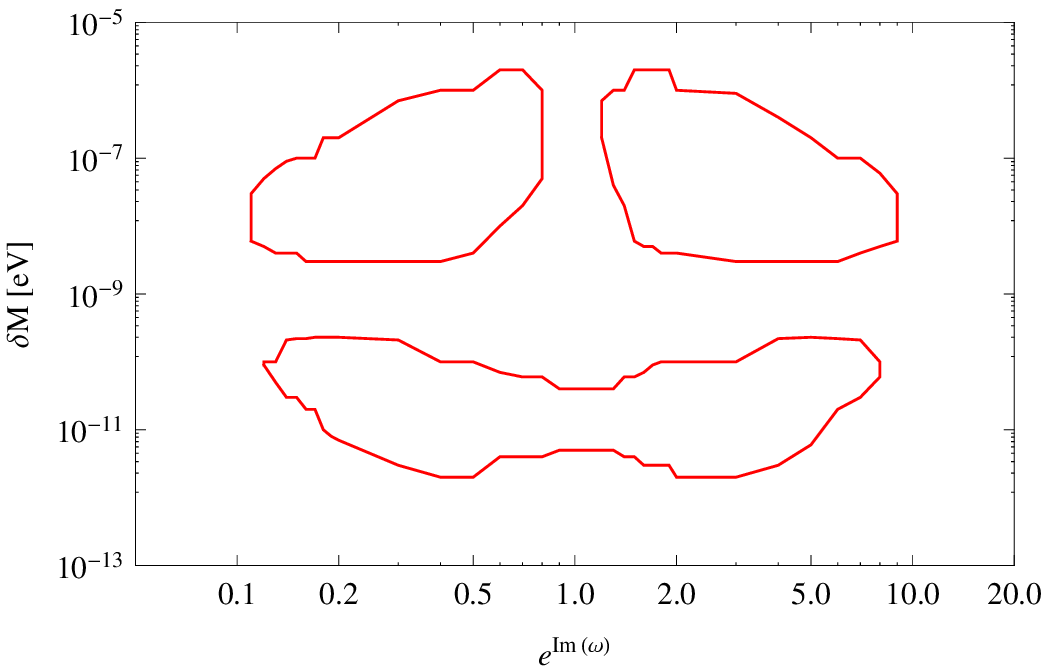} \\
\includegraphics[width=6.5cm]{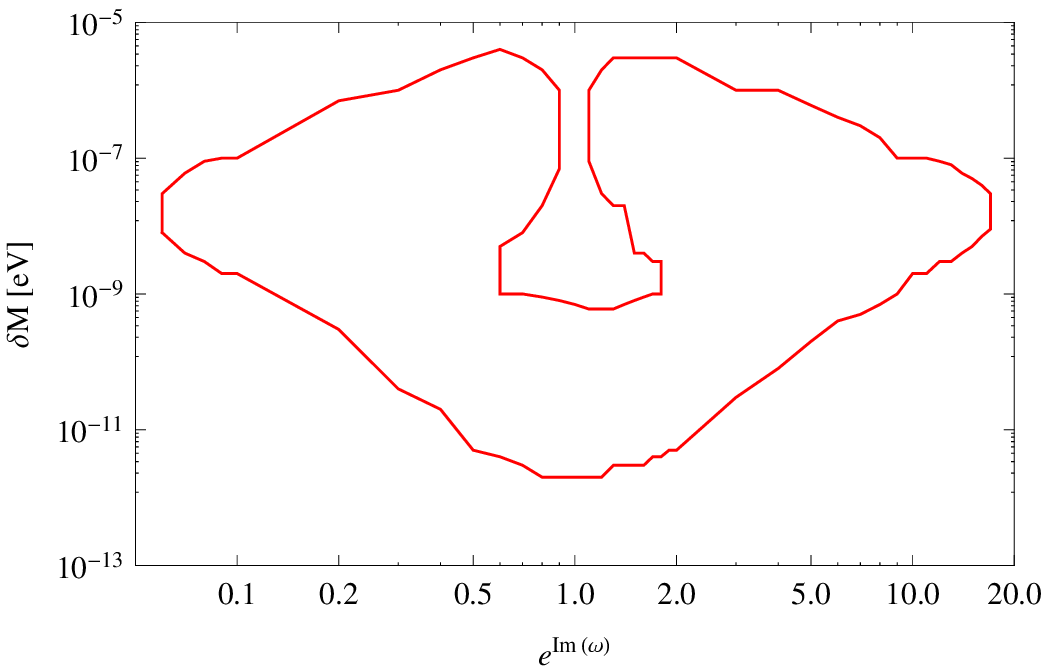} &
\includegraphics[width=6.5cm]{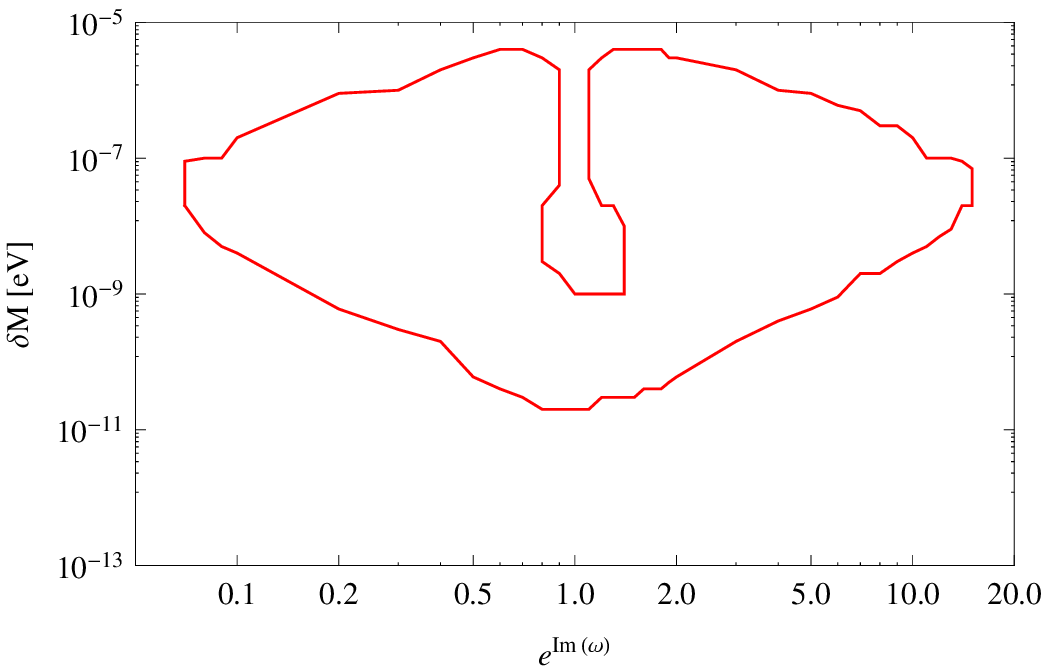}
\end{tabular}
\caption{Values of $\delta M$ and $Im\upomega$ that lead to the
lepton asymmetry required for dark matter production in scenario I for
different singlet fermion masses, $M = 2.5$, $4$, $7$ and $10$ GeV and
inverted hierarchy. The upper left panel corresponds to $M = 2.5$ GeV,
the upper right panel to $M = 4$ GeV, the lower left panel to $M=7$
GeV and the lower right panel to $M = 10$ GeV. The phases that
maximize the asymmetry differ significantly for $Im\upomega \approx 0$
and away from that region.  We chose  $\alpha_2-\alpha_1=\frac{7}{5}$
and $\delta=\frac{3}{5}\pi$ in the region $0.5<e^{{\rm
Im}\upomega}<1.5$ and  $\alpha_2-\alpha_1=0$, 
$\delta=\frac{9}{10}\pi$ everywhere else.
\label{DM_Max_contour_inv}}
\end{figure}

The results are shown in figures \ref{DM_Max_contour_nor} and
\ref{DM_Max_contour_inv}. The required asymmetry can be created when
the sterile neutrinos have masses in the GeV range. For small masses
of $M=2-4$ GeV the CP violation contained in $\alpha_1$, $\alpha_2$
and $\delta$ alone is not sufficient for normal hierarchy and barely
sufficient for inverted hierarchy;  a non-zero ${\rm Im}\upomega$ is
required and the allowed region consists of two disjoint parts along
the ${\rm Im}\upomega$ axis which are separated by the ${\rm
Im}\upomega\simeq 0$ region. For larger masses ($M\gtrsim7$ GeV for
normal hierarchy, $M\gtrsim 4$ GeV for inverted hierarchy ), the
regions merge, but ${\rm Im}\upomega$ continues to be the most
relevant source of CP violation in most of the parameter space. In
addition, one can also observe disjoint regions along the $\delta M$
axis. These can be identified with the {\it decay scenario} and {\it
freezeout scenario}. In the upper part of the figures, the asymmetry
is mainly created during the freezeout of $N_{2,3}$, in the lower part
during the decay.  At $T_-$, $\Gamma_N$ has considerably larger
entries than at $T_d$. Thus, the resonance condition (\ref{resoncond})
requires a smaller mass splitting in the decay scenario.  For larger
masses, both regions merge. However, freezeout and decay are always
two separated processes, i.e. $T_-\gg T_d$. As in figure
\ref{BA_Max_contour}, there is an approximate symmetry under a change
of sign for ${\rm Im}\upomega$, which is related to the symmetry of
the Lagrangian under exchange of $N_2$ and $N_3$ and becomes exact
when also changing $\xi$.


\begin{figure}[!h]
\centering
\begin{tabular}{cccc}
\includegraphics[width=10cm]{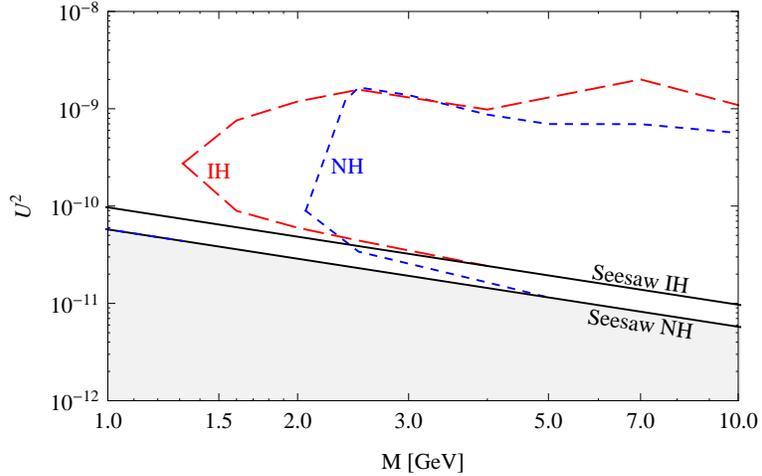}
\end{tabular}
\caption{Constraints on the $N_{2,3}$ masses $M_{2,3}\simeq M$ and
mixing $U^2={\rm tr}(\theta^\dagger\theta)$ in scenario I. The lepton asymmetry at
$T=100$ MeV can be large enough that the resonant enhancement of $N_1$
production is sufficient to explain the observed $\Omega_{DM}$ inside
the dashed blue and red lines for normal and inverted neutrino mass
hierarchy, respectively. The regions below the ``seesaw'' lines are
excluded by neutrino oscillation experiments for the indicated choice
of hierarchy.
\label{DM_Max_mixingangle}}
\end{figure}

For experimental searches for sterile neutrinos, the most relevant
parameters are the mass $M$ and the mixing between active and sterile
species. As in section \ref{baryogenesissection}, we translate our
results for the parameters in the Lagrangian into bounds on the mass
and mixing. For each mass, we chose $\delta M$ in a way that maximizes
the allowed region in the $U^2 - M$-plane. The results are shown in
figure \ref{DM_Max_mixingangle}.

\begin{figure}[!h]
\centering
\begin{tabular}{cccc}
\includegraphics[width=10cm]{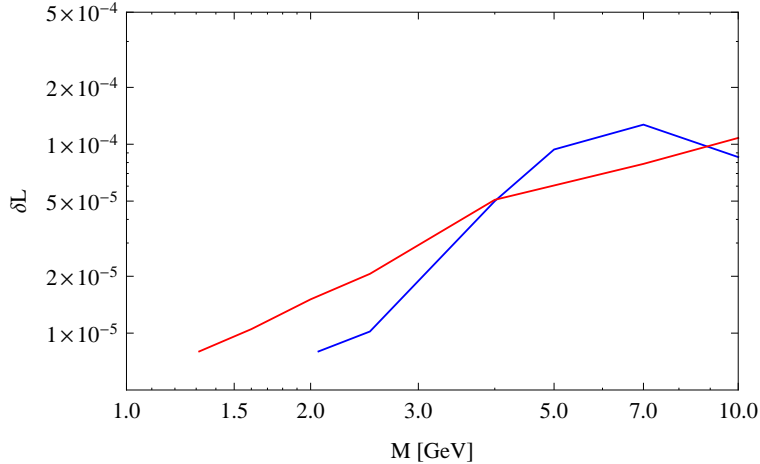}
\end{tabular}
\caption{Estimate of the maximal lepton asymmetry that can be created
in scenario I around $T=100$ MeV. The blue small dashed line line
corresponds to normal hierarchy, the red long dashed line to inverted
hierarchy. This plot was generated using the maximal values found in
the data files used to produce figures \ref{DM_Max_contour_nor} and
\ref{DM_Max_contour_inv}.\label{DM_Max_asymmetry}}
\end{figure}

Finally, we estimate the maximal asymmetry that can be generated at
$T\sim 100$ MeV as a function of $M$ by its largest value within the
data files we used to create figures \ref{DM_Max_contour_nor} and
\ref{DM_Max_contour_inv}. The maximal asymmetry allows to impose a
lower bound on the $N_1$ mixing; bigger lepton asymmetries make the
resonant DM production more efficient and allow for smaller $N_1$
mixing, displayed in figure \ref{DMexclusion}.  Furthermore, the
maximal $|\mu_\alpha|$ is of interest because in \cite{Schwarz:2009ii}
it was pointed out that a large lepton asymmetry may lead to a first
order phase transition during hadronisation. The maximal asymmetries
we found are shown in figure \ref{DM_Max_asymmetry}.  For both
hierarchies they remains well below cosmological bounds (see
\cite{Canetti:2012zc}) at all masses of consideration and are about a
factor $5$ smaller than the value $7\cdot 10^{-4}$ estimated in
\cite{Shaposhnikov:2008pf}. However, given the uncertainties
summarized in appendix \ref{Uncertainties}, they can easily change by
a factor $\mathcal{O}[1]$.


\section{DM, BAU and Neutrino Oscillations in the $\nu$MSM}
\label{Combined}
In the previous sections \ref{baryogenesissection} and
\ref{DMproductionSection} we have studied independently the conditions
for successful baryogenesis on one hand and sufficient dark matter
production on the other. The most interesting question is of course in
which part of the $\nu$MSM parameter space scenario I can be realized, i.e. both can be achieved
simultaneously. This region cannot be found by simply superposing the
figures from the previous sections because the phases that maximize
the asymmetry are different for $T\gtrsim T_{EW}$ and $T\lesssim T_-$.
The requirement to produce enough DM imposes the stronger constraint. 
We therefore fix the CP-violating phases in a way that is consistent
with $\sum_\alpha |\mu_\alpha|>8\cdot 10^{-6}$ at $T=100$ GeV in some
significant region in the $M-U^2$-plane.  We then check for which
combination of $M$ and $U^2$ the correct BAU is created by these
phases.

\begin{figure}[!h]
\centering
\includegraphics[width=10cm]{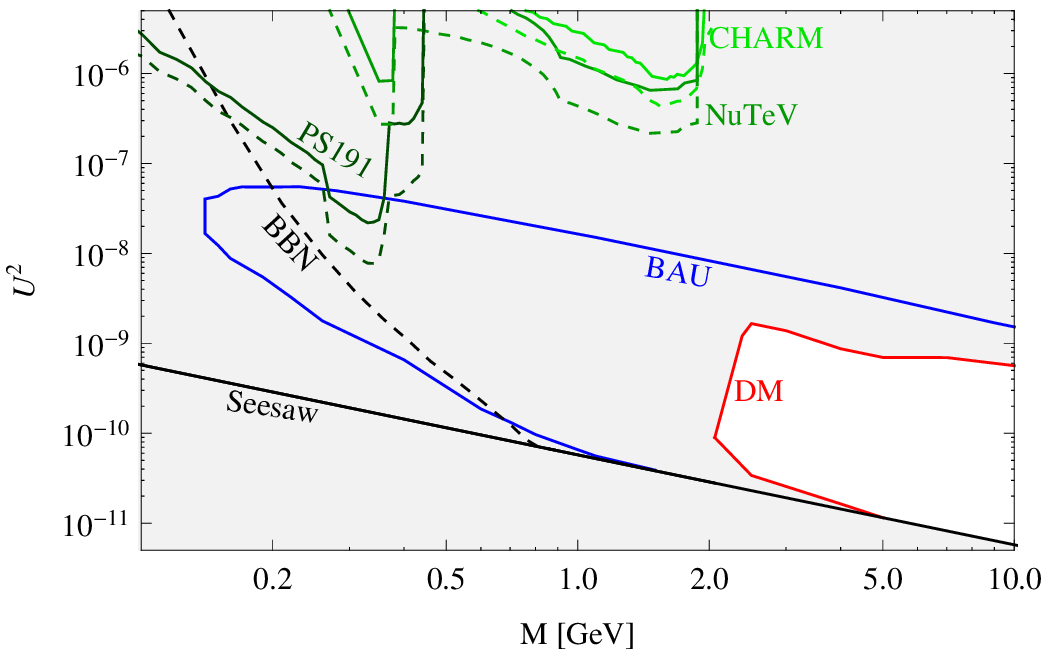} 
\includegraphics[width=10cm]{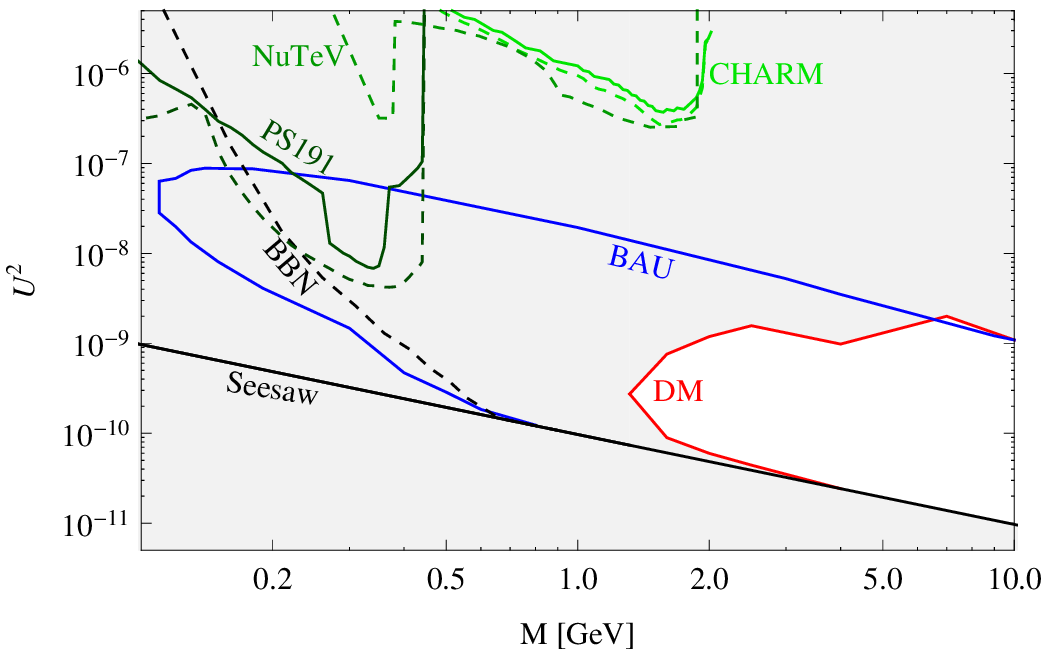}
\caption{Constraints on the $N_{2,3}$ masses $M_{2,3}\simeq M$ and
mixing $U^2={\rm tr}(\theta^\dagger\theta)$ in the constrained
$\nu$MSM (scenario I); upper panel - normal hierarchy, lower panel - inverted
hierarchy. In the region between the solid blue ``BAU'' lines, the
observed BAU can be generated. The lepton asymmetry at $T=100$ MeV can
be large enough that the resonant enhancement of $N_1$ production is
sufficient to explain the observed $\Omega_{DM}$ inside the solid red
``DM'' line. The CP-violating phases were chosen to maximize the
asymmetry at $T=100$ MeV. The regions below the solid black ``seesaw''
line and dashed black ``BBN'' line are excluded by neutrino
oscillation experiments and BBN, respectively. The areas above the
green lines of different shade are excluded by direct search
experiments, as indicated in the plot. The solid lines are {\it
exclusion plots} for all choices of $\nu$MSM parameters, for the
dashed lines the phases were chosen to maximize the late time
asymmetry, consistent with the red line.
\label{BA_DM_U}}
\end{figure}

\begin{figure}[!h]
\centering
\begin{tabular}{cccc}
\includegraphics[width=10cm]{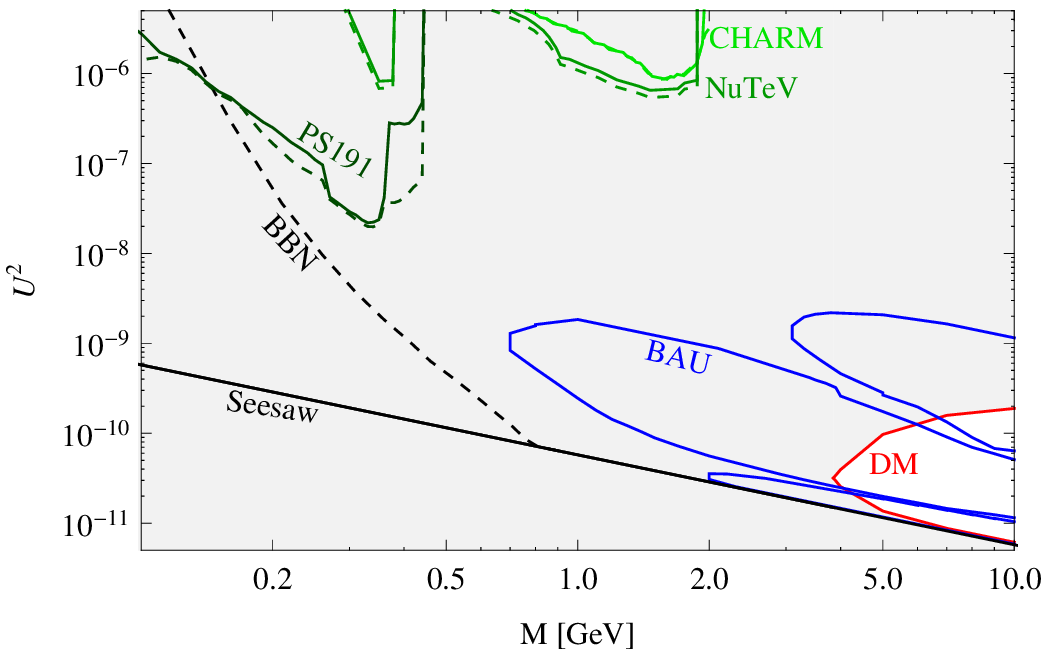}
\end{tabular}
\caption{
Same as figure \ref{BA_DM_U}, but with a different set of CP-violating
phases. The plot illustrates how the ``BAU'' line moves when the
phases are changed; it can cross through points deep inside the
maximal ``BAU'' region while the phase still allow for DM production. 
The sign of the BAU is opposite in the two disjoint ``BAU'' regions.
\label{BA_SP_U}}
\end{figure}

We start with the phases used in figure \ref{DM_Max_mixingangle},
which were chosen to maximize the area in the $M-U^2$-plane where
$\sum_\alpha |\mu_\alpha|>8\cdot 10^{-6}$ at $T=100$ MeV.  The result
is shown in figure \ref{BA_DM_U}. The blue line corresponds to the
points where the asymmetry at $T_{EW}$ corresponds to the observed
BAU. While the requirement to produce enough DM only imposes a lower
bound on the asymmetry at $100$ MeV, the value of the BAU is known to
be given by (\ref{BAUvalue}), i.e. has a fixed value.   Thus, only the
points on the blue line that lie within the region encircled by the
red line (DM region) form the allowed parameter space.  The shape of
the blue BAU line can be modified by changing the phases, see figure
\ref{BA_SP_U}, but this also changes the shape of the red line (DM
region).   Solving the kinetic equations for different phases reveals
that the BAU line can be brought to most points within the DM region.
This region therefore gives a good estimate of the allowed parameter
space.

\begin{figure}[!h]
\centering
\includegraphics[width=10cm]{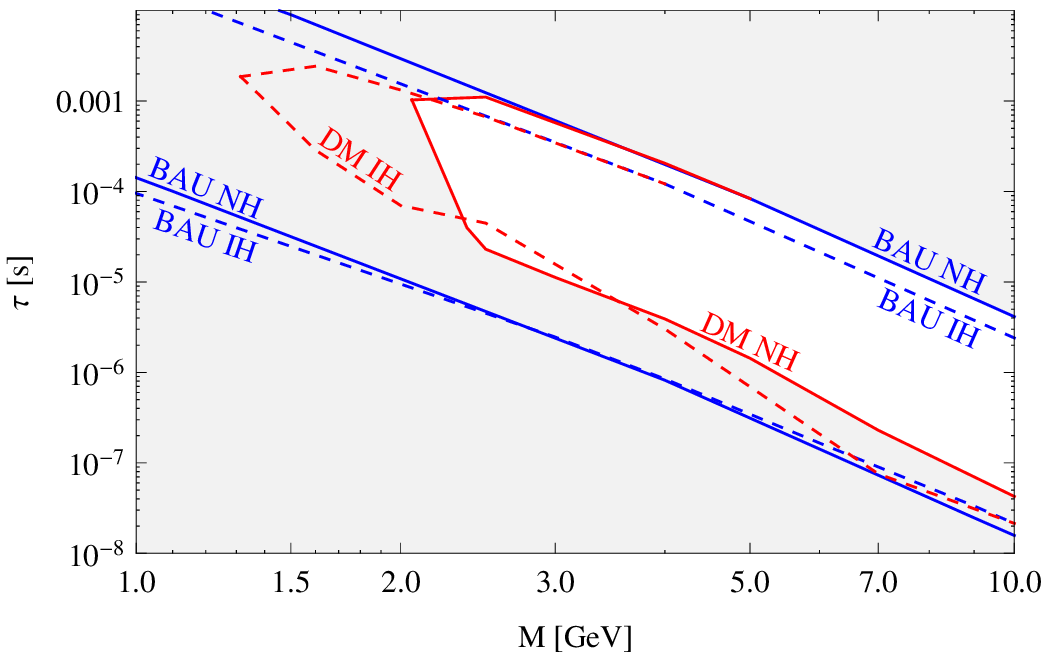} 
\caption{Constraints on the $N_{2,3}$ masses $M_{2,3}\simeq M$ and
lifetime $\tau^{-1}\simeq\frac{1}{2}{\rm tr}\Gamma_N|_{T=1 {\rm MeV}}$
in the constrained $\nu$MSM (scenario I). In the region between the blue ``BAU''
lines, the observed BAU can be generated. The lepton asymmetry at
$T=100$ MeV can be large enough that the resonant enhancement of $N_1$
production is sufficient to explain the observed $\Omega_{DM}$ inside
the red ``DM'' line. The CP-violating phases were chosen to maximize
the asymmetry at $T=100$ MeV. Solid lines - normal hierarchy, dotted
lines - inverted hierarchy.
\label{BA_DM_Lifetime}}
\end{figure}

The constraints derived on the mixing angle $U^2$ are translated into
constraints on the neutrino lifetime $\tau^{-1}\simeq\frac{1}{2}{\rm
tr}\Gamma_N$ (at $T=1$ MeV) in figure \ref{BA_DM_Lifetime}.

\begin{figure}[!h]
\centering
\includegraphics[width=10cm]{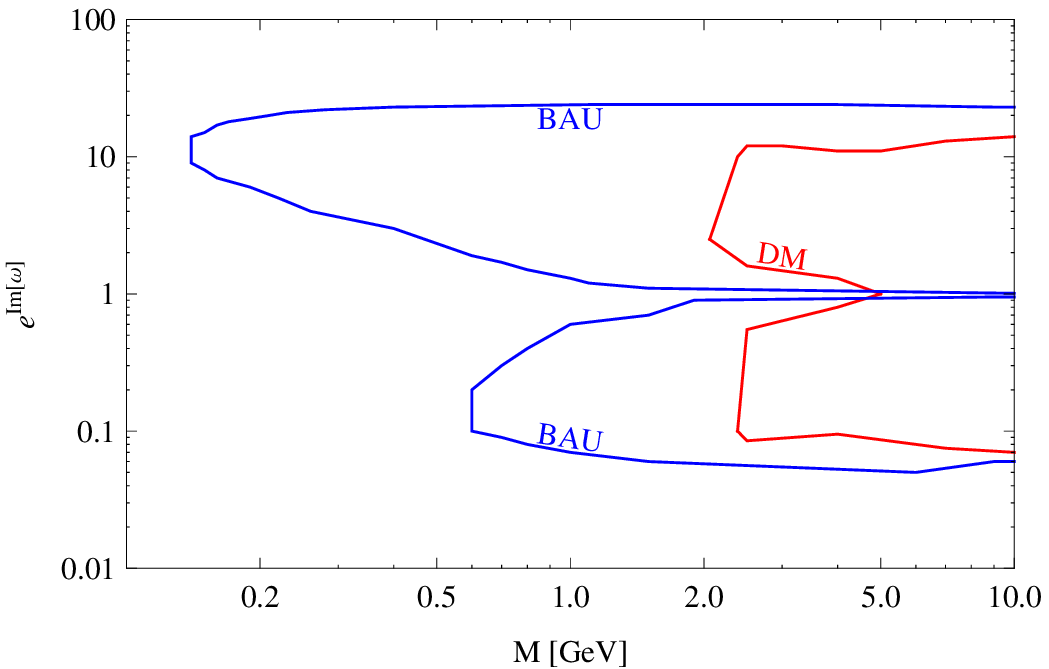} 
\includegraphics[width=10cm]{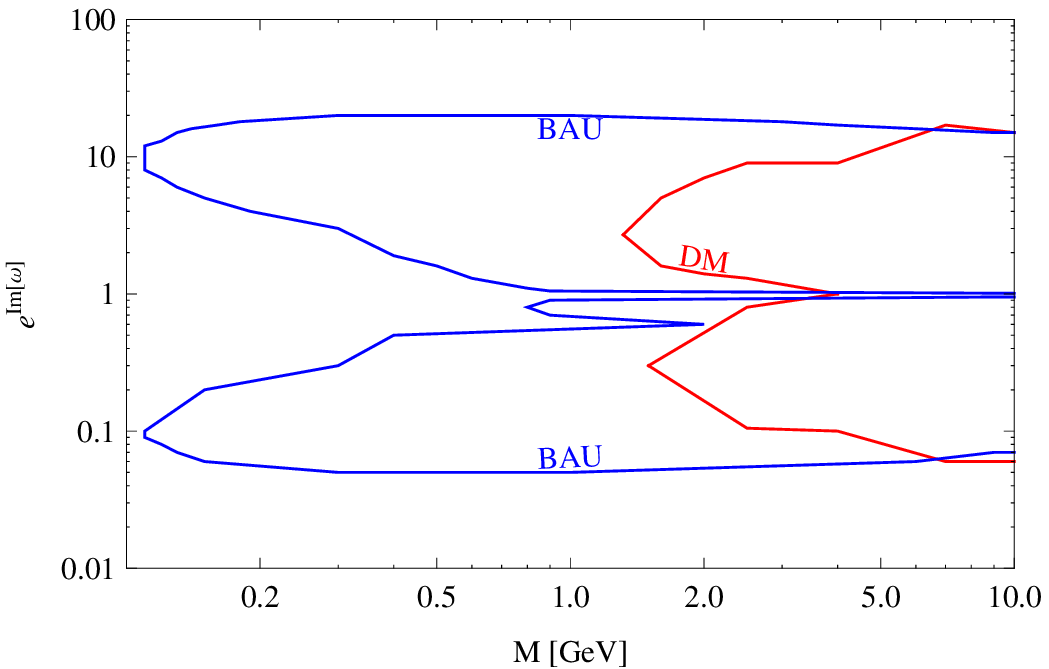}
\caption{Constraints on the $N_{2,3}$ masses $M_{2,3}\simeq M$ and
parameter ${\rm Im}\upomega$ in the constrained $\nu$MSM (scenario I); upper panel -
normal hierarchy, lower panel - inverted hierarchy. In the region
between the solid blue ``BAU'' lines, the observed BAU can be
generated. The lepton asymmetry at $T=100$ MeV can be large enough
that the resonant enhancement of $N_1$ production is sufficient to
explain the observed $\Omega_{DM}$ inside the solid red ``DM'' line.
The CP-violating phases were chosen to maximize the asymmetry at
$T=100$ MeV. 
\label{BA_DM_ImOmega}}
\end{figure}

In the plots of figure \ref{BA_SP_U}, there are two regions where the 'BAU' and 'DM' lines are close, leading to the successful baryogenesis and dark matter production. One is near the seesaw line and the other is for higher mixing. These regions are easier to identify in the ${\rm Im}\upomega-M$ plane shown in figure \ref{BA_DM_ImOmega}.
 The baryon asymmetry almost vanishes for ${\rm Im}\upomega$ really
close to $0$, but this is not the case for dark matter production.
Therefore, there is a region near ${\rm Im}\upomega=0$ which produce
the right amount of baryon asymmetry and enough dark matter. This is
the region where the blue 'BAU' line is inside the red 'DM' line in
figure \ref{BA_DM_ImOmega}. For large value of $|{\rm Im}\upomega|$,
there also are regions where the two constraint are close.

\section{Conclusions and Discussion}
\label{section:concl}
We tested the hypothesis that three right handed neutrinos with
masses below the electroweak scale can be the common origin of the
observed dark matter, the baryon asymmetry of the universe and
neutrino flavor oscillations. 
This possibility can be realized in the
$\nu$MSM, an extension of the SM that is based on the type-I seesaw
mechanism with three right handed neutrinos $N_I$. 
Centerpiece of our analysis is the study of sterile and active neutrino abundances in the early universe, which allows to determine the range of sterile neutrino parameters in which DM, baryogenesis and all known data from active neutrino experiments can be explained {\it simultaneously} within the $\nu$MSM.
We combined our results with astrophysical constraints and re-analyzed bounds from past experiments in the face of recent data from neutrino oscillation experiments. 
We found that all these requirements can be fulfilled for a wide range of sterile neutrino masses and mixings, see figures \ref{BA_DM_U},
\ref{BA_SP_U} in section \ref{Combined}.  
In some part of this parameter space, all three new particles may be found in experiment or observation, using upgrades to existing facilities.

This is the first complete quantitative study of the above scenario (scenario I), in which no physics beyond the $\nu$MSM is required. 
We found that the $\nu$MSM can explain all
experimental data if one sterile neutrino ($N_1$), which composes the
dark matter, has a mass in the keV range, while the other two ($N_{2,3}$)
have quasi-degenerate masses in the GeV range.
The heavier particles $N_{2,3}$ generate neutrino
masses via the seesaw mechanism and create flavored lepton asymmetries from CP-violating
oscillations in the early universe.  
These lepton asymmetries are crucial on two occasions in the early universe.
On one hand they create the BAU via flavored leptogenesis.
One the other hand they affect the rate of thermal DM production via the MSW effect.
The second point allows to derive strong constraints on the $N_{2,3}$ properties from the requirement to explain the observed $\Omega_{DM}$ by thermal $N_1$ production, see section \ref{DMproductionSection}. 
This can be achieved by resonant production, caused by the presence of lepton
asymmetries in the primordial plasma at $T\sim 100$ MeV. The
required asymmetries can be created 
when $N_{2,3}$ are heavier than $1-2$ GeV 
and the physical mass splitting between the $N_2$ and $N_3$ masses is comparable to the active neutrino mass differences. 
This can be achieved in a subspace of the $\nu$MSM
parameter space that is defined by fixing two of the unknown
parameters (the Majorana mass splitting $\Delta M$ and a mixing angle ${\rm
Re}\upomega$ in the sterile sector). This choice, in which scenario I can be realized, is dubbed ``constrained $\nu$MSM''.

We also studied systematically how the parameter constraints relax if one allows $N_1$ DM to be produced by some unspecified mechanism beyond the $\nu$MSM (scenario II), see section \ref{baryogenesissection}. 
In this case the strongest constraints come from baryogenesis and the required mass degeneracy is much weaker, $\Delta M/M\lesssim 10^{-3}$.
We found that successful baryogenesis is possible for $N_{2,3}$ masses
as low as $10$ MeV. These results are based on an extension of the
analysis performed in \cite{Canetti:2010aw} that accounts for a
non-zero value of the neutrino mixing angle $\uptheta_{13}$ and a
temperature dependent Higgs expectation value. While the low mass
region is severely constrained by BBN and experiments, the allowed
parameter space becomes considerably bigger for masses in the GeV
range. Detailed results for the allowed sterile neutrino masses and
mixings are shown in figures \ref{BA_Max_contour} -
\ref{BA_Max_mixingangle}.

If one completely drops the requirement that DM is composed of $N_1$ and considers the $\nu$MSM as a theory of baryogenesis and neutrino oscillations only (scenario III),
no degeneracy in masses is required.
Note that this also implies that no degeneracy is required in scenario II if more than three right handed neutrinos are added to the SM.

For masses below $5$ GeV, the heavier sterile
neutrinos can be searched for in experiments using present day
technology. This makes the $\nu$MSM one of the few truly testable
theories of baryogenesis. The parameter space for the DM candidate
$N_1$ is bound in all directions, see figure  \ref{DMexclusion}, and
can be tested using observations of cosmic X-rays and the large scale
structure of the universe.  Since the model does not require new
particle physics up to the Planck scale to be consistent with
experiment, the hierarchy problem is absent in the $\nu$MSM.



We conclude that neutrino physics can explain all confirmed detections
of physics beyond the standard model except accelerated cosmic
expansion.\\
\newline

{\large \textbf{Acknowledgements}}
We are grateful to Mikko Laine for providing the numerical data shown
in figure \ref{RRMplots}. We also would like to thank Oleg Ruchayskiy,
Alexey Boyarsky and Artem Ivashko for sharing their expertise on
experimental bounds.  This work was supported by the Swiss National
Science Foundation, the Gottfried Wilhelm Leibniz program of the
Deutsche Forschungsgemeinschaft, the Project of Knowledge Innovation
Program   of the Chinese Academy of Sciences grant KJCX2.YW.W10 and
the IMPRS-PTFS.


\begin{appendix}

\newpage
\section{Kinetic Equations}
\label{HeffSec}
In the following we sketch the derivation of the kinetic equations (\ref{kinequ1})-(\ref{kinequ3}). 
Our basic assumptions can be summarized as follows:
\begin{enumerate}
\item[1)] Coherent states containing more than one $N_{I}$ quantum are not relevant. Their contributions are suppressed by additional powers of the small mixing $\theta_{\alpha I}$. 
Processes involving one sterile neutrino include decays of $N_I$ particles, their scatterings with SM particles and flavor oscillations. 
\item[2)] Screened one-particle states are the only relevant propagating neutrino degrees of freedom. In particular, we do not consider any collective excitations, which are infrared effects and only give a small contribution when the typical neutrino momenta are hard $\sim T$.
\item[3)] The interactions that keep the SM fields in equilibrium act much faster than interactions involving $N_I$ at all times due to the smallness of $F$.
\item[4)] $T_{-}\gtrsim T_d$, i.e. the lifetime of the $N_I$ is sufficiently long that freezeout and decay are two well-separated events. This is the case in the parameter space we study.
\item[5)] The typical momentum of $N_I$ particles is $\bar{p} \sim T$ even when they are out of equilibrium. This is justified because they are produced from a thermal bath and freeze out from a thermal state, hence their distribution functions should mimic those of kinetic equilibrium even when out of equilibrium.
\item[6)] We neglect the effect of the $N_{2,3}$ on the time evolution of the entropy (or temperature). This is justified as their contribution to the total entropy and energy densities are always small.  
\item[7)] We neglect the effect of the lepton asymmetry on hadronization. This aspect has e.g. been discussed in \cite{Schwarz:2009ii}. 
\end{enumerate}
\subsection{How to characterize the Asymmetries}\label{howtocharaceriseasymmetries}
The leptonic charges that we are interested in can be expressed in terms of field bilinears. This and 1) imply that this is sufficient as we only need to deal with a reduced density matrix, in which all states including more than one sterile neutrino have been removed by partial tracing. 
Instead of bilinears in the field operators themselves we consider expectation values of bilinears in the ladder operators $a_{I}$, $a_I^{\dagger}$ for sterile and $a_{\alpha}$, $a_\alpha^{\dagger}$ for active neutrinos. 
To be explicit, we decompose $N_I$ as\footnote{The sterile neutrinos may be described as Weyl, Dirac or Majorana spinors. Here we have chosen to write $\nu_{R}$ as a right chirality four-spinor and pulled the $P_R$ out of the definition (\ref{NIdef}) so that the $N_I$ are Majorana spinors.} 
\begin{equation}
N_I=\int\frac{d^3\textbf{p}}{(2\pi)^{3}}\sum_{s}\frac{1}{\sqrt{2\omega_{\textbf{p}}}}
\left( u_{I,\textbf{p}}^{s}e^{-i\textbf{p}\textbf{x}}a_{I,s}(\textbf{p},t)+ v_{I,\textbf{p}}^{s}e^{i\textbf{p}\textbf{x}}a_{I,s}^{\dagger}(\textbf{p},t)
\right).
\end{equation}
Here $\textbf{p}$ is momentum, $s$ the spin index and $u$, $v$ are the usual plane wave solutions to the Dirac equation,
\begin{eqnarray}
\begin{tabular}{c c c}
$(\slashed{p}-M_I)u^s_{I,\textbf{p}}=0$ & & $(\slashed{p}+M_I)v^s_{I,\textbf{p}}=0$.
\end{tabular}
\end{eqnarray}
We will in the following always assume that the spacial momentum is directed along the $z$-axis, which is also the axis of angular momentum quantization. 
We chose the convention that $\mathfrak{h}u_{\textbf{p}}^{s}=(-1)^{s+1}u_{\textbf{p}}^{s}$ while $\mathfrak{h}v_{\textbf{p}}^{s}=(-1)^{s}v_{\textbf{p}}^{s}$, where $\mathfrak{h}$ is the helicity matrix 
\begin{eqnarray}
\mathfrak{h}&=&\frac{1}{2}\frac{p_i}{|\textbf{p}|}\gamma^i\gamma^0\gamma^5
=\frac{1}{2}\frac{p_i}{|\textbf{p}|}\Upsigma_i, \phantom{X}\Upsigma_i=\frac{i}{2}\epsilon_{ijk}\gamma^j\gamma^k\label{helicitymatrix}.
\end{eqnarray}
All relevant matrix elements of the density matrix $\hat{\rho}$ can be identified with expectation values of bilinears in the ladder operators. Because of this the matrix $\rho$ of bilinears, to be defined in (\ref{rhodef}), is often referred to as ``density matrix'' (rather than $\hat{\rho}$ itself).  In principle there is a large number of such bilinears. A complete characterization of the system requires knowledge of all their expectation values at all times. However, it can be simplified dramatically, and for our purpose it will be sufficient to follow the time evolution of two $2\times 2$ matrices $\rho_{N}$ and $\rho_{\bar{N}}$ and three chemical potentials.

The only term in (\ref{nuMSM_lagrangian}) that violates lepton number is $M_M$. For $T\gg M$, it is negligible and lepton number is approximately conserved. There is no total lepton asymmetry at $T_{EW}$ in the $\nu$MSM, but there can be asymmetries of opposite sign for fermions with different chirality. Baryogenesis occurs because sphalerons only couple to left handed fermions.
As far as the (Majorana) neutrinos are concerned, the two helicity states act as ``particle'' and ``antiparticle''. 
Terms containing two creation or two annihilation operators such, as $\langle a_Ia_J\rangle$ or $\langle a_\alpha^\dagger a_\beta^\dagger\rangle$, can be related to processes that violate lepton number and are suppressed at $T>M$. For $T\lesssim M$ they in principle could contribute, but the leading order contribution in the Yukawa coupling $F$ to the corresponding rates $\frac{d}{dt}\langle a_Ia_J\rangle$ etc.\footnote{Whether the ladder operators or $\hat{\rho}$ or both change with time depends on whether one choses the Heisenberg, Schr\"odinger or interaction picture. The expectation value, however, always has the same time dependence.} oscillate fast. We therefore only consider terms that contain exactly one creation and one annihilation operator. Since only two of the sterile neutrinos are relevant here, these form a $10\times10$ matrix that can be written as 
\begin{eqnarray}\label{rhodef}
\rho=
\left(
\begin{tabular}{c c c c}
$\rho_{NN}$ & $\rho_{N\bar{N}}$ & $\rho_{NL}$ & $\rho_{N\bar{L}}$ \\
$\rho_{\bar{N}N}$ & $\rho_{\bar{N}\bar{N}}$ & $\rho_{\bar{N}L}$ & $\rho_{\bar{N}\bar{L}}$ \\
$\rho_{LN}$ & $\rho_{L\bar{N}}$ & $\rho_{LL}$ & $\rho_{L\bar{L}}$ \\
$\rho_{\bar{L}N}$ & $\rho_{\bar{L}\bar{N}}$ & $\rho_{\bar{L}L}$ & $\rho_{\bar{L}\bar{L}}$ 
\end{tabular}
\right),
\end{eqnarray}
with
\begin{eqnarray}\label{rhodefinitions}
\begin{tabular}{c c}
$(\rho_{NN})_{IJ}=\langle a_{I,1}^{\dagger} a_{J,1}\rangle/V$ & $(\rho_{N\bar{N}})_{IJ}=\langle a_{I,1}^{\dagger} a_{J,2}\rangle/V$ \\ 
$(\rho_{NL})_{I\beta}=\langle a_{I,1}^{\dagger} a_{\beta,1}\rangle/V$ & $(\rho_{N\bar{L}})_{I\beta}=\langle a_{I,1}^{\dagger} a_{\beta,2}\rangle/V$ \\
$(\rho_{\bar{N}N})_{IJ}=\langle a_{I,2}^{\dagger} a_{J,1}\rangle/V$ & $(\rho_{\bar{N}\bar{N}})_{IJ}=\langle a_{I,2}^{\dagger} a_{J,2}\rangle/V$ \\
$(\rho_{\bar{N}L})_{I\beta}=\langle a_{I,2}^{\dagger} a_{\beta,1}\rangle/V$ & $(\rho_{\bar{N}\bar{L}})_{I\beta}=\langle a_{I,2}^{\dagger} a_{\beta,2}\rangle/V$ \\
$(\rho_{LN})_{\alpha J}=\langle a_{\alpha,1}^{\dagger} a_{J,1}\rangle/V$ & $(\rho_{L\bar{N}})_{\alpha J}=\langle a_{\alpha,1}^{\dagger} a_{J,2}\rangle/V$ \\
 $(\rho_{LL})_{\alpha\beta}=\langle a_{\alpha,1}^{\dagger} a_{\beta,1}\rangle/V$ & $(\rho_{L\bar{L}})_{\alpha\beta}=\langle a_{\alpha,1}^{\dagger} a_{\beta,2}\rangle/V$ \\
$(\rho_{\bar{L}N})_{\alpha J}=\langle a_{\alpha,1}^{\dagger} a_{J,1}\rangle/V$ & $(\rho_{\bar{L}\bar{N}})_{\alpha J}=\langle a_{\alpha,2}^{\dagger} a_{J,1}\rangle/V$ \\
 $(\rho_{\bar{L}L})_{\alpha\beta}=\langle a_{\alpha,2}^{\dagger} a_{\beta,1}\rangle/V$ & $(\rho_{\bar{L}\bar{L}})_{\alpha\beta}=\langle a_{\alpha,2}^{\dagger} a_{\beta,2}\rangle/V$ 
\end{tabular},
\end{eqnarray}
where we have suppressed time and momentum indices (all momenta are $\textbf{p}$ and all times $t$). $V$ is the overall spacial volume, which will always drop out of the computations in the end.

\subsection{Effective Kinetic Equations}
The time evolution of $\rho$ is governed by an effective Hamiltonian. In absence of Hubble expansion, which we will add later, it follows the kinetic equation 
\begin{equation}\label{effkineq}
i\frac{d\rho}{dt}=[{\rm H},\rho]-\frac{i}{2}\{\Gamma^>,\rho\}+\frac{i}{2}\{\Gamma^<,1-\rho\}.
\end{equation}
${\rm H}$ can be viewed as the \textit{dispersive part} of the effective Hamiltonian. The \textit{absorbtive part} given by the matrices $\Gamma^\gtrless$ arises because the system is coupled to the background plasma formed by all other degrees of freedom of the SM. 
Note that (\ref{effkineq}) is valid for each momentum mode separately. The different modes are coupled by ${\rm H}$ and $\Gamma^\gtrless$, which in principle depend on $\rho$ and the lepton chemical potentials.
The smallness of the sterile neutrino couplings $F$ allows to simplify (\ref{effkineq}) due to a separation of time scales: The time scale associated with the $N_I$ dynamics and the time scale on which chemical equilibration of the lepton asymmetries occurs are much longer than the typical relaxation time to kinetic equilibrium in the SM plasma.  
This allows to employ a \textit{relaxation time approximation} and 
relate $\Gamma^>$ and $\Gamma^<$ by a detailed balance (or Kubo-Martin-Schwinger) relation \cite{Sigl:1992fn,Asaka:2005pn,Shaposhnikov:2008pf},
\begin{equation}\label{relaxtimeapprox}
i\frac{d\rho}{dt}=[{\rm H},\rho]-\frac{i}{2}\{\Gamma,\rho-\rho^{eq}\},
\end{equation}
with $\Gamma=\Gamma^>-\Gamma^<$. $\rho^{eq}$ is $\rho$ evaluated with an equilibrium density matrix, $\hat{\rho}=Z/{\rm tr}Z$, $Z=\exp(-\hat{H}/T)$, where $\hat{H}$ is the Hamiltonian corresponding to (\ref{nuMSM_lagrangian}).
The matrices ${\rm H}$ and $\Gamma$ are Hermitian.

The effective masses of active and sterile neutrinos are very different and fast oscillations between them play no role. We thus put to zero $\rho_{NL}$, $\rho_{LN}$, $\rho_{\bar{N}L}$, $\rho_{\bar{L}N}$, $\rho_{N\bar{L}}$, $\rho_{L\bar{N}}$, $\rho_{\bar{N}\bar{L}}$ and $\rho_{\bar{L}\bar{N}}$. The time evolution of the asymmetries is related to the relaxation time scales of the $N_I$. Since interactions of the active neutrinos amongst themselves and with other SM fields are much faster, coherent effects in the active sector are negligible on this time scale. This allows to furthermore neglect $\rho_{L\bar{L}}$ and $\rho_{\bar{L}L}$. $\rho_{LL}$ and $\rho_{\bar{L}\bar{L}}$ are taken diagonal with equilibrium occupation numbers and are thus characterized by the temperature $T$ and three slowly varying chemical potentials. Thus, we can entirely describe the active sector by four numbers. Instead of the chemical potentials, we will in the following use $n_{\alpha}=(\rho_{LL})_{\alpha\alpha}-(\rho_{\bar{L}\bar{L}})_{\alpha\alpha}$, i.e. the number of particles minus number of antiparticles, to characterize the asymmetries\footnote{We work in the $\tilde{F}=FU_N$ base in flavor space. This is the mass base of sterile neutrinos in vacuum, but the flavor base for active neutrinos. Hence, the diagonal elements of $\rho_{NN}$ and $\rho_{\bar{N}\bar{N}}$ have a straightforward interpretation as number densities for physical particles in vacuum, while the ladder operators $a^\dagger_\alpha$ create linear combinations of physical particles, and the matrices $\rho_{LL}$ and $\rho_{\bar{L}\bar{L}}$ are strictly speaking not diagonal in thermal equilibrium. This adds a subtlety to the interpretation of $n_\alpha$ as ``particles minus antiparticles'', which, however, is of no practical relevance due to the smallness of the neutrino masses.}. The relation between both can be found in the appendix of \cite{Laine:2008pg}
.
In the sterile sector we have to keep track of coherences. The system can then be described by the following set of kinetic equations,
\begin{eqnarray}
i \frac{d\rho_{NN}}{dt}&=& [{\rm H}, \rho_{NN}]
-\frac{i}{2}\{\gamma_N, \rho_{NN} - \rho^{eq}_{NN}\} +
\frac{i}{2} n_\alpha{\tilde\gamma^\alpha}_N~,\label{Kineq1}\\
i \frac{d\rho_{\bar{N}\bar{N}}}{dt}&=& [{\rm H}^*, \rho_{\bar{N}\bar{N}}]
-\frac{i}{2}\{\gamma^*_N, \rho_{\bar{N}\bar{N}} - \rho^{eq}_{\bar{N}\bar{N}}\} -\frac{i}{2} n_\alpha{
\tilde\gamma^{\alpha *}}_N~,\label{Kineq2}
\\
i \frac{dn_\alpha}{dt}&=&-i\gamma^\alpha_L n_\alpha +
i {\rm Tr}\left[{\tilde \gamma^\alpha}_L(\rho_{NN} -\rho^{eq}_{NN})\right] -
i {\rm Tr}\left[{\tilde \gamma^{\alpha*}}_L(\rho_{\bar{N}\bar{N}}  -\rho^{eq}_{\bar{N}\bar{N}})\right]\label{Kineq3}
~.
\end{eqnarray}
Here $\gamma_{N}$, $\gamma_{\bar{N}}$ are the appropriate block-diagonal submatrices of $\Gamma$, for the corresponding submatrix of ${\rm H}$ we used the same symbol 
as for the full matrix to simplify the notations.
These equations do not take into account the expansion of the universe. 
As usual, it can be included by using abundances (or ``yields'') instead of number densities.
It is also convenient to introduce the variable $X=M/T$ rather than time $t$.

All quantities appearing in the above equations depend on momentum. 
The different momentum modes are coupled by the scattering and decay processes. We have suppressed this momentum dependence. 
We define the abundances $\rho_{N}=\int d^3\textbf{p}/(2\pi)^3 \rho_{NN}/s$, $\bar{\rho}_{N}=\int d^3\textbf{p}/(2\pi)^3 \rho_{\bar{N}\bar{N}}/s$, $\rho^{eq}=\int d^3\textbf{p}/(2\pi)^3\rho^{eq}_{NN}/s\approx \int d^3\textbf{p}/(2\pi)^3\rho^{eq}_{\bar{N}\bar{N}}/s$ and $\mu_{\alpha}=\int d^3\textbf{p}/(2\pi)^3 n_{\alpha}/s$\footnote{Note that the $\mu_{\alpha}$ defined this way are dimensionless and basically abundances, not chemical potentials, cf. appendix \ref{asymmconv}.}. 
Assumption 5) is justified if the common \textit{kinetic equilibrium assumption} 
\begin{equation}
\frac{(\rho_{NN})_{IJ}}{(\rho^{eq}_{NN})_{IJ}}=\frac{(\rho_{N})_{IJ}}{(\rho^{eq})_{IJ}}\label{KinEqEq}
\end{equation}
holds.
We can use (\ref{KinEqEq}) to rewrite the anticommutator in (\ref{Kineq1}) as $\{\Gamma_N, \rho_{N} - \rho^{eq}\}$ with 
\begin{eqnarray}
\Gamma_N&=&\uptau\phantom{i} \int d^3\textbf{p} \gamma_N \frac{\rho_{NN}^{eq} s}{\rho^{eq}}=\uptau\phantom{i} \int d^3\textbf{p} \gamma_N \frac{f_F(\omega_{\textbf{p}})}{n_{F}},\label{Gammadefinition}\\
n_F&=&\int d^3\textbf{q} f_F(\omega_{\textbf{q}}).
\end{eqnarray}
We again emphasize that $\rho_{NN}$, $\gamma_N$ etc. appearing in (\ref{Kineq1})-(\ref{Kineq3}) depend on momentum while $\Gamma_N$, $\rho_N$, $\rho^{eq}$ etc. do not.

For $T\ll M$, almost all particles have the momentum $\bar{p}\sim T$ and $\Gamma_N$ is essentially obtained by evaluating $\gamma_N$ at $\textbf{p}=\bar{p}$. Practically we compute the rates as described in section \ref{DMproductionrates}. Similarly, we can use $H=\uptau\phantom{i}  {\rm H}|_{\textbf{p}=\bar{p}}$ for the Hermitian part of the effective Hamiltonian.
For $|\textbf{p}|\sim T\gtrsim M$, $n_F$ can be approximated by $n_F\approx \frac{3}{2}\zeta(3)T^3\approx 1.8 T^3$, but $\gamma_N$ has to be computed numerically. 

Using the above considerations, we can write down the following effective kinetic equations:
\begin{eqnarray}
\label{kineq1}
i \frac{d\rho_{N}}{d X}&=& [H, \rho_{N}]-\frac{i}{2}\{\Gamma_N, \rho_{N} - \rho^{eq}\} +\frac{i}{2} \mu_\alpha{\tilde\Gamma^\alpha}_N~,\\
i \frac{d\rho_{\bar{N}}}{d X}&=& [H^*, \rho_{\bar{N}}]-\frac{i}{2}\{\Gamma^*_N, \rho_{\bar{N}} - \rho^{eq}\} -\frac{i}{2} \mu_\alpha{\tilde\Gamma^{\alpha *}}_N~,\label{kineq2}\\
i \frac{d\mu_\alpha}{d X}&=&-i\Gamma^\alpha_L \mu_\alpha +
i {\rm Tr}\left[{\tilde \Gamma^\alpha}_L(\rho_{N} -\rho^{eq})\right] -
i {\rm Tr}\left[{\tilde \Gamma^{\alpha*}}_L(\rho_{\bar{N}}  -\rho^{eq})\right]
~.
\label{kineq3}
\end{eqnarray}
They are equivalent to the ones used in \cite{Asaka:2005pn}. Their interpretation is straightforward. 
In the mass base, the diagonal elements of $\rho_N$ and $\rho_{\bar{N}}$ are the abundances of sterile neutrinos and antineutrinos, respectively.
The off-diagonal elements are flavor coherences.
$\rho_N$ thus gives the abundances for ``particles'' and $\rho_{\bar{N}}$ those for ``antiparticles'', defined as the helicity states of the Majorana fields $N_I$. This interpretation holds in vacuum, while at finite temperature the effective mass matrix rotates due to the interplay between the (temperature dependent) Higgs expectation value, the Majorana mass $M_M$ and thermal masses in the plasma.

The first two terms in (\ref{kineq1}) and (\ref{kineq2}) are due to sterile neutrino oscillations and dissipative effects, respectively, either by scatterings or by decays and inverse decays of sterile neutrinos.
More precisely, the Hermitian $2\times 2$ matrix $H$ in (\ref{kineq1}) and (\ref{kineq2}) is the dispersive part of the effective Hamiltonian for $\rho_N$ and $\bar{\rho}_N$. 
The matrix $\Gamma_{N}$ is the dissipative part of the effective Hamiltonian for $\rho_N$ and $\bar{\rho}_N$ that arises because the sterile neutrinos are coupled to the SM. 
$\rho^{eq}$ is the common equilibrium value of $\rho_N$ and $\bar{\rho}_N$ in absence of an asymmetry. 
All these terms appeared already in earlier studies \cite{Akhmedov:1998qx}. 
The equations of motion (\ref{kineq3}) for the asymmetries in the active sector follow from consistency consideration and the symmetries of the $\nu$MSM.
The terms containing ${\tilde \Gamma^{\alpha}}_L$ in (\ref{kineq3}) are their counterparts in the active sector.
The last term is due to backreaction and has been discussed in \cite{Asaka:2011wq}.

\subsection{The Effective Hamiltonian}\label{HamiltonianAppendix}   
We follow the approach used in \cite{Asaka:2006rw} and split the Hamiltonian in the Heisenberg picture into a free part $\hat{H}_0$ and interaction $\hat{H}_{int}$.
We perform the computation in Minkowski spacetime and for the moment omit the factor $\partial t / \partial X$ included in the definition (\ref{Gammadefinition}). The same rates, multiplied by this factor, can be used in the early universe when abundances are considered instead of number densities.  
Starting point of the computation is the von Neumann equation in the interaction picture,
\begin{equation} 
 i \frac{{\rm d} \hat \rho_\rmi(t)}{{\rm d} t} =
 [\hat H_\rmi(t),\hat\rho_\rmi(t)]
 \;, \label{liuv1}
\end{equation}
where 
$ 
 \hat \rho_\rmi \equiv \exp(i \hat H_0 t)\hat \rho \exp(-i \hat H_0 t)
$
is the density matrix in the interaction picture and
$
 \hat H_\rmi= \exp(i \hat H_0 t) \hat H_{int} \exp(-i \hat H_0 t)
$, where $\hat{\rho}$ is the (time independent) density matrix in the Heisenberg picture. 
Equation (\ref{liuv1}) can be solved perturbatively, 
\begin{eqnarray}
 \hat \rho_\rmi(t) = \hat\rho_0
 - i \int_0^t \! {\rm d} t' \, 
 [\hat H_\rmi(t'), \hat\rho_0]
 + (-i)^2 
 \int_0^t \! {\rm d} t' \,
 \int_0^{t'} \! {\rm d} t'' \,
 [\hat H_\rmi(t'),[\hat H_\rmi(t''), \hat\rho_0]]
 + ... \;,
 \label{pert}
\end{eqnarray}
where $\hat\rho_0 \equiv \hat\rho(0) = \hat\rho_\rmi(0)$.

We use (\ref{pert}) to compute the expectation values $\langle a^\dagger_{I,r}(\textbf{p},t)a_{J,s}(\textbf{p},t)\rangle/V$ by insertion into (\ref{expvalue}). 
For $\hat{\rho}_0$ we chose a product density matrix $\hat{\rho}_0=\hat{\rho}_{\rm N}\otimes\hat{\rho}_{SM}^{eq}$, where $\hat{\rho}_{SM}^{eq}$ is an equilibrium density matrix for the SM fields and
\begin{equation}\label{InItIaLrho}
\hat{\rho}_{\rm N}=
\sum_I{\rm P}_{I,s} a^\dagger_{I,s}|0\rangle\langle0|a_{I,s} .
\end{equation}
This is not the most general density matrix that can be build from one-particle $N_I$ states, but it is sufficient to derive the effective Hamiltonian. 
The formula (\ref{pert}) formally gives expressions for the bilinears at all times. These are strictly valid only at times much shorter than the sterile neutrino relaxation time because the perturbative expansion at some point breaks down due to secular terms. 
In the relaxation time approximation (\ref{relaxtimeapprox}) we can use a trick to deal with this problem.  We differentiate $\langle a^\dagger_{I,r}(\textbf{p},t)a_{J,s}(\textbf{p},t)\rangle/V$ with respect to time to obtain a "rate".
We then send $t$ to infinity to eliminate its explicit appearance from the rate. This last step is allowed because all correlation functions of SM fields are damped 
on time scales much shorter than the sterile neutrino relaxation time by thermal damping rates due to the gauge interactions.
Thus, the late time part of the integrand in (\ref{pert}) does not contribute significantly to $d\langle a^\dagger_{I,r}(\textbf{p},t)a_{J,s}(\textbf{p},t)\rangle/(V dt)$. This way we obtain the rate of change of the matrices $\rho_{NN}$ and $\rho_{\bar{N}\bar{N}}$ at initial time. In the relaxation time approximation, these rates can also be used at later times because backreaction is accounted for in the $\rho_{N} -\rho^{eq}$ term.

Repeating literally all steps in section 2.2 of reference \cite{Asaka:2006rw} for the two flavor case and the initial density matrix (\ref{InItIaLrho}), we obtain 
\begin{eqnarray}
\lefteqn{\frac{d}{dt}\frac{\langle a^\dagger_{I,r} a_{J,s}\rangle}{V}=\sum_{A=1,2}{\rm P}_A\bigg[ -i(\delta_{AI}-\delta_{AJ})\frac{M}{\omega_{\textbf{p}}}{\rm Re}(\Delta\tilde{M}_M)_{IJ}}\nonumber\\
&&\frac{1-2f_F(\omega_{\textbf{p}})}{2\omega_{\textbf{p}}}\Big[  (\delta_{AI}+\delta_{AJ}) {\rm tr}\left(P_R(P_u^{rs})_{IJ}P_L \Pi^-_{IJ}(p)+P_R(P_v^{sr})_{JI}P_L\Pi^-_{JI}(p)\right)     \nonumber\\
&&+i(\delta_{AI}-\delta_{AJ}) \mathcal{P}\int\frac{dq_0}{2\pi}\frac{1}{\omega_{\textbf{p}}-q_0}{\rm tr}\left( P_R(P_u^{rs})_{IJ}P_L\Pi^-_{IJ}(q_0,\textbf{p})+ P_R(P_v^{sr})_{JI}P_L\Pi^-_{JI}(q_0,\textbf{p})\right)\Big]\bigg],\nonumber\\\label{Heff1}
\end{eqnarray}
with 
\begin{eqnarray}
\Delta\tilde{M}_M=\tilde{M}_M-\mathbbm{1}\bar{M}\ ,\ \tilde{M}_M=U_N^T M_M U_N
\end{eqnarray}
and
\begin{equation}
(P_u^{rs})_{IJ}= u_{I,r}(\textbf{p}) \bar{u}_{J,s}(\textbf{p}) ,\phantom{X}(P_v^{rs})_{IJ}= v_{I,r}(\textbf{p}) \bar{v}_{J,s}(\textbf{p})\label{PuPvDef}
.\end{equation}
In the limit $\delta M\rightarrow 0$ the projectors are independent of the sterile flavor indices and reduce to
\begin{eqnarray}
(P_u)^{ss}_{IJ}&=&\left(\slashed{p}+\bar{M}\right)\left(\frac{1}{2}-(-1)^s\mathfrak{h}\right),\phantom{X}(P_v)_{IJ}^{ss}=\left(\slashed{p}-\bar{M}\right)\left(\frac{1}{2}+(-1)^s\mathfrak{h}\right)\label{pRojectordef},
\end{eqnarray}
where $\mathfrak{h}$  is the helicity matrix (\ref{helicitymatrix}). 
We have used $u^c=C\bar{u}^T=v$ and introduced the self energies 
\begin{equation}\label{SelfEnergies}
\Pi^{\gtrless}_{IJ}(p)=\tilde{F}_{\alpha I}\tilde{F}_{\beta J}^*\int\frac{d^4k}{(2\pi)^4}\left(v^2\delta(p-k)+\Delta^\gtrless(p+k)\right)S^\lessgtr_{\alpha\beta}(k),
\end{equation} 
with $\tilde{F}=FU_N$.
The thermal Wightman functions appearing therein are defined as 
\begin{eqnarray}
\Delta^>(x_1-x_2)&=&\langle\phi(x_1)\phi(x_2)\rangle\\
\Delta^<(x_1-x_2)&=&\langle\phi(x_2)\phi(x_1)\rangle\\
S^>_{\alpha\beta ij}(x_1-x_2)&=&\langle\nu_{\alpha i}(x_1)\bar{\nu}_{\beta j}(x_2)\rangle\\
S^<_{\alpha\beta ij}(x_1-x_2)&=&-\langle\bar{\nu}_{\beta j}(x_2)\nu_{\alpha i}(x_1)\rangle.
\end{eqnarray}
Here $_i , _j$ are spinor indices, which we suppress in the following.
Transitions with $r\neq s$ do not contribute at leading order in $\theta_{\alpha I}$ due to the projectors. This justifies our description of the sterile sector by two $2\times 2$ matrices $\rho_N$ and $\rho_{\bar{N}}$ rather than a $4\times 4$ matrix including elements $\propto \rho_{N\bar{N}}$ etc.
Transitions with $\alpha\neq\beta$ are suppressed by the small active neutrino masses $m_i/T\ll 1$.
The above expressions are written in the $\tilde{F}$-base (vacuum mass base). They can be translated into the $F$-base used in (\ref{nuMSM_lagrangian}) by the replacements $\tilde{F}\rightarrow F$ and $\Delta\tilde{M}_M\rightarrow\Delta M\sigma_3$. Note that $\bar{M}$ is defined at $T=0$.

 With (\ref{InItIaLrho}), the initial value for $\rho_N$ can be written as $\rho_N\propto{\rm diag}({\rm P_1}, {\rm P}_2)$. 
The RHS of (\ref{Heff1}) has a real and an imaginary part. They allow to extract the dispersive and dissipative parts  ${\rm H}$ and $\gamma_N$ of the effective Hamiltonian.

\subsubsection{Dispersive Part $H$}\label{DispersivePart}
\begin{figure}
  \centering
    \includegraphics[width=12cm]{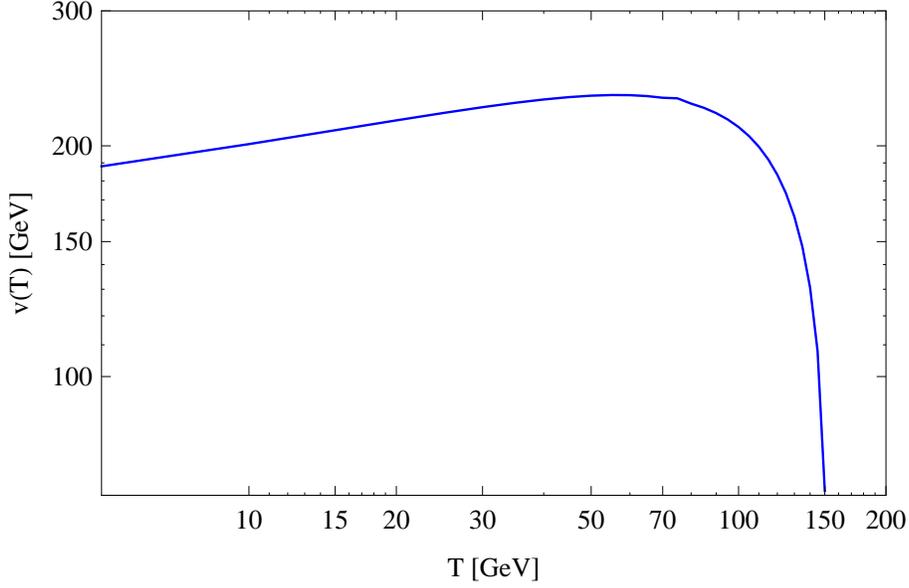}
    \caption{The Higgs expectation value as a function of temperature for $m_H=126$ GeV.\label{HiggsV}}
\end{figure}

Comparison of (\ref{Heff1}) and (\ref{Kineq1}) in absence of active lepton asymmetries (since we chose $\hat{\rho}_{SM}^{eq}$ without chemical potentials) allows to define the dispersive part of the effective Hamiltonian  appearing in (\ref{kinequ1new}),
\begin{eqnarray}
H_{IJ}&=&\uptau(\bar{M}^2+\bar{p}^2)^{\frac{1}{2}}\delta_{IJ}+
\uptau\int d^3\textbf{p}\frac{f_F(\omega_{\textbf{p}})}{n_F}\bigg[-\frac{M}{\omega_{\textbf{p}}}{\rm Re}(\Delta\tilde{M}_M)_{IJ}\label{Heff2}\\
&&+\frac{1-2f_F(\omega_{\textbf{p}})}{2\omega_{\textbf{p}}}\Big[\mathcal{P}\int\frac{dq_0}{2\pi}\frac{1}{\omega_{\textbf{p}}-q_0}{\rm tr}\left({\rm P}_u\Pi^-_{IJ}(q_0,\textbf{p})+{\rm P}_v\Pi^-_{JI}(q_0,\textbf{p})\right)\Big]\bigg]\nonumber
,\end{eqnarray}
where we have introduced the short notation
\begin{equation}
{\rm P}_u=P_R(P_u^{11})_{IJ}P_L=(\frac{1}{2}+\mathfrak{h})P_L \  ,\ {\rm P}_v=P_R(P_v^{11})_{JI}P_L = (\frac{1}{2}-\mathfrak{h})P_L\label{ShortProj}.
\end{equation}
The additional factor $f_F(\omega_{\textbf{p}})/n_F$ and the momentum integral come from the momentum averaging, cf (\ref{Gammadefinition}).
One can distinguish between three contributions. The term involving $\Delta\tilde{M}_M$ comes from the splitting of the Majorana masses and remains present in vacuum. 
The term involving $\Pi^-_{IJ}$ is due to the Yukawa interactions.
It contains two contributions, see (\ref{SelfEnergies}), which are related to the Feynman diagrams shown in Figure \ref{Nselfenergies}. The part $\propto v(T)^2$ is due to the interaction with the Higgs condensate and produces the Dirac mass at $T<T_{EW}$. The part involving $\Delta^\gtrless$ comes from scatterings with Higgs particles.
The Higgs expectation value as a function of temperature can be calculated for a given Higgs mass. We used $m_H=126$ GeV, as suggested by recent LHC data \cite{:2012gu,:2012gk}, to obtain the dependence shown in figure \ref{HiggsV}. However, we checked that varying $m_H$ within the allowed window $115-130$ GeV does not have a big effect on the results. 

Evaluation of (\ref{Heff2}) requires knowledge of the dressed active neutrino and Higgs propagators $S^\gtrless_{\alpha\beta}(p)$ and $\Delta^\gtrless$, respectively. These are in principle complicated functions of $p$ and $T$. However, we are mostly interested in very high or low temperatures, $T\gtrsim T_{EW}\gg M$ during baryogenesis and $T\lesssim M$ in the context of DM production. This allows to simplify the expressions. 
It is convenient to dissect the self energy $\Pi^-$ into the Lorentz components 
\begin{equation}
P_L\Pi^-_{IJ}(p)=P_L(A_{IJ}(p)\Slash{p}+B_{IJ}(p)\Slash{u}),
\end{equation} 
where $u=(1,0,0,0)$ is the four-velocity of the primordial plasma, and write
\begin{eqnarray}
\lefteqn{ H_{IJ}=\uptau(\bar{M}^2+T^2)\delta_{IJ}+
\uptau\int d^3\textbf{p}\frac{f_F(\omega_{\textbf{p}})}{n_F}\bigg[-\frac{M}{\omega_{\textbf{p}}}{\rm Re}(\Delta\tilde{M}_M)_{IJ}
 + \frac{1-2f_F(\omega_{\textbf{p}})}{2\omega_{\textbf{p}}}}\label{Heff3}\\
&&\times\Big[\mathcal{P}\int\frac{dq_0}{2\pi}\frac{1}{\omega_{\textbf{p}}-q_0}
\left(\omega_{\textbf{p}}(B_{IJ}+B_{JI})+\textbf{p}(B_{IJ}-B_{JI})+M^2(A_{IJ}+A_{JI})
\right)\Big]\bigg].\nonumber
\end{eqnarray} 
Here the momentum dependence of $B_{IJ}(q_o,\textbf{p})$, $A_{IJ}(q_0,\textbf{p})$ has been suppressed and $\p$ has to be read as $|\p|$.
At temperatures $T\gg M$, the integral is dominated by hard momenta $\sim T$ and the term involving $B_{IJ}$ dominates $H_{IJ}$. 
For $T\lesssim v(T)$, the interaction with the Higgs condensate dominates the $N_I$ self energy and $B_{IJ}$ can be estimated as 
\begin{equation}
B_{IJ}(p)\simeq v(T)^2\tilde{F}_{\alpha I}\tilde{F}_{\alpha J}^*\frac{\pi}{|\textbf{p}|} b_{\alpha\beta}(p) \left(\delta(\omega-|\textbf{p}|+b)-\delta(\omega+|\textbf{p}|+b)\right) \ {\rm for} \ T\lesssim v(T)
.\end{equation}
Here $b$ is the so-called ``potential contribution'' to the active neutrino propagator \cite{Shaposhnikov:2008pf},
obtained by decomposing the retarded active neutrino self energy as  
\begin{equation}
{\rm Re}\Sigma^R_{\alpha\beta}(p)=a_{\alpha\beta}(p)\Slash{p}+b_{\alpha\beta}(p)\Slash{u}
.\end{equation}
Since active neutrinos mainly scatter via weak gauge interactions, the coefficients are in good approximation flavor independent in the primordial plasma, we can define $b\delta_{\alpha\beta}\equiv b_{\alpha\beta}(p)$, where $b_{\alpha\beta}(p)$ is to be evaluated on-shell.
For hard momenta, $b$ gives \cite{Weldon:1982bn,Notzold:1987ik}:
\begin{eqnarray}
b =
\left\{\begin{array}{c c}
-\frac{\pi\alpha_W T^2}{8p}
\left(2+\frac{1}{\cos^2\theta_W}\right)~,&~~~ T \gg M_W\\
\frac{16 G_F^2}{\pi\alpha_W}\left(2+\cos^2\theta_W\right)
\frac{7 \pi^2T^4p}{360}~,&~~~ T \ll M_W~.
\end{array}\right.~
\label{bdef}
\end{eqnarray}
Here $\theta_W$ is the Weinberg angle and $\alpha_W$ the weak gauge coupling.
This leads to a contribution to $H_{IJ}$ of the form 
\begin{equation}\label{Hamiltonian_highT}
v(T)^2\tilde{F}_{\alpha I}\tilde{F}_{\alpha J}^*\frac{\omega_\p+\p}{2\p}\frac{b\omega_\p}{M^2+2b\omega_\p}. \nonumber
\end{equation}
For $T>T_{EW}$, scatterings with Higgs bosons dominate and the hard thermal loop result 
$H\simeq \tilde{F}^\dagger\tilde{F} T/8 $ from \cite{Asaka:2005pn}  can be used. Combining the two contributions, we approximate $H$ by
\begin{equation}\label{BAUH}
H\simeq -\frac{M}{T} \Delta \tilde{M} + (\tilde{F}^\dagger \tilde{F})^* \left(\frac{T}{8}+\frac{v^2(T)}{T}\right)
\end{equation}
during the calculations in section \ref{baryogenesissection}. In practice it is more convenient to work in the $F$-base, where the Hamiltonian reads $-\sigma_3\frac{M}{T}\Delta M + F^\dagger F (\frac{T}{8}+\frac{v^2(T)}{T})$.

At temperatures $T\ll T_{EW}$, there are no Higgs particles in the plasma and the Higgs expectation value is constant, thus $B_{IJ}=v^2\tilde{F}_{\alpha I}^*\tilde{F}_{\beta J}b_{\alpha\beta}$, $A_{IJ}=v^2\tilde{F}_{\alpha I}^*\tilde{F}_{\beta J}a_{\alpha\beta}$. 
In \cite{Shaposhnikov:2008pf} it has been estimated that thermal corrections to the active neutrino propagator are small below a temperature 
\begin{equation}
T_{pot}=13\left(\frac{M}{{\rm GeV}}\right)^{\frac{1}{3}}{\rm GeV}
.\end{equation} 
For the masses under consideration in this work, we can approximately use free active neutrino propagators in section \ref{DMproductionSection} because $T_+<T_{pot}$. Furthermore, due to the considerations in section \ref{finetunings}, we are mainly interested in the case $U_N\simeq\mathbbm{1}$ for DM production, thus $\tilde{F}\simeq F$. 
Then $b_{\alpha\beta}(p_0,|\textbf{p}|)\simeq 0$ and $a_{\alpha\beta}(p_0,|\textbf{p}|)\simeq
(U_\nu)_{\alpha i}(U_\nu)_{\beta i}^*
\frac{\pi}{\omega_i}
(\delta(p_0-\omega_i)-\delta(p_0+\omega_i))$, with $\omega_i=(\textbf{p}^2+m_i^2)^{1/2}$ \cite{LeB}, where $m_i$ are the active neutrino masses. 
This recovers the vacuum result for the mass matrix at $|\textbf{p}|=0$, cf (\ref{sterileneutrinomasses}). 
 $H$ can be approximated by
\begin{equation}
 H=\uptau(\bar{p}^2+M_N^2)^{1/2}
.\end{equation}
In the basis of vacuum mass eigenstates
it has the form ${\rm H}={\rm diag}((\textbf{p}^2+M_2^2)^{1/2},(\textbf{p}^2+M_3^2)^{1/2})$. Since $\delta M\ll T, M$ we can expand in $\delta M$ and obtain for $\bar{p}=T$
\begin{equation}\label{hamiltonian_lowT}
H\simeq \uptau\phantom{i}  \delta M (X^{-2}+1)^{-1/2}\sigma_3,
\end{equation}
with $X=M/T$ and the third Pauli matrix $\sigma_3$. The part of $H$ that is proportional to the identity matrix has been dropped as it always cancels out of the commutators in the kinetic equations.

\subsubsection{Dissipative Part $\Gamma_N$}\label{DissipativePartSec}
Again comparing (\ref{Heff1}) and (\ref{Kineq1}), we define
\begin{eqnarray}
(\Gamma_N)_{IJ}&=&
\uptau\int d^3\textbf{p}\frac{f_F(\omega_{\textbf{p}})}{n_F}
\frac{1-2f_F(\omega_{\textbf{p}})}{2\omega_{\textbf{p}}} {\rm tr}\left({\rm P}_u \Pi^-_{IJ}(p)+{\rm P}_v\Pi^-_{JI}(p)\right) 
\label{GammaNFinalAppendix}
.\end{eqnarray}
The rate for the ``antiparticles''  $\rho_{\bar{N}}$ can be found by using projectors analogue to (\ref{ShortProj}), but with helicity index $_2$.
It is given by $(\Gamma_N^<)^*$ as expected\footnote{This can be seen by noticing that the traces are real and $P_RP_u^{22} P_L=P_R P_v^{11} P_L$, $P_R P_v^{22} P_L=P_R P_u^{11} P_L$ under the trace.}. 
For what follows, it is useful to pull the Yukawa matrices out of the self energies and define
\begin{equation}
\bar{\Pi}^{\gtrless}_{\alpha\beta}(p)=\int\frac{d^4k}{(2\pi)^4}\left(v^2\delta(p-k)+\Delta^\gtrless(p+k)\right)S^\lessgtr_{\alpha\beta}(k).\label{barPi}
\end{equation} 
Obviously $\Pi^\gtrless=\tilde{F}_{\alpha I}^*\tilde{F}_{\beta J}\bar{\Pi}^{\gtrless}_{\alpha\beta}$. 
For the computation of $\Gamma_N$ according to (\ref{rates1})
we can now define the matrices 
\begin{eqnarray}
R(T,M)_{\alpha\beta}&=&\int d^3\textbf{p}\frac{f_F(\omega_{\textbf{p}})}{n_F}
\frac{1-2f_F(\omega_{\textbf{p}})}{2\omega_{\textbf{p}}} {\rm tr}\left({\rm P}_u \bar{\Pi}^-_{\alpha\beta}(p)\right) \label{Rdef}\\
R_M(T,M)_{\alpha\beta}&=&\int d^3\textbf{p}\frac{f_F(\omega_{\textbf{p}})}{n_F}
\frac{1-2f_F(\omega_{\textbf{p}})}{2\omega_{\textbf{p}}} {\rm tr}\left({\rm P}_v\bar{\Pi}^-_{\alpha\beta}(p)\right) \label{RMdef}
.\end{eqnarray}
They in general have to be computed numerically. We discuss their properties in section \ref{ratecompsec}.

As usual in thermal field theory, the sterile neutrino self energies $\Pi^<$ and $\Pi^>$ can be associated with the gain and loss rate.
Their difference $\Pi^-=\Pi^>-\Pi^<$ gives the total relaxation rate $\Gamma_N$ for the sterile neutrinos. It acts as thermal production rate when their occupation numbers are below their equilibrium values and as dissipation rate in the opposite case. 
In configuration space, the self energy $\Pi^-(x)$ is related to the retarded self energy by $\Pi^R(x)=\theta(x_0)\Pi^-(x)$. 
This implies  $\bar{\Pi}^-(p)= 2i{\rm Im}\bar{\Pi}^R(p)$ in momentum space.
As usual in field theory, the imaginary part of $\bar{\Pi}^R$ can be related to the total scattering cross section by the optical theorem (or its finite temperature generalization) \cite{Bedaque:1996af}, while the real part is responsible for the mass shift (or modified dispersion relation in the plasma). Both are related by the Kramers-Kronig relations.
The appearance of $\Pi^-$ in (\ref{GammaNFinalAppendix}) is in accord with the optical theorem, and the contributions to the dispersive and dissipative parts of the effective Hamiltonian are indeed related by a Kramers-Kronig relation, cf (\ref{Heff2}) and (\ref{GammaNFinalAppendix}). This provides a good cross-check for our result.

\subsubsection{The remaining Rates}\label{OtherRates}
The remaining rates (\ref{rates2}) and (\ref{rates3}) appearing in (\ref{kinequ1})-(\ref{kinequ3}) in principle have to be calculated independently. The precise computation is considerably more involved than in the case of $\Gamma_N$. $\Gamma_N$ is related to the discontinuities of the $N_I$ self energies, which to leading order in the tiny Yukawa couplings $F_{\alpha I}$ only contain propagators of SM-fields as internal lines. Due to the fast gauge interactions these are in equilibrium in the relaxation time approximation and the RHS of (\ref{GammaNFinalAppendix}) can be computed by means of thermal (equilibrium) field theory. This is not possible in the computation of the damping rates for the SM-lepton asymmetries, which are related to self energies where the out of equilibrium fields $N_I$ appear as internal lines. For simplicity we follow the approach taken in \cite{Shaposhnikov:2008pf}, see section 6 therein, and use the symmetries of the $\nu$MSM in certain limits to fix the structure of the rates. 

We first consider the limit $(M_M)_{IJ}\rightarrow 0$, that is the absence of a Majorana mass term. In this case the ``total lepton number'' is conserved,
\begin{eqnarray}
0&=&\partial_\mu \left(\sum_I J_I^\mu + \sum_\alpha J_\alpha^\mu \right)_{M_M=0}\\
J_I^\mu&=&\bar{\nu}_{R,I}\gamma^\mu\nu_{R,I}\label{JI}\\
J_\alpha^\mu&=&\bar{\nu}_{L,\alpha}\gamma^\mu\nu_{L,\alpha}+\bar{e}_{L,\alpha}\gamma^\mu e_{L,\alpha}.\label{Jalpha}
\end{eqnarray}
To leading order in the small mixing $\theta_{\alpha I}$ this implies
\begin{equation}
\frac{d}{dt}\left({\rm tr}\rho_- + \sum_\alpha \mu_\alpha\right)_{M_M=0}\simeq 0\label{symmetry1}
.\end{equation}
This situation is in good approximation realized for $T\gg M$, when baryogenesis takes place - the total lepton number is not violated during this process and a non-zero baryon number is only realized because sphalerons couple exclusively to left handed fields.
Equation (\ref{symmetry1}) implies 
\begin{eqnarray}
\begin{tabular}{c c c c}
$\Gamma_N\simeq\sum_\alpha \tilde{\Gamma}_L^{\alpha}$ & & ${\rm tr}\tilde{\Gamma}_N^{\alpha}\simeq\Gamma_L^\alpha$ & for $(M_M)_{IJ}=0$.
\end{tabular}
\end{eqnarray}
Other interesting limits considered in \cite{Shaposhnikov:2008pf} include $F_{\alpha I}\rightarrow 0$ with fixed $\alpha$ for all $I$ (leading to conservation of $J_\alpha^\mu$) and  $ F_{\alpha I}\rightarrow 0$ with fixed $I$ for all $\alpha$ (leading to individual conservation of the combination $J_I^\mu+\sum_\alpha J_\alpha^\mu$ and the remaining current $J_{J\neq I}^{\mu}$). These limits allow to fix the basic structure of equations (\ref{rates2}), (\ref{rates3}). For a general choice of parameters some corrections of $\mathcal{O}[1]$ to these relations may be necessary, the determination of which we postpone until the precision of experimental data on the $\nu$MSM requires it.

\subsection{Uncertainties}\label{Uncertainties}
Our study is the most complete quantitative study of bounds on the $\nu$MSM parameter space from cosmology to date.
However, the various assumptions made in the derivation of the kinetic equations lead to uncertainties that may be of order one. 
These can be grouped into three categories:
\begin{itemize}
\item We only consider momentum averaged quantities. Since the sterile neutrinos can be far from thermal equilibrium, one in principle has to study the time evolution of each mode separately. Our treatment is a reasonable approximation as long as the kinetic equilibrium assumption (\ref{KinEqEq}) holds. A study of this aspect published in \cite{Asaka:2011wq} suggests that deviations from kinetic equilibrium are indeed only of order one .
\item The rates (\ref{rates2}) and (\ref{rates3}) have been calculated in a rather crude way in section \ref{OtherRates}, leading to another source of uncertainties of order one. In addition, a precise calculation of the BAU requires knowledge of the sphaleron rate throughout the electroweak transition. Including this is expected to yield a slightly bigger value for the BAU \cite{D'Onofrio:2012jk}.
\item Though they are matrix valued and allow to study flavor oscillations, the equations (\ref{kineq1})-(\ref{kineq3}) are of the Boltzmann type. They assume that the system can be described as a collection of (possibly entangled) individual particles that move freely between isolated scatterings and carry essentially no knowledge about previous interactions ("molecular chaos").
\end{itemize}
The first two issues can be fixed by more precise computations. However, with the current experimental data, order one uncertainties are small compared to the experimental and observational bounds on the model parameters.  The corrections will only slightly change the boundaries of the allowed regions in parameter space found in this work. We therefore postpone more precise calculation to the time when such precision is required from the experimental side.

In contrast to that, the third point is more conceptual. In a dense plasma, multiple scatterings, off-shell and memory effects may affect the dynamics.
The effect of these cannot be estimated within the framework of Boltzmann type equations, it requires a derivation from first principles that either confirms (\ref{kineq1})-(\ref{kineq3}) and allows to estimate the size of the corrections or replaces them by a modified set of equations.
In the past years, much progress has been made in the derivation of effective kinetic equations from first principles \cite{Sigl:1992fn,Konstandin:2004gy,Konstandin:2005cd,Buchmuller:2000nd,Prokopec:2003pj,Prokopec:2004ic,FillionGourdeau:2006hi,Lindner:2005kv,Lindner:2007am,DeSimone:2007rw,Anisimov:2008dz,Drewes:2010pf,Anisimov:2010aq,Gagnon:2010kt,Garbrecht:2010sz,Beneke:2010dz,Garny:2009qn,Beneke:2010wd,Anisimov:2010gy,Anisimov:2010dk,Kiessig:2010pr,Kiessig:2011fw,Kiessig:2011ga,Garny:2009rv,Cirigliano:2009yt,Garny:2010nz,Herranen:2011zg,Herranen:2010mh,Laine:2011pq,Garbrecht:2011aw,Garny:2011hg,Drewes:2012qw,Garbrecht:2012pq}. Recent studies suggest that kinetic equation of the Boltzmann type are in principle applicable to study leptogenesis \cite{Anisimov:2010aq,Garbrecht:2011aw,Garny:2011hg,Drewes:2012qw}, but the resonant amplification may be weaker than found in the standard Boltzmann approach \cite{Garny:2011hg}. It remains to be seen which effects possible corrections have in the $\nu$MSM, where baryogenesis and dark matter production both crucially rely on the resonant amplification. A first principles study is difficult in the $\nu$MSM due to the various different time dependent scales related to production, oscillations, freezeout and decay of the sterile neutrinos and the vast range of relevant temperatures. However, at this stage it seems likely that a first principles treatment is, if at all, only of phenomenological interest in the region around $\Gamma_N\sim\delta M$, which makes up only a small fraction of the relevant parameter space, cf. figure \ref{BA_Max_contour}.

\section{Connection to Pseudo-Dirac Base}
\label{DiracConnection}
In our notation, the elements of $\rho_N$ and $\rho_{\bar{N}}$ are bilinears in ladder operators that create quanta of the fields $N_I$, i.e. mass eigenstates in vacuum. This has the advantage that the diagonal elements can be interpreted as abundances of physical particles. 
The rates $R$ and $R_M$ have been introduced in \cite{Shaposhnikov:2008pf}, to which we regularly refer in this article. The basis in the field space right handed neutrinos there differs from the one we use in (\ref{nuMSM_lagrangian}) and corresponds to $U\nu_R$ with 
\begin{equation}
U=\frac{1}{\sqrt{2}}\left(
\begin{tabular}{c c}
$i$ & $1$\\
$-i$ & 1
\end{tabular}
\right)
.\end{equation} 
In this basis $M_M$ is not diagonal and the Yukawa couplings should be rewritten as $hU=F$. The computation of the rates is then performed by defining a Dirac-spinor
\begin{equation}
\Psi=U_{2 I}\nu_{R,I}+\left(U_{3 I}\nu_{R,I}\right)^c
.\end{equation}
This is possible when the small mass splitting between the sterile neutrinos is neglected (or viewed as a perturbation and placed in the interacting part of $\mathcal{L}$). 
The fields $\nu_{R,I}$ can be recovered from this as $\nu_{R,I}=U^*_{2I}P_R\Psi+U^*_{3I}P_R\Psi^c$. In terms of $\Psi$, the $\nu$MSM Lagrangian  reads 
\begin{eqnarray}
\mathcal{L}&=&\mathcal{L}_{SM}+\mathcal{L}_0+\mathcal{L}_{int}\\
\mathcal{L}_0&=&\bar{\Psi}(i\slashed{\partial}-\bar{M})\Psi\nonumber\\
\mathcal{L}_{int}&=&-v\left(h_{\alpha 3}\bar{\nu}_{L \alpha}\Psi^c-h_{\alpha 3}^*\bar{\Psi^c}\nu_{L \alpha}-h_{\alpha 2}\bar{\nu}_{L \alpha}\Psi-h_{\alpha 2}^*\bar{\psi}\nu_{L \alpha}\right)\nonumber\\
&&-\frac{1}{2}\Delta M\left(\bar{\Psi^c}\Psi+\bar{\Psi}\Psi^c\right).
\end{eqnarray}
The analogue of our matrix $\rho_N$ (which is also called $\rho_N$ in \cite{Shaposhnikov:2008pf}) is defined as 
\begin{equation}
\rho_N^\Psi=\int\frac{d^3\textbf{p}}{V(2\pi)^3}
\left(\begin{tabular}{c c}
$\langle c^\dagger_{1} c_{1} \rangle$ & $\langle c^\dagger_{1} d_{1} \rangle$\\
$\langle d^\dagger_{1} c_{1} \rangle$ & $\langle d^\dagger_{1} d_{1} \rangle$
\end{tabular}\right),
\end{equation}
where $c_s$ ($c_s^\dagger$) and $d_s$ ($d_s^\dagger$) are the annihilation (creation) operators for $\Psi$ particles (antiparticles) with momentum $\textbf{p}$ and helicity $s$. 
The corresponding rate $\Gamma_N^\Psi$ in the kinetic equations is given by\footnote{Note that there were errors in the corresponding expressions (5.19), (5.20) in \cite{Shaposhnikov:2008pf} and that there only the case $T\gg M$ was considered.}
\begin{eqnarray}
\Gamma_N^{\Psi}&=&
\uptau\int d^3\textbf{p}\frac{f_F(\omega_{\textbf{p}})}{n_F}v^2\frac{f_F(p_{0})}{2\omega_{\textbf{p}}}\bigg(h_{\alpha I} h_{\beta J}^* R(T,M)_{\alpha\beta}
+(\sigma_1h^\dagger)_{I \alpha}(h\sigma_1)_{\beta J} R_M(T,M)_{\alpha\beta} \bigg),
\end{eqnarray}
where $h=F U^\dagger$,  $\sigma_1$ is the first Pauli matrix and we have neglected flavor off-diagonal elements. In the high temperature regime that was considered in \cite{Shaposhnikov:2008pf} this simplifies to
\begin{eqnarray}
\Gamma_N^{\Psi}\uptau^{-1}= (h^\dagger h)^* R^{(S)}(T,M) + \sigma_1h^\dagger h\sigma_1R_M^{(S)}(T,M) .\label{RdefPhi}
\end{eqnarray}
$\sigma_1h^\dagger h\sigma_1$ is $(h^\dagger h)^*$ with the diagonal elements swapped. The equation (\ref{Rdef}) defines the quantities $R(T,M)$ and $R_M(T,M)$\footnote{Our definitions of $R$ and $R_M$ differ from those in \cite{Shaposhnikov:2008pf} by a constant factor $F_0^2$ with $F_0=2\times 10^{-9}$.}. Ignoring the small mixing between active and sterile neutrinos,  $\Gamma_{N}$ is related to $\Gamma_N^\Psi$ by
\begin{equation}
\Gamma_N\simeq (U U_N)^T\Gamma_N^\Psi(U U_N)^*
\end{equation}
for $T\gg m_i$.

Finally, the Yukawa matrix in \cite{Shaposhnikov:2008pf} and \cite{Canetti:2010aw} is expressed in terms of the parameters $\epsilon_{di}$, $\eta_{di}$, $\phi_{di}$, which differ from those we use here. In the limit $\epsilon_{di} >> 1$, these can be related to our parameters by
\begin{equation}
	\sqrt{\epsilon_{di}} = e^{-\Im\upomega},\ \eta_{di} = 2 \Re\upomega,\ \alpha_{di} = \frac{\alpha_2}{2},\ \phi_{di} = \delta.\label{OldParam}
\end{equation}
We here prefer to use the parameterization fixed in section \ref{Parametrisationsection} because the expressions in therms of (\ref{OldParam}) given in \cite{Shaposhnikov:2008pf} are only approximate. 
\section{How to characterize the lepton asymmetries}\label{asymmconv}
In the $\nu$MSM neither the individual lepton numbers, related to the currents (\ref{Jalpha}), nor their sum are conserved. However, since the rates of all processes that violate them are suppressed by the small Yukawa couplings $F$, they evolve on a much slower time scale than other processes in the primordial plasma and are well-defined. 
For practical purposes the magnitude of flavoured lepton asymmetries in the primordial plasma can be characterized in different ways.
In this article we describe them by the ratio between the number densities (particles minus antiparticles) and the entropy density $
 s \equiv {2 \pi^2 T^3} g_* / {45} 
$,  
\begin{equation}
\mu_\alpha=\frac{n_\alpha}{s} \ \ \rm{cf.} \ \ (\ref{muDef}) \nonumber
.\end{equation}
This quantity is convenient because it is not affected by the expansion of the universe as long as the expansion is adiabatic.
In the following we relate $\mu_\alpha$ to other quantities that are commonly used in the literature, using the relations given in the appendix of \cite{Laine:2008pg}. 

In quantum field theory calculations it is common to parameterize the asymmetries by chemical potentials $\upmu_\upalpha$, which can be extracted from the distribution functions that appear in the free propagators at finite temperature. In the massless limit $T\gg m_i$ these are related to $n_\alpha$ by
\begin{equation}
 n_\alpha = \frac{\upmu_\upalpha T^2}{6} + \frac{\upmu_\upalpha^3}{6\pi^2},
\end{equation}
leading to 
\begin{equation}
 \mu_\alpha \approx \frac{15}{4\pi^2 g_*}
 \frac{\upmu_\upalpha}{T}.
\end{equation}

Alternatively one can normalize the lepton numbers $n_\alpha$ (``particles minus antiparticles'') by the total density of ``particles plus untiparticles'' in the plasma,
\begin{eqnarray}
 \Delta_\alpha  
 = \frac{n_\alpha}{{\rm n}_\alpha^{eq}}
\end{eqnarray}
where 
${\rm n}_\alpha^{eq} \equiv 2 \int {\rm d}^3\textbf{q} /(2\pi)^3/(e^{|\textbf{q}|/T}+1) = 3\zeta(3)T^3/2\pi^2
$
and 
\begin{equation}
 \mu_\alpha = \frac{135\zeta(3)}{4\pi^4 g_*}
 \Delta_\alpha.
\end{equation}
Finally, one can normalize with respect to the photon density, 
\begin{equation}
 L_\alpha \equiv \frac{n_\alpha}{n_\gamma}
\end{equation}
where 
$
 n_\gamma
 \equiv 2 \int {\rm d}^3\textbf{q} /(2\pi)^3/(e^{|\textbf{q}|/T}-1) = 2 \zeta(3)T^3/\pi^2
$, 
which yields
\begin{equation}
 \mu_\alpha = \frac{45\zeta(3)}{\pi^4 g_*}
  L_\alpha.
\end{equation}

\section{Low temperature Decay Rates for sterile Neutrinos}
\label{DecayRatesAppendix}
Most of the rates relevant for this work have been computed in \cite{Gorbunov:2007ak}, where they are listed in the appendix. 
Here we only list those rates that are needed in addition to those or require refinement. This was necessary for the decay rates into leptons, where masses of the final state particles had been neglected in the original computation.
\subsection{Semileptonic decay}
Decay into up-type quarks through neutral current :
\begin{align}
\Gamma_{N_I \rightarrow \nu_{\alpha} u \overline {u}}=\frac{G_F^2 \left|\Theta_{\alpha I}\right|^2 M^5}{192\pi^3} \Bigg( f^{(u)}(x_q)S(x_q,x_q)+  
&x_q^4 \left( 3-\frac{16}{3}C_1 x_q^2+(3-8C_1)x_q^4 \right) \notag \\ &\times \log\left[\frac{1-4x_q^2+2x_q^4+S(x_q,x_q)(1-2x_q^2)}{2x_q^4}\right] \Bigg)
\end{align}
where $x_q=m_q/M$.

Decay into down-type quarks through neutral current :
\begin{align}
\Gamma_{N_I \rightarrow \nu_{\alpha} d \overline {d}}=\frac{G_F^2 \left|\Theta_{\alpha I}\right|^2 M^5}{192\pi^3}\Bigg( f^{(d)}(x_q)S(x_q,x_q)+& x_q^4\left(3-\frac{8}{3}C_2x_q^2-(1-\frac{4}{3}C_2)x_q^4\right) \notag \\ &\times \log\left[ \frac{1-4x_q^2+2x_q^4+S(x_q,x_q)(1-2x_q^2)}{2x_q^4}\right]\Bigg)
\end{align}

Decay into quarks through charged current :
\begin{align}
\Gamma_{N_I \rightarrow e_{\alpha} u_n \overline{d}_m}=&\frac{G_F^2\left|V_{nm}\right|^2 \left|\Theta_{\alpha I} \right|^2 M^5}{192 \pi^3} \Bigg( g(x,y)S(x,y) \notag   \\ & -12x^4\log \left[\frac{1-S(x,y)(1+x^2-y^2)-2y^2+(x^2-y^2)^2}{2x^2}\right] \notag \\ &-12y^4\log\left[\frac {1-S(x,y)(1-x^2+y^2)-2x^2+(x^2-y^2)^2}{2y^2} \right]\notag\\
& +12x^4y^4\log \left[\frac{1-2x^2-2y^2+x^4+y^4-S(x,y)(1-x^2-y^2)}{2x^2y^2}\right] \Bigg)
\end{align}
where $\min(m_\alpha,m_{u_n},m_{d_m})$ is neglected, and $x$ and $y$ are the two heavier masses divided by $M$.

\subsection{Leptonic decay}
\begin{align}
\Gamma_{N_I \rightarrow e^-_{\alpha\neq\beta} e^+_{\beta} \nu_\beta}= & \frac{G_F^2 \left|\Theta_{\alpha I}\right|^2M^5}{192\pi^3}\Bigg( S(x_\alpha,x_\beta)g(x_\alpha,x_\beta) \notag  \\ & -12x_\alpha^4\log \left[ \frac{1-S(x_\alpha,x_\beta)(1+x_\alpha^2-x_\beta^2)-2x_\beta^2+(x_\alpha^2-x_\beta^2)^2}{2x_\alpha^2}\right] \notag \\
&-12x_\beta^4\log\left[\frac {1-S(x_\alpha,x_\beta)(1-x_\alpha^2+x_\beta^2)-2x_\alpha^2+(x_\alpha^2-x_\beta^2)^2}{2x_\beta^2} \right]\notag\\
& +12x_\alpha^4x_\beta^4\log \left[\frac{1-2x_\alpha^2-2x_\beta^2+x_\alpha^4+x_\beta^4-S(x_\alpha,x_\beta)(1-x_\alpha^2-x_\beta^2)}{2x_\alpha^2x_\beta^2}\right] \Bigg)
\end{align}

with

\begin{align*}
S(x,y)&=\sqrt{(1-(x+y)^2)(1-(x-y)^2)} \\
C_1&=\operatorname{sin}^2\theta_W\left( 3-4\sin^2\theta_W\right) \\
f^{(u)}(x)&=\frac{1}{4} -\frac{2}{9}C_1 -\left( \frac{7}{2}-\frac{20}{9}C_1 \right)x^2-\left(\frac{1}{2}+4C_1\right)x^4 +\left( -3+8C_1\right)x^6 \\
C_2&=\sin^2\theta_W(3-2\sin^2\theta_W) \\
f^{(d)}(x)&=\frac{1}{4}-\frac{1}{9}C_2+(\frac{10}{9}C_2-\frac{2}{7})x^2-(\frac{1}{2}+2C_2)x^4-(3-4C_2)x^6 \\
g(x,y)&=1-7x^2-7y^2-7x^4-7y^4+12x^2y^2-7x^2y^4-7x^4y^2+x^6+y^6 .\\
\end{align*}

\end{appendix}
\bibliographystyle{JHEP}
\bibliography{all}

\end{document}